\documentclass[fleqn,usenatbib]{mnras}

\usepackage{newtxtext,newtxmath}

\usepackage[T1]{fontenc}
\usepackage{ae,aecompl}

\usepackage{graphicx}	
\usepackage{amsmath}	
\usepackage{multicol}
\usepackage{subfig}
\usepackage{array}
\usepackage{xcolor}
\usepackage{hyperref}

\newcommand{\PreserveBackslash}[1]{\let\temp=\\#1\let\\=\temp}
\newcolumntype{C}[1]{>{\PreserveBackslash\centering}p{#1}}
\newcommand{\arcs}{\hbox{$^{\prime\prime}$}}

\title[Dust RM and LC Modelling of \mbox{Zw229-015}]{Dust Reverberation Mapping and Light-Curve Modelling of \mbox{Zw229-015}}

\author[E. Guise et al.]{
E. Guise,$^{1}$\thanks{E-mail: ella.guise@soton.ac.uk}
S. F. H{\"o}nig,$^{1}$
V. Gorjian,$^{2}$
A. J. Barth,$^{3}$
T. Almeyda,$^{1,4}$
L. Pei,$^{3}$
S. B. Cenko,$^{5,6}$
R. Edelson,$^{7}$
\newauthor
A. V. Filippenko,$^{8}$,
M. D. Joner,$^{9}$
C. D. Laney,$^{9,10}$
W. Li,$^{8,11}$
M. A. Malkan,$^{12}$
M. L. Nguyen,$^{13}$
W. Zheng$^{8}$
\newauthor
\\
$^{1}$Department of Physics and Astronomy, University of Southampton, Southampton, SO17 1BJ, UK \\
$^{2}$Jet Propulsion Laboratory, M/S 169-327, 4800 Oak Grove Drive, Pasadena, CA 91109, USA \\
$^{3}$Department of Physics and Astronomy, 4129 Frederick Reines Hall, University of California, Irvine, CA 92697-4575, USA \\
$^{4}$Department of Biological and Physical Sciences, South Carolina State University, Orangeburg, SC 29117, USA\\ 
$^{5}$Astrophysics Science Division, NASA Goddard Space Flight Center, MC 661, Greenbelt, MD 20771, USA \\
$^{6}$Joint Space-Science Institute, University of Maryland, College Park, MD 20742, USA \\
$^{7}$Department of Astronomy, University of Maryland, College Park, MD 20742-2421, USA \\
$^{8}$Department of Astronomy, University of California, Berkeley, CA 94720-3411, USA \\
$^{9}$Department of Physics and Astronomy, N283 ESC, Brigham Young University, Provo, UT 84602, USA \\
$^{10}$Department of Physics and Astronomy, Western Kentucky University, 1906 College Heights Boulevard, Bowling Green, KY 42101, USA \\
$^{11}$Deceased 12 December 2011 \\
$^{12}$Department of Physics and Astronomy, University of California, Los Angeles, CA 90095-1547, USA \\
$^{13}$Department of Physics and Astronomy, University of Wyoming, Laramie, WY 82071, USA \\
}

\date{Accepted XXX. Received YYY; in original form ZZZ}

\pubyear{2022}

\begin{document}
\label{firstpage}
\pagerange{\pageref{firstpage}--\pageref{lastpage}}
\maketitle


\begin{abstract}

Multiwavelength variability studies of active galactic nuclei (AGN) can be used to probe their inner regions which are not directly resolvable. Dust reverberation mapping (DRM) estimates the size of the dust emitting region by measuring the delays between the infrared (IR) response to variability in the optical light curves. We measure DRM lags of \mbox{Zw229-015} between optical ground-based and \textit{Kepler} light curves and concurrent IR \textit{Spitzer} 3.6 and 4.5~$\mu$m light curves from 2010--2015, finding an overall mean rest-frame lag of {\color{black}18.3 $\pm$ 4.5 days}. Each combination of optical and IR light curve returns lags that are consistent with each other within 1$\sigma$, which implies that the different wavelengths are dominated by the same hot dust emission. The lags measured for Zw229-015 are found to be consistently smaller than predictions using the lag-luminosity relationship. Also, the overall IR response to the optical emission actually depends on the geometry and structure of the dust emitting region as well, so we use Markov chain Monte Carlo (MCMC) modelling to simulate the dust distribution to further estimate these structural and geometrical properties. We find that a large increase in flux between the 2011--2012 observation seasons, which is more dramatic in the IR light curve, is not well simulated by a single dust component. When excluding this increase in flux, the modelling consistently suggests that the dust is distributed in an extended flat disk, and finds a mean inclination angle of \color{black}49$^{+3}_{-13}$ degrees. 

\end{abstract}




\section{Introduction}

Active galactic nuclei (AGN) are the very luminous central regions of galaxies that are powered by accretion onto a supermassive black hole (SMBH). Their emission is widely detectable over the entire observable electromagnetic spectrum and is variable over different timescales, including intraday variability ($\sim$~minutes to hours), short-term variability ($\sim$~days to a few months), and long-term variability (several months to years).

Initially, it was proposed that the infrared (IR) emission solely originated in a circumnuclear dusty region, often referred to as the "dusty torus," that surrounds the accretion disk (AD) and SMBH. The torus was originally introduced to unify AGN through its obscuration of optical emission depending on the orientation of the AGN with respect to the observer \citep[e.g.,][]{Antonucci1993}, where type 1 and type 2 AGN are respectively unobscured and obscured. This unification model has been successful in explaining many observed properties of AGN, however there has been evidence provided since that suggests the classical circumnuclear torus model is not solely able to explain the observed behaviour of all AGN. For example, a torus-like structure might be missing from AGN that are observed with little to no obscuration detected but are still missing the broad optical emission lines (dubbed "true" type 2 AGN), though such unobscured type 2 AGN are believed to be rare objects \citep{Shi2010}. Furthermore, a second component which takes the form of an extended polar structure has also been observed to contribute to the overall IR emission \citep[e.g.,][]{Raban2009, Honig2012, Honig2013, Leftley2018}. \cite{Honig2012} suggest that the polar dust structure is the result of a radiatively driven dusty wind from the inner region of the torus. IR interferometry has been used to show that the mid-IR (MIR) emission predominantly originates in this polar region \citep[e.g.,][]{Raban2009, Honig2012, Honig2013, Leftley2018}, and the hottest dust closest to the dust sublimation region in the equatorial disk component dominates the near-IR (NIR) emission \citep[e.g.,][]{Swain2003, Kishimoto2009b, Pott2010, Weigelt2012}. However, IR interferometry is only possible for nearby, relatively bright AGN, and therefore alternative indirect methods are necessary to study the location of the IR emission in the inner regions of more AGN.

Reverberation mapping uses the assumption that changes in the variability of the emission from shorter wavelength ranges can drive changes in longer wavelength emission, therefore allowing the size of different components in the AGN to be estimated from time delays between variability in the light curves \cite[e.g.,][]{Blandford1982, Peterson1993}. The inner radius of the dust emitting region, which is set by the dust sublimation temperature, can be estimated via dust reverberation mapping (DRM) as some of the optical emission produced in the AD is reprocessed by the surrounding dust and reemitted in the IR after a delay corresponding to the light-travel time between the AD and dust emitting region \citep[e.g.,][]{Barvainis1987}. Several studies of the dust reverberation radii of different AGN have been performed, most using the NIR $K$ band ($\sim$ 2.2\,$\mu$m) to define the dust sublimation radius of AGN with redshift $z < 1$ \citep[e.g.,][]{Nelson1996, Minezaki2004, Suganuma2006, Koshida2014, Yoshii2014, Nunez2015, Mandal2018, Ramolla2018, Minezaki2019, Mandal2020} as it traces the peak of the hot dust emission at the inner radius with a dust sublimation temperature $T_\text{sub} \approx 1500$\,K \citep[e.g.,][]{Rieke1981, Barvainis1987}. Additionally, studies of DRM using longer wavelength IR light curves have been performed to further constrain the spatial information of the dust emitting region; for example, \cite{Lyu2019} combine MIR DRM lags from multiple quasars with $z < 0.5$ using WISE W1 ($\sim 3.4$\,$\mu$m) and W2 ($\sim 4.5$\,$\mu$m) bands with $K$-band results from the literature to infer the dust emission size ratios of \mbox{$R_k:R_{W1}:R_{W2}~=$~0.6 : 1 : 1.2}. Similarly, \cite{Vazquez2015} performed DRM on the optical $B$ and $V$ and IR \textit{Spitzer} channel 1 (3.6\,$\mu$m) and \textit{Spitzer} channel 2 (4.5\,$\mu$m)  light curves of NGC~6418, and measured lags between the optical continuum and the IR 3.6\,$\mu$m and 4.5\,$\mu$m with values of $37.2^{+2.4}_{-2.2}$ and $47.1^{+3.1}_{-3.1}$ days, respectively. They also measured a lag between the IR 3.6\,$\mu$m and 4.5\,$\mu$m light curves with a value of $13.9^{+0.5}_{-0.1}$ days. These studies find that the emission detected in the longer IR wavelengths of these AGN is likely dominated by the black-body peak of the emission from the cooler dust at larger radii; however, contributions to the MIR emission are also expected from the Rayleigh-Jeans tail of the hotter dust from the inner regions. In compact systems that lack the extended dust, it is therefore suggested by \cite{Honig2011} that the emission at these wavelength ranges could predominately come from the hot dust Rayleigh-Jeans tail.

While DRM gives a good indication of the size of the dust emitting region, the overall delayed IR response to the optical emission actually depends on the structure and geometry of the dust emission region. While initially a smooth distribution of dust was assumed, it is now suggested to be distributed in clumps, both theoretically  \citep[e.g.,][]{Krolik1988,Tacconi1994} and with evidence from observations \citep[e.g.,][]{Shi2006,Tristram2007}. By modelling this clumpy dust distribution, the overall reprocessed IR light curves can be simulated. The IR response to a delta-function input continuum pulse represents the dust transfer function (DTF), which can be convolved with the driving optical light curve to give the portion of the IR light that corresponds to the reemitted optical continuum. Though the DTF is not directly measurable, simulated DTFs can be convolved with optical light curves and the resulting simulated IR light curve can be compared with the IR observations to constrain properties, including the dust distribution, inclination angle, and dust reverberation radius of the AGN. Several such models of the dust emitting region have been used to constrain the inner regions of AGN; for example, \cite{Honig2011} showed that the observed NIR $K$-band light curves of NGC~4151 could be well reproduced using observed optical $V$-band light curves convolved with simulated DTFs. Furthermore, \cite{Almeyda2017, Almeyda2020} modelled the IR response light curves to variations in the UV/optical using a torus reverberation mapping code (TORMAC), to explore geometrical and structural properties of the dusty torus.

Zw229-015\footnote{$\alpha = 19^{\rm h}05^{\rm m}25.939^{\rm s}$, $\delta = +42^\circ 27' 39.65''$ (J2000)} is a nearby Seyfert 1 Galaxy at $z = 0.028$ \citep{Falco1999}. It is one of the brightest and most highly varying AGN in the \textit{Kepler} field, and was therefore selected as a target to be observed with \textit{Kepler} during Quarters 4--17 (2010--2013), resulting in one of the highest quality, most complete optical light curves of any AGN. Concurrent IR observations with \textit{Spitzer} (2010--2015), along with additional optical observations from ground-based telescopes (2010--2015), make it ideal for dust reverberation mapping analysis over 5\,yr. Multiple previous studies of \mbox{Zw229-015} have been performed. \cite{Barth2011} measured a broad-line region (BLR) reverberation lag of $3.86^{+0.69}_{-0.24}$ days between the H$\beta$ and $V$-band continuum light curves from 2010, and estimated a black hole mass of $\sim 10^7$\,M$_{\odot}$. \cite{Smith2018}
report a bolometric luminosity of log($L_\text{Bol}$\,[erg\,s$^{-1}$]) = 44.11 and an Eddington ratio of $L/L_\text{Edd} = 0.125$. \cite{Williams2018} modelled the BLR geometry and dynamics of \mbox{Zw229-015} to estimate an inclination angle of the AD of $i = 32.9^{+6.1}_{-5.2}$ degrees. Similarly, \cite{Raimundo2020} estimated an inclination angle of $i = 36.4^{+6.7}_{-6.4}$ degrees, consistent with \cite{Williams2018} within the 1$\sigma$ uncertainties. Furthermore, \cite{Mandal2020} performed dust reverberation mapping of \mbox{Zw229-015} with observations between July 2017 and December 2018, finding significant rest-frame lags between the $V$ and $K_s$ light curves of $20.36^{+5.82}_{-5.68}$ days.

The structure of this paper is as follows. In Section~\ref{Sect:Data} we describe the observations and data reduction. We present the dust reverberation mapping analysis of \mbox{Zw229-015} in Section~\ref{Sect:DustRM}, and in Section~\ref{Sect:IRModelling} we use MCMC modelling of the light curves to further constrain the dust reverberation lag and the geometry of the inner regions of the AGN. Section~\ref{Sect:Discuss} discusses the results, and in Section~\ref{Sect:Concl} we present the summary and conclusions.

\section{Observations and Data Reduction}
\label{Sect:Data}

\begin{figure*}
    \begin{minipage}{\textwidth}
        \centering
        \includegraphics[width=\textwidth]{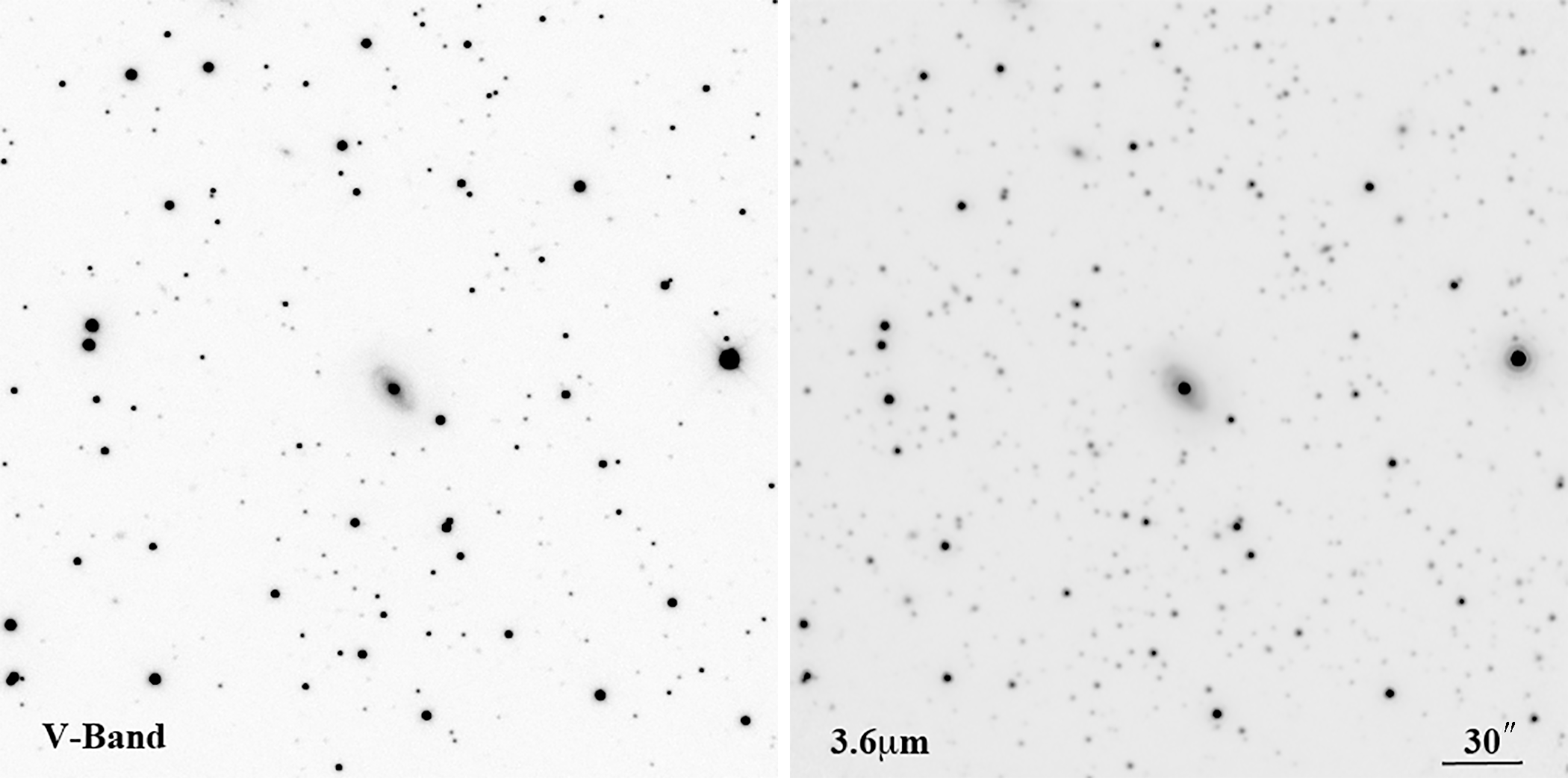}
        \caption{\centering Stacked images of \mbox{Zw229-015} from ground-based optical ($V$-band) imaging from the West Mountain Observatory \citep{Barth2011} and the IR (3.6\,$\mu$m) image from the {\it Spitzer Space Telescope.}}
        \label{fig:zw229images}
    \end{minipage}
\end{figure*}

\subsection{Ground-Based Optical}

Ground-based $V$-band photometric monitoring of Zw229-015 was carried out from 2010 June until 2014 December. \citet{Barth2011} presented the 2010 data, while the first five months of the 2011 data were included in the light-curve compilation presented by \citet{Pancoast2019}. For this work, we carried out measurements of the 2010--2014 photometric data, including new measurements for the images used in these earlier studies, in order to ensure uniformity across the full time series.

The data included here were obtained from four facilities: (1) The 0.9\,m telescope of the BYU West Mountain Observatory (WMO) in Utah, (2) the 0.76\,m Katzman Automatic Imaging Telescope (KAIT) at Lick Observatory \citep{Filippenko2001}, (3) the 1\,m telescopes of the Las Cumbres Observatory (LCO) located at McDonald Observatory \citep{Brown2013}, and (4) the 2\,m Faulkes Telescope North (also a part of the LCO network) located at Haleakala Observatory using the Spectral camera. Exposure times ranged from 60\,s to 300\,s for individual integrations, and two or more exposures per night were usually taken at a given facility in order to increase the signal-to-noise ratio (S/N) and facilitate cosmic-ray rejection. The previously published measurements of the 2010 data \citep{Barth2011} included a small number of additional observations obtained at other facilities, but those images were not included for the light-curve measurements presented here.

Images were processed by the standard reduction pipeline for each facility, including bias subtraction and flat-fielding. Photometry was carried out using the IDL-based aperture photometry pipeline described by \citet{Pei2014}. This procedure automatically identifies the position of the AGN and a set of comparison stars in each image based on their coordinates, and uses routines in the IDL Astronomy Users' Library \citep{Landsman1995} to measure their instrumental magnitudes. Photometry was obtained in a circular aperture of radius $4''$, with a sky annulus between $10''$ and $20''$. For each image, a shift was applied to the measured instrumental magnitudes in order to minimise the scatter in the comparison star light curves. When multiple exposures were taken on a single night at a given facility, the resulting measurements were combined using a weighted average to obtain a single photometric data point. Small discrepancies between the magnitude scales of the different telescopes were removed by designating the WMO light curve as the reference, and applying additive offsets to the light curves of Zw 229-015 from the other telescopes to yield the best average match to the WMO magnitudes on nights with observations in common between WMO and the other facilities. Finally, the light curve was placed on a calibrated $V$ magnitude scale (in the Vega system) using the comparison-star calibrations from \citet{Barth2011}. Since the photometry was carried out using different methods than those applied for the previous measurements of the 2010 and 2011 data, there are small differences between the light-curve data presented here and the measurements previously presented by \citet{Barth2011} and \citet{Pancoast2019}.

\subsection{\textit{Kepler}}

The pipeline processed {\it Kepler} light curves are not appropriate for AGN monitoring  \citep[e.g.,][]{Edelson2014,Kasliwal2015}. This is broadly due to the fact that the {\it Kepler} pipeline processing is optimized for small flux changes due to exoplanet transits and so tends to minimize large scale variations that can appear in AGN light curves, and so separate reductions need to be undertaken for AGN. We have used the light curve produced by \citet{Edelson2014} which is extracted directly from the two dimensional pixel data using different extraction masks than the pipeline. This process gives a significantly different result than the pipeline. For details of the process please see \citet{Edelson2014}. The $\sim$ 30 minute cadences were of much finer resolution than was necessary for the upcoming analysis, so the multiple measurements from each epoch were combined to provide a single photometric data point. The following sections show that the {\it Kepler} light curve matches the ground-based optical light curve very well both in direct comparison and the reverberation mapping results.

\begin{figure*}
    \begin{minipage}{\textwidth}
        \centering
        \includegraphics[width=\textwidth]{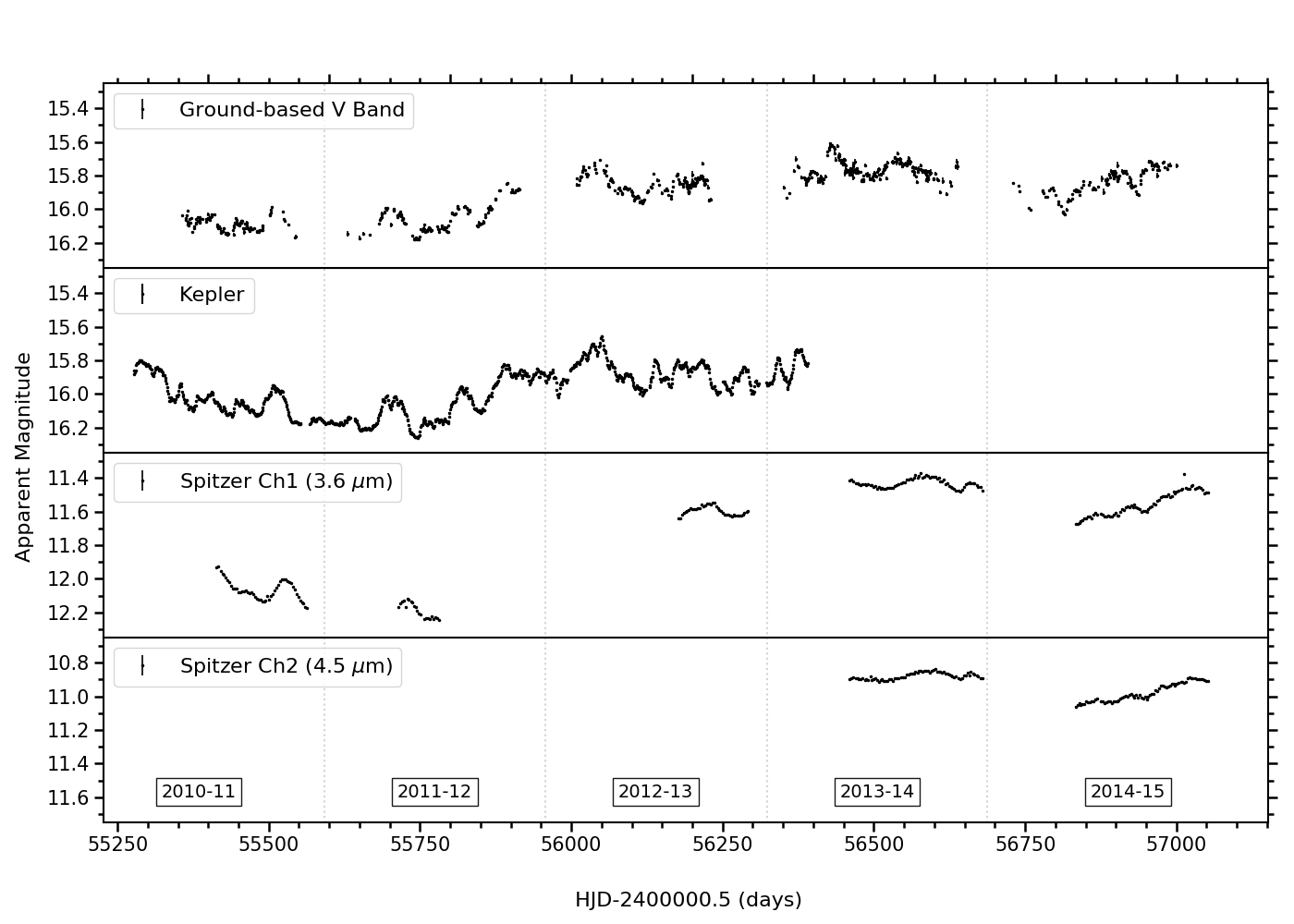}
        \caption{\centering Light curves of \mbox{Zw229-015} from the ground-based optical telescopes and \textit{Kepler}, and from IR \textit{Spitzer} Channel 1 (3.6\,$\mu$m) and \textit{Spitzer} Channel 2 (4.5\,$\mu$m). The points represent observations, and the grey dotted lines separate each observation season.}
        \label{fig:zw229lc}
    \end{minipage}
\end{figure*}

\subsection{\textit{Spitzer}}

The {\it Spitzer Space Telescope} \citep{Werner2004} observations were taken with the Infrared Array Camera (IRAC) \citep{fazio2004} scheduled between August 6, 2010 and January 26, 2015 (UTC) with an average spacing of 3 days between observations; moreover, observations were not taken $< 2.5$ days apart nor $> 3.5$ days apart. These limits were set to allow maximum flexibility of scheduling for {\it Spitzer's} other programs; a narrower time window would force an ``either/or"  choice by the schedulers, thereby endangering the continuous coverage of Zw229-015.
   
For the first two epochs of data  (PID 70119, 80148), we requested data only at the 3.6\,$\mu$m band, thinking that the 4.5\,$\mu$m-band data would show variability on a longer timescale than our observing window. We then added the 4.5\,$\mu$m data in the third and fourth epochs (PID 90144, 10125), as preliminary analysis of the time variability at the shorter wavelength showed the variability timescales to be significantly less than our observing window.
   
In order to attain the highest precision in the light-curve data, we followed the procedures in the IRAC  Instrument Handbook \footnote{\url{ https://irsa.ipac.caltech.edu/data/SPITZER/docs/irac/ }} to reduce interpixel and intrapixel systematic errors.  Based on analysis of standard stars used for calibration, it was determined that having at least 16 dithers with IRAC at the medium spacing would provide $\sim 1$\%  photometry for a large field around the AGN. 
   
Figure~\ref{fig:zw229images} shows a stacked image obtained at 3.6\,$\mu$m by {\it Spitzer} and another at $V$ obtained by the WMO and first presented by \citet{Barth2011}. IR photometry was acquired in a 4 pixel (2\farcs4) radius aperture with a background annulus  of 12--20 pixels (7\farcs2--12\arcs). The aperture radius was chosen to be the region of the image most dominated by the AGN, and the sky annulus was chosen to subtract a significant portion of the host-galaxy light without subtracting any of the extended point-spread function (PSF) from the AGN. The quality of the data was such that there was no need to make relative measurements to the stars near the AGN to generate the light curves.
 
The uneven gaps between the {\it Spitzer} light curves (Figure~\ref{fig:zw229lc}) arose because of a mismatch between the visibility of Zw229-015 to the telescope and the proposal cycles for {\it Spitzer}.

\subsection{Light-Curve Variability}

Figure~\ref{fig:zw229lc} shows the optical and IR light curves of \mbox{Zw229-015} in 2010--2015, with observation seasons separated by the dotted grey lines, and with overall magnitude variations (brightest $-$ dimmest magnitude) of each light curve of $|\Delta \text{ground}| \ \approx 0.6$\,mag, $|\Delta \textit{Kepler}| \ \approx 0.6$\,mag, $|\Delta \textit{Spitzer 1}| \ \approx 0.9$\,mag, and $|\Delta \textit{Spitzer 2}| \ \approx 0.2$\,mag. The short-term variability (i.e., variability over timescales of days to months) of the individual seasons in each light curve are approximately consistent for the entire light curves, with mean overall magnitude variations of $\sim 0.3$, $\sim 0.4$, $\sim 0.2$, and $\sim 0.1$\,mag in the ground-based optical, \textit{Kepler}, and \textit{Spitzer} channel 1 and 2 light curves, respectively. Additionally, underlying long-term variability (i.e., variability over timescales of several months to years) can be seen in the light curves, especially those that are observed in 2011--2013 as the object gets brighter, with a change of mean apparent magnitude of $\sim 0.2$, $\sim 0.2$, and $\sim 0.6$\,mag in the ground-based optical, \textit{Kepler}, and \textit{Spitzer} channel 1 light curves, respectively.

The light curves in Figure~\ref{fig:zw229lc} were then converted into fluxes for the upcoming analysis. It is worth noting that the observations of \mbox{Zw229-015} also likely contain a substantial amount of flux from the host galaxy, the amount of which varies for each survey; however, this host-galaxy flux is assumed to be constant over the entire observational period. For consistency between the optical light curves in Figure~\ref{fig:zw229lc}, a constant flux of 3.7\,mJy corresponding to the mean of the difference between the overlapping ground-based $V$-band and \textit{Kepler} observations was subtracted from the \textit{Kepler} observations.

\section{Dust Reverberation Mapping}
\label{Sect:DustRM}

Potential reverberation lags between the optical and IR light curves of \mbox{Zw229-015} were recovered using the cross-correlation function (CCF), which computes a correlation value between the light curves for a range of potential lags (e.g., \citealt{Peterson1993}). This requires the light curves to be continuous; therefore, they were interpolated to simulate data between observations using the structure function (SF), which measures the fractional change in flux of the observations that are separated by given time intervals (e.g., \citealt{Suganuma2006}, \citealt{Emmanoulopoulos2010}).

Interpolation can lead to large portions of simulated data between observation seasons, which in turn creates uncertainties in the cross-correlation analysis. Therefore, to reduce the impact of the interpolations, two methods of interpolated cross correlation function (ICCF; e.g., \citealt{Peterson1998}) were compared. The first method used a standard-ICCF (\mbox{S-ICCF}), in which both light curves were interpolated with 1 day cadences before cross correlating. This method utilised all available observations in both light curves, but also treated the interpolated data equally with observations and could therefore decrease the reliability of the result. The second method only interpolated one of the light curves at a time, and the data corresponding to the epochs of observation of the second light curve plus a range of potential lags were extracted and cross correlated \citep{Guise2022}. This meant less interpolated data was included, especially when cross correlating multiple observation seasons with months-long gaps between them, but did not use all available data. This method would therefore need to be performed twice, interpolating each of the light curves, which will be referred to as the modified-ICCF (\mbox{M-ICCF}) and reverse modified-ICCF (\mbox{RM-ICCF}) methods, respectively.

\begin{figure*}
    \centering
    \begin{minipage}{\textwidth}
        \subfloat[figure][CCFs and ACFs of the entire ground (gr) and \textit{Spitzer} 1 (sp1) light curves. \label{fig:gr_sp1_CCF_all}]{
            \begin{minipage}{0.32\textwidth}
                \includegraphics[width=\textwidth]{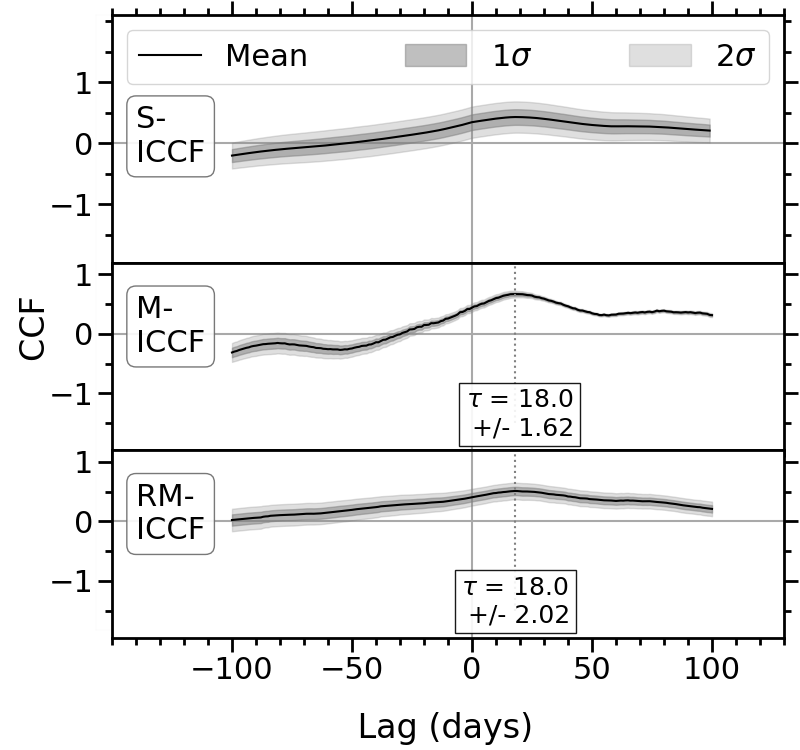} 
                \\
                \includegraphics[width=\textwidth]{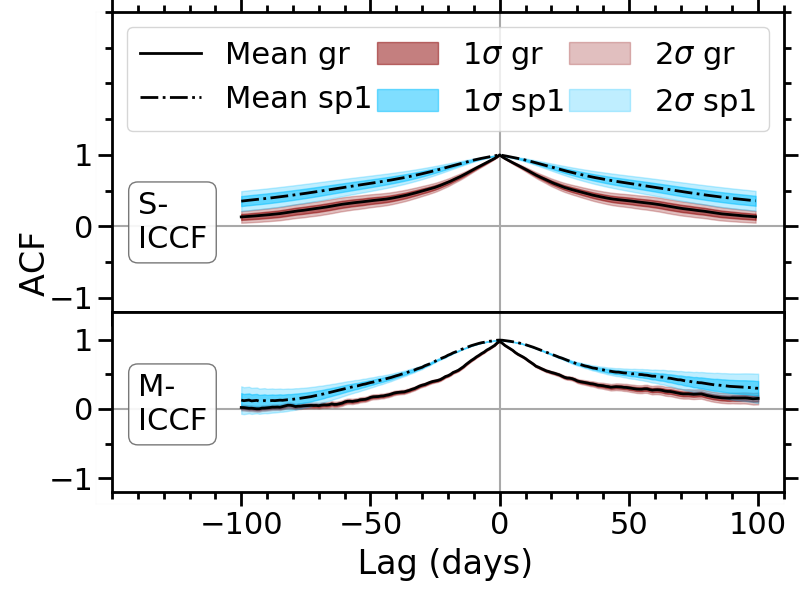}
            \end{minipage}}
            \hfill
        \subfloat[figure][CCFs and ACFs of the entire \textit{Kepler} (kep) and \textit{Spitzer} 1 (sp1) light curves. \label{fig:kep_sp1_CCF_all}]{
            \begin{minipage}{0.32\textwidth}
                \includegraphics[width=\textwidth]{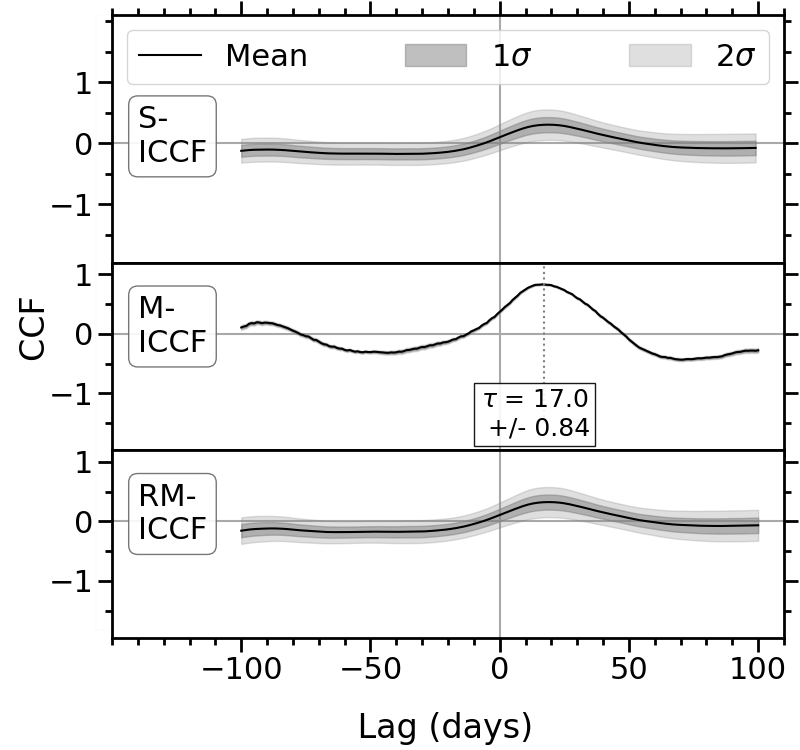}
                \includegraphics[width=\textwidth]{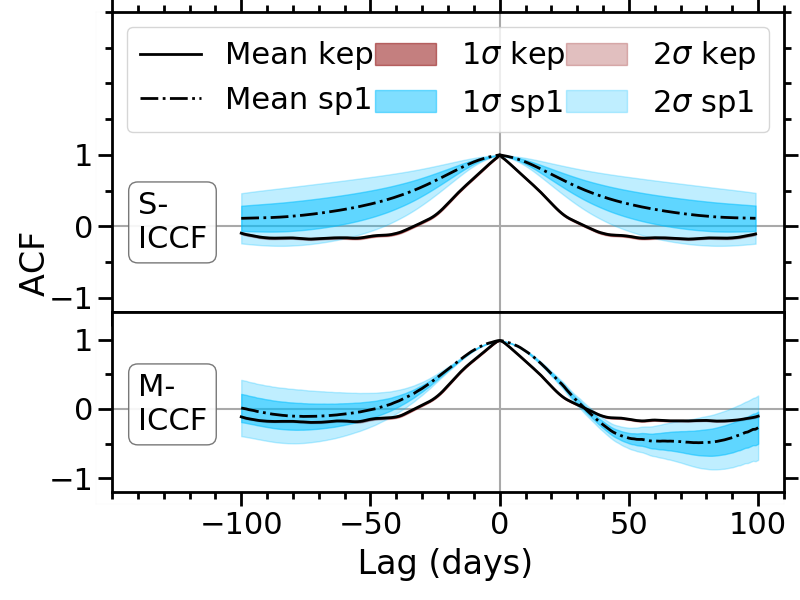}
            \end{minipage}}
        \hfill
        \subfloat[][CCFs and ACFs of the entire ground (gr) and \textit{Spitzer} 2 (sp2) light curves. \label{fig:gr_sp2_CCF_all}]{
            \begin{minipage}{0.32\textwidth}
                \includegraphics[width=\textwidth]{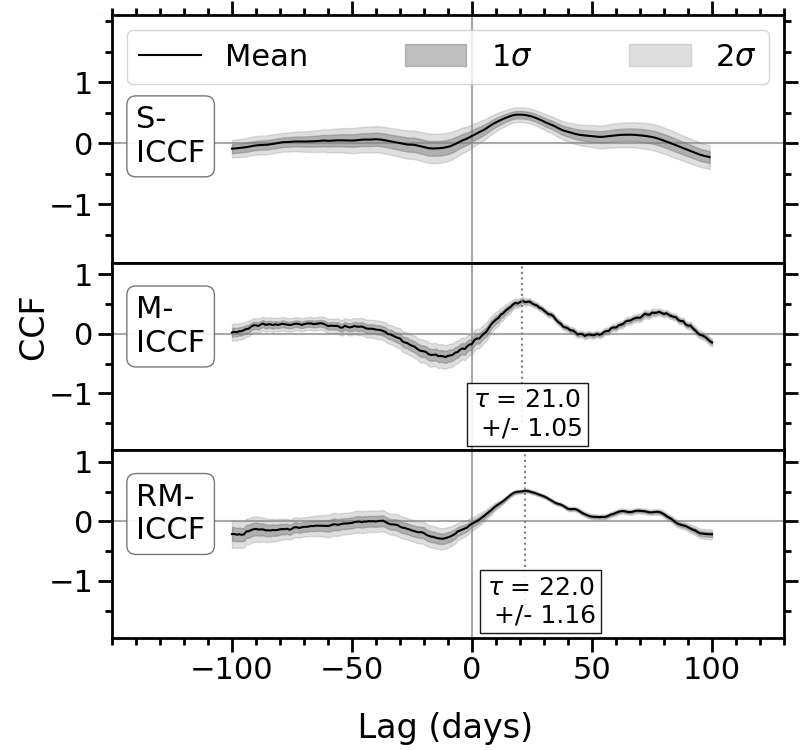} \\
                \includegraphics[width=\textwidth]{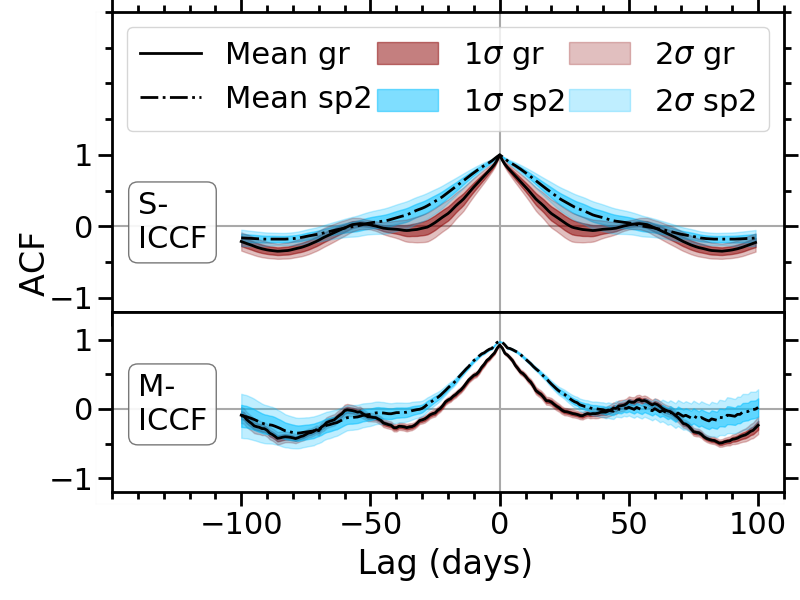} 
            \end{minipage}}
        \caption{CCF and ACF of each combination of the optical and IR light curves of \mbox{Zw229-015} over the entire overlapping observational periods, after subtraction of long-term variability. Here, \mbox{M-ICCF} refers to interpolating the optical light curve and \mbox{RM-ICCF} refers to interpolating the IR light curve.  \label{fig:entire_lc_CCFs}}
    \end{minipage}
    \vfill
    \begin{minipage}{\textwidth}
        \subfloat[figure][CCFs and ACFs of the 2010 ground and \textit{Spitzer} 1 light curves. \label{fig:gr_sp1_CCF_S1}]{
            \begin{minipage}{0.32\textwidth}
                \includegraphics[width=\textwidth]{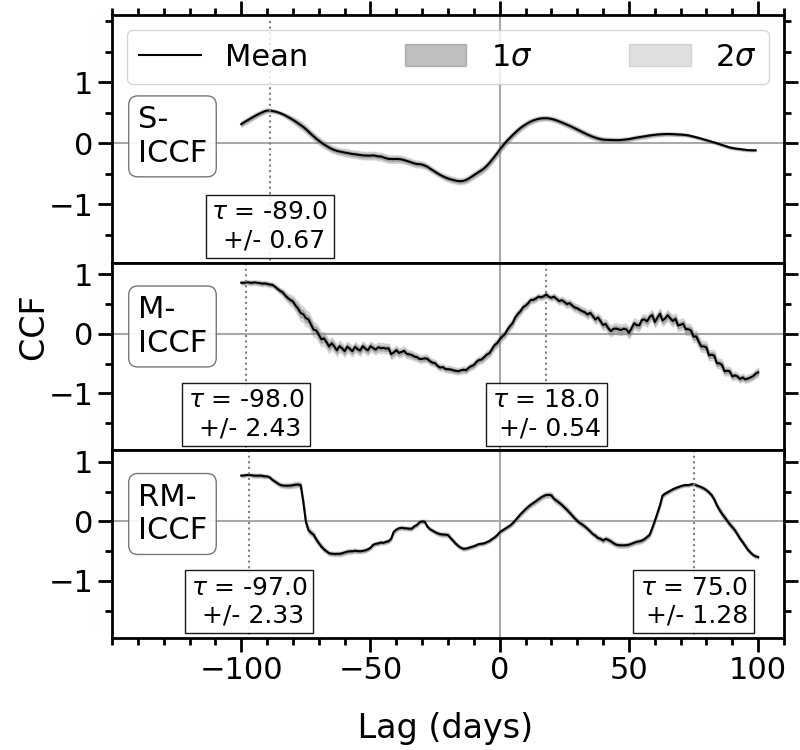} \\
                \includegraphics[width=\textwidth]{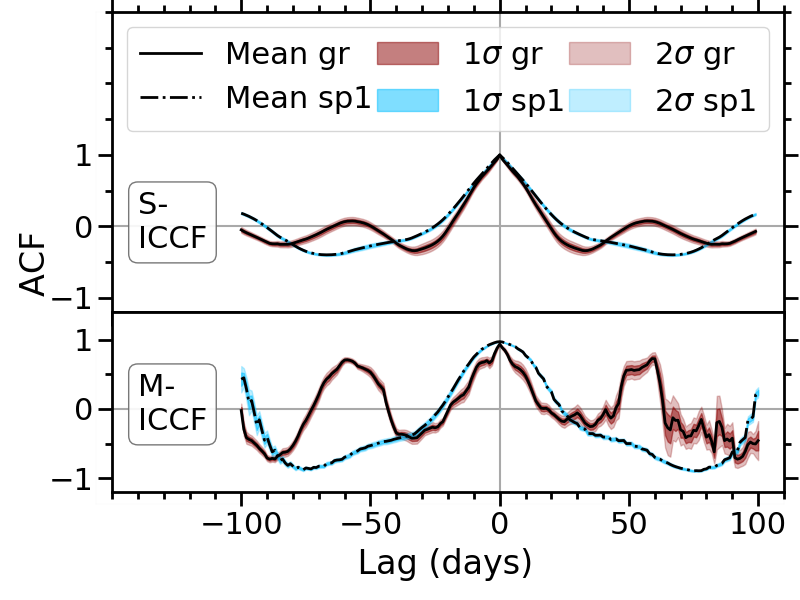}
            \end{minipage}}
        \hspace{1mm}
        \subfloat[figure][CCFs and ACFs of the 2013 ground and \textit{Spitzer} 1 light curves. \label{fig:gr_sp1_CCF_S4}]{
            \begin{minipage}{0.32\textwidth}
                \includegraphics[width=\textwidth]{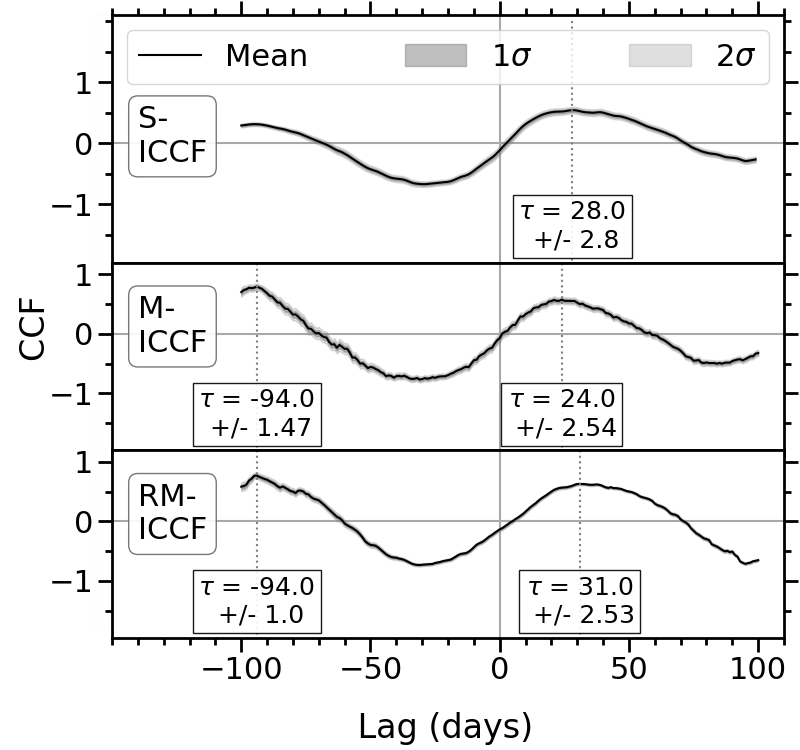} \\
                \includegraphics[width=\textwidth]{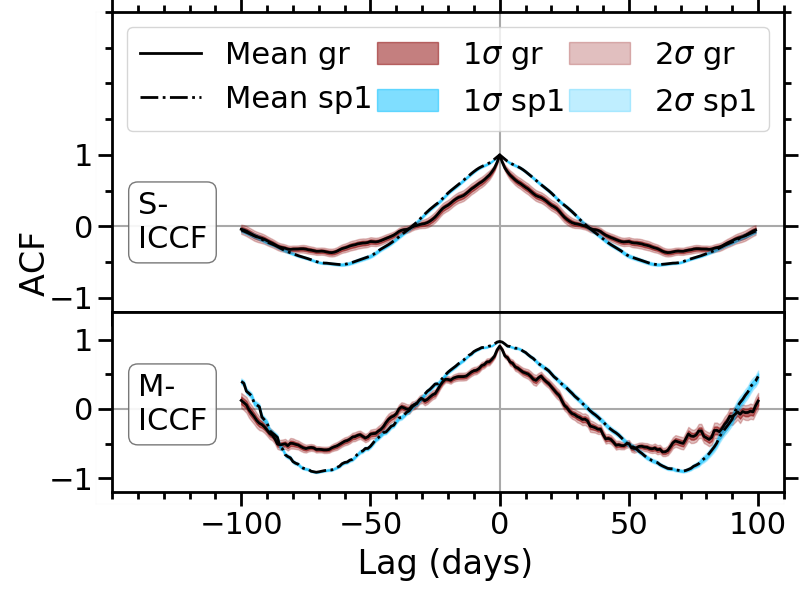}
            \end{minipage}}
        \hspace{2mm}
        \subfloat[figure][CCFs and ACFs of the 2014 ground and \textit{Spitzer} 1 light curves. \label{fig:gr_sp1_CCF_S5}]{
            \begin{minipage}{0.32\textwidth}
                \includegraphics[width=\textwidth]{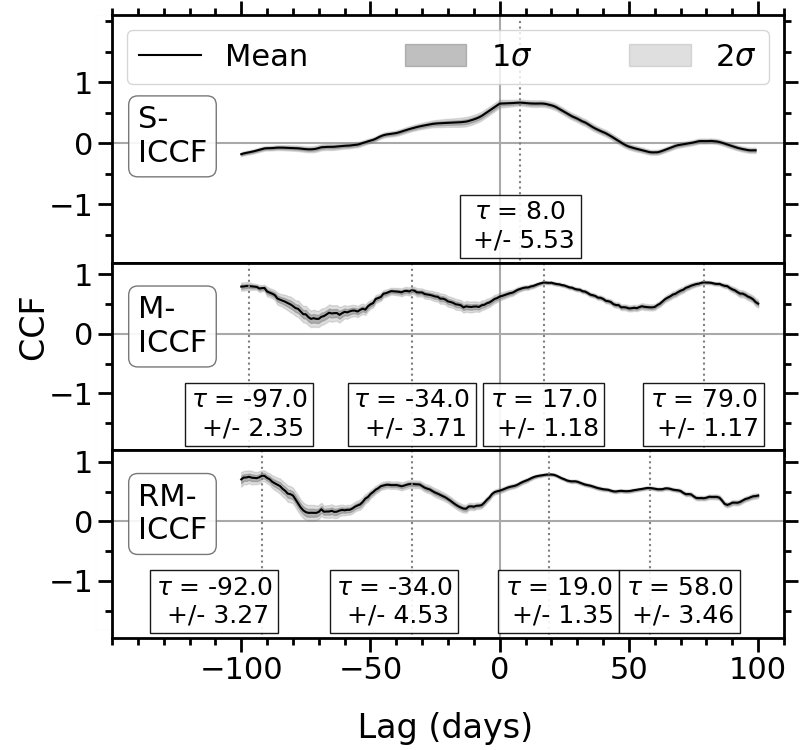} \\
                \includegraphics[width=\textwidth]{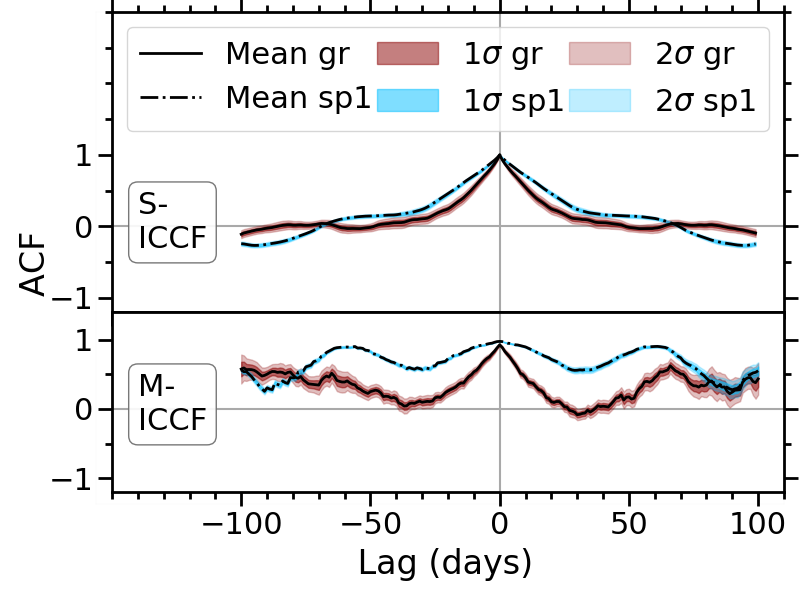}
            \end{minipage}}
        \caption{CCF and ACF of the ground optical (gr) and \textit{Spitzer} 1 (sp1) light curves for some of the individual observation season light curves of \mbox{Zw229-015}. Here, \mbox{M-ICCF} refers to interpolating the ground and \mbox{RM-ICCF} refers to interpolating the \textit{Spitzer} 1 light curve. \label{fig:gr_sp1_CCFs_years}}
    \end{minipage}
\end{figure*}

\begin{figure*}
    \begin{minipage}{\columnwidth}
        \centering
        \includegraphics[width=0.85\textwidth]{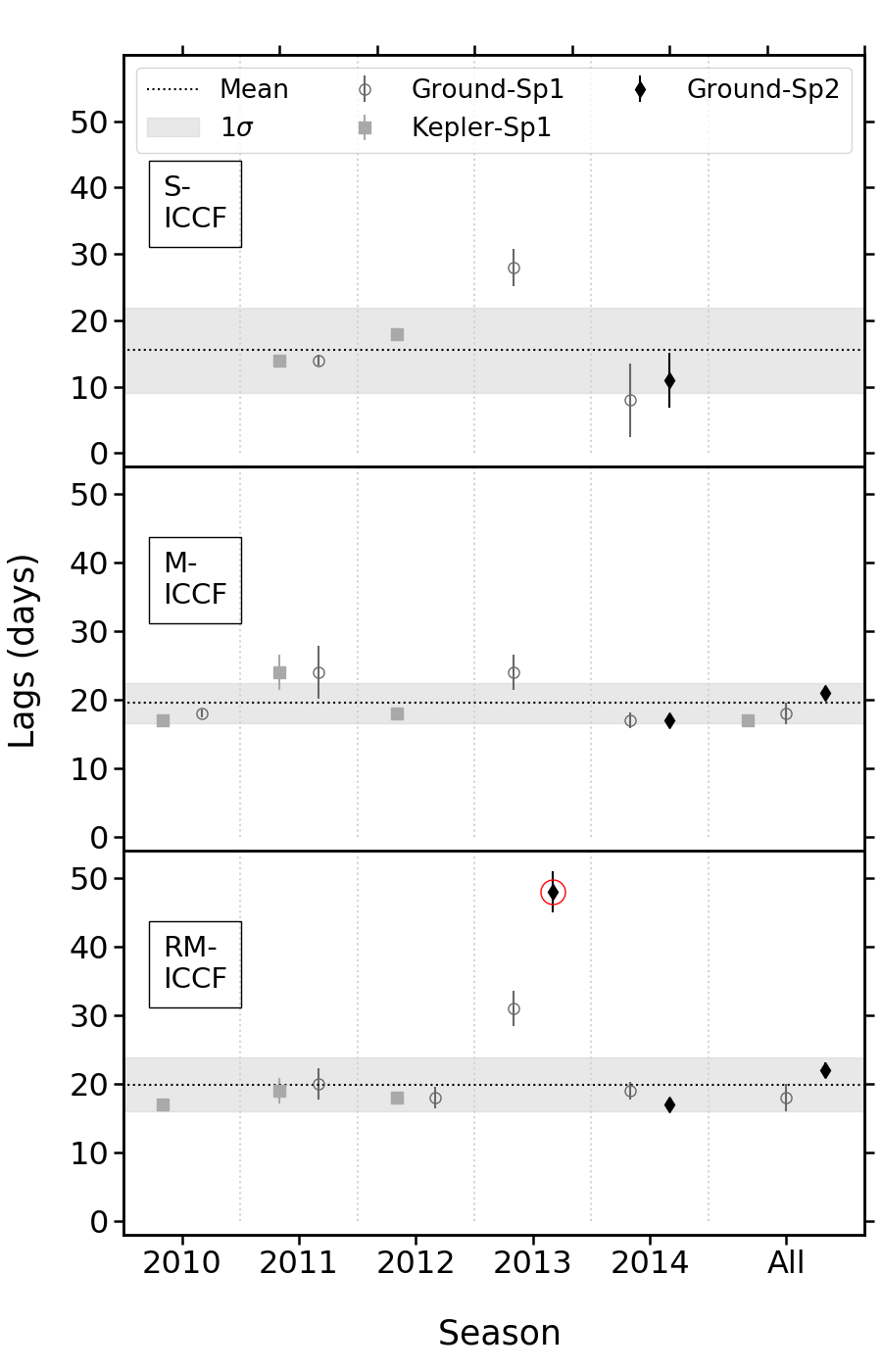}
        \captionof{figure}{The lags measured between the optical and IR light curves of Zw229-015 between 0 and 50 days for each combination of optical (ground and \textit{Kepler}) and IR (\textit{Spitzer} 1 (sp1) and \textit{Spitzer} 2 (sp2)) for each season and using each CCF method (\mbox{S-ICCF}, \mbox{M-ICCF} and \mbox{RM-ICCF}). The ground-\textit{Spitzer} 2 \mbox{RM-ICCF} is circled in red as a possible outlier, and is excluded from the mean calculation.  \label{fig:AllLags}}
        \vspace{3mm}
    \end{minipage}
    \hfill
    \begin{minipage}{0.49\textwidth}
        \centering
        \includegraphics[width=0.7\textwidth]{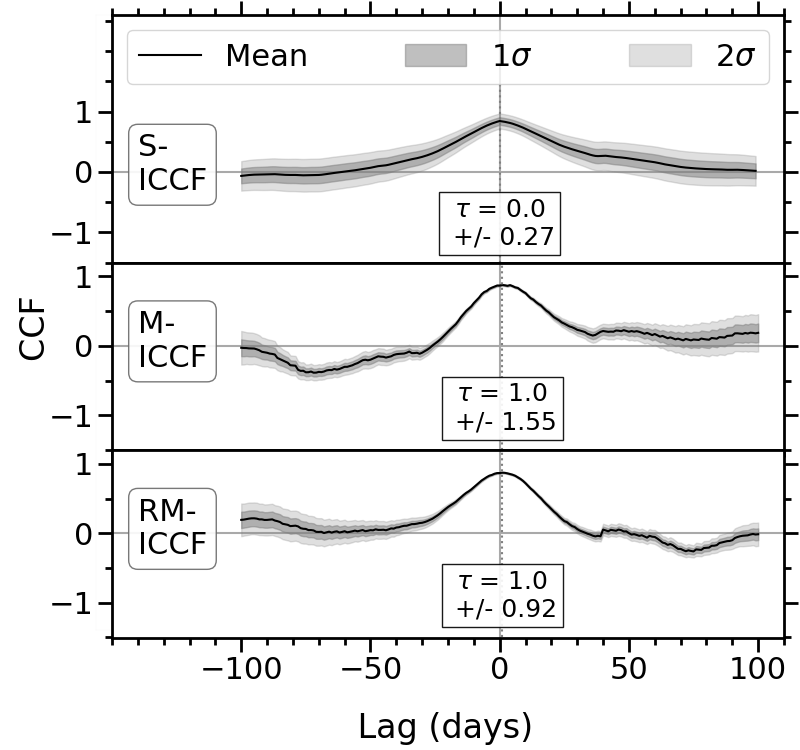} \\
        \includegraphics[width=0.7\textwidth]{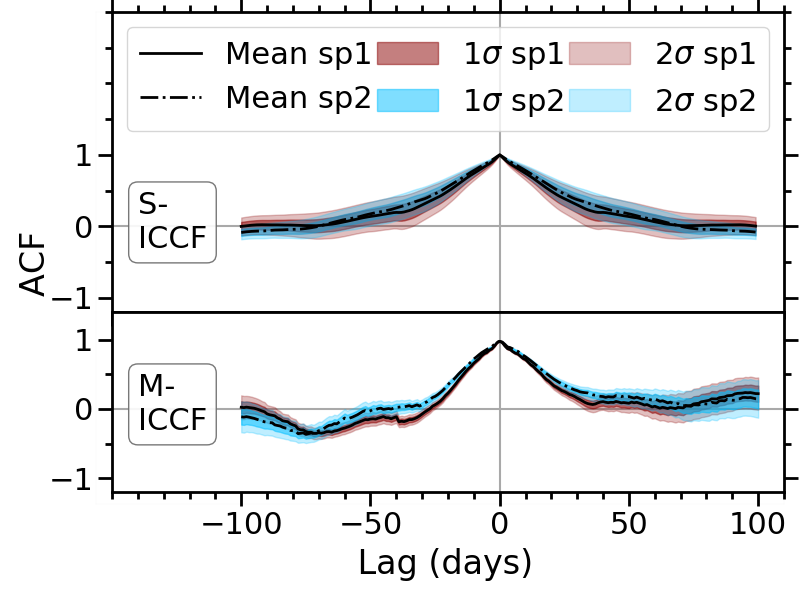}
        \caption{CCF and ACF of the \textit{Spitzer} 1 (sp1) and \textit{Spitzer} 2 (sp2) light curves for the entire overlapping light curves of \mbox{Zw229-015}. \label{fig:sp1_sp2_CCFs_all}}
    \end{minipage}
\end{figure*}

\subsection{Optical-IR Cross-Correlation Results}
\label{sect:CCFResults}

Each individual season of each combination of the optical and IR light curves, as well as the entire overlapping light curves of \mbox{Zw229-015}, were interpolated and cross correlated using the methods discussed above. The light curves were interpolated using the SF, filling in the missing data in a random order to estimate the variability from different time separations, therefore this method was repeated 10,000 times for each combination of light curves to decrease the impact of the interpolated data in each instance on the cross correlation results. When cross correlating the light curves covering more than one observation season, the light curves were first detrended as explained in Appendix~\ref{ap:detrend}, to remove the influence of long-term variations (that were over timescales much greater than expected for DRM \citep[e.g.,][]{Clavel1989, Glass1992, Suganuma2006}) from the CCFs \citep[e.g,][]{Welsh1999}. 

Each CCF was tested with a lag range of $\pm 100$ days owing to the length of the individual observation seasons, except for the season starting 2011, in which the \textit{Spitzer} 1 light curve only covers a period of 67 days and hence was only tested with a range of $\pm 60$ days, and the season starting in 2012 in the ground-\textit{Spitzer} 1 CCF, in which the light curves only have overlapping periods of observation of 54 days, and hence was only tested with lags between $\pm 50$ days. Potential lags that were measured from peaks on the CCFs were considered positive detections if the CCF values were $> 0.5$, as most nonzero peaks in the autocorrelation functions (ACFs) had values smaller than this, with the exception of the edges of some ACFs, and the \mbox{M-ICCF} ACFs of seasons starting 2010, 2012, and 2014 which display secondary peaks between $\pm 55$ and 70 days as discussed below. These possible lags were labelled on the plot, with the uncertainties calculated as the standard deviations of the peak of the CCF for each interpolation around the mean CCF. 

Figures~\ref{fig:entire_lc_CCFs} and \ref{fig:gr_sp1_CCFs_years} contain the mean CCFs and ACFs of the entire overlapping periods of the ground-\textit{Spitzer} 1, \textit{Kepler}-\textit{Spitzer} 1 and ground-\textit{Spitzer} 2 light curves, and the CCFs and ACFs of some of the individual observation seasons of the ground-\textit{Spitzer} 1 light curves, respectively. Appendix~\ref{ap:more_ccfs} contains the remaining individual season CCFs of each combination of optical and IR light curve. 

The most consistent possible lag detected occurs between \mbox{$\sim$ 5 and 30 days}. As there are no corresponding peaks in the ACFs at these lags, it is likely not affected by the variability of each light curve individually and instead is a potential delay between the optical and IR light curves. A positive detection of this lag is recorded in at least one method in each season and in the entire combined light curves (with the exception of the ground-\textit{Spitzer} 2 2013 season CCF which is discussed in Appendix~\ref{ap:S4_CCF}); however, it is not detected in all methods of each observation season. For example, this lag is not detected in the \mbox{S-ICCF} method in the entire overlapping ground-\textit{Spitzer} 1 CCFs in Figure~\ref{fig:gr_sp1_CCF_all}, or in the \mbox{S-ICCF} and \mbox{M-ICCF} methods of the entire overlapping \textit{Kepler}-\textit{Spitzer} 1 CCFs in Figure~\ref{fig:kep_sp1_CCF_all}; however, this is likely due to interpolations of the \textit{Spitzer} 1 light curve, including in the large gap in observations between HJD 55800 and 56150. 

Furthermore, potential lags between \mbox{$\sim$ 55 and 80 days} are detected in multiple plots, although less frequently than the $\sim 20$~day lag. For example, they are present in either the \mbox{M-ICCF} or \mbox{RM-ICCF} methods of each combination of optical-IR in the 2010 and 2014 CCFs as shown in Figures~\ref{fig:gr_sp1_CCF_S1}~and~\ref{fig:gr_sp1_CCF_S5}, respectively. The shape of the light curves in these seasons could be impacting the CCFs, however, as the corresponding optical ACF for each CCF that detects this lag has a secondary peak at $\sim 55$--70 days. For example, the lag is detected in the \mbox{RM-ICCF} method of the 2010 season ground-\textit{Spitzer} 1 CCF and in the corresponding ground ACFs there is a relatively high peak at $\sim \pm 60$ days, with an ACF value of $\sim 0.7$. This means that the \textit{Spitzer} 1 light curve would correlate with the ground light curve at the lag $\tau$ and $\tau + 60$\,days (i.e., if $\tau \approx 20$\,days, then $\tau + 60 \approx 80$\,days). Furthermore, the \mbox{M-ICCF} and \mbox{RM-ICCF} methods of the 2014 season CCFs display multiple peaks separated by $\sim 50$--60 days, as shown in Figure~\ref{fig:gr_sp1_CCF_S5} for example, which could also be due to the shape of the optical and IR light curves in this season as they follow an increase with multiple distinct bumps separated by $\sim 50$--60 days that correlate with each other. This correlation can be seen in the \mbox{M-ICCF} ACFs of both light curves by the peaks at $\sim \pm 60$ days with ACF values of $\sim 0.6$ in the ground optical and 0.9 in the \textit{Spitzer} 1 and 2. Therefore, as this $\sim 55$--80 day lag has corresponding peaks in the optical ACFs, it is unlikely to be due to a delay between optical and IR light curves, but instead aliasing in the light curves of certain seasons. Appendix~\ref{ap:periodicty} explores whether this is a periodicity in the optical light curves, but finds that it appears to be a sampling effect as it only occurs for the overlapping light-curve periods and not on longer light curves. 

Finally, possible lags between $\sim -100$ and $-90$ days are displayed in multiple methods and seasons --- for example, in all methods of the ground-\textit{Spitzer} 1  2010 season, and the \mbox{M-ICCF} and \mbox{RM-ICCF} methods of ground-\textit{Spitzer} 1 in the 2013 and 2014 seasons displayed in Figure~\ref{fig:gr_sp1_CCFs_years}. These peaks in the CCFs occur only owing to a small number of overlapping data points, so are deemed less significant than other possible lags. Additionally, similar correlations are present in at least one of the corresponding ACFs, which further implies that the peak is not necessarily due to a lag between light curves but impacted by aliasing. 

As the $\sim 20$ day lag was deemed to be most significant, it was further compared in each season and for each combination of optical and IR light curves in Figure~\ref{fig:AllLags}. This lag is found consistently with values between 8 and 30 days, with a mean of {\color{black} $18.8 \pm 4.6$ days (or $19.8 \pm 7.1$ days} including the 2013 season \mbox{RM-ICCF} lag). The mean of the individual methods, \mbox{S-ICCF}, \mbox{M-ICCF} and \mbox{RM-ICCF}, are {\color{black} $15.5 \pm 6.4$, $19.5 \pm 2.9$, and $19.9 \pm 4.0$ days (or $22.5 \pm 8.9$} days including the 2013 season \mbox{RM-ICCF} lag) respectively, which are all consistent within 1 standard deviation of each other, and the mean of each combination of optical and IR light curve are {\color{black} $19.8 \pm 5.7$, $18.0 \pm 2.5$, and $17.6 \pm 3.9$ days (or $22.7 \pm 11.9$} days including the 2013 season \mbox{RM-ICCF} lag) for the ground-\textit{Spitzer} 1, \textit{Kepler}-\textit{Spitzer} 1 and ground-\textit{Spitzer} 2 respectively, which are also within 1 standard deviation of each other. The 2013 season in the \mbox{RM-ICCF} method shows a longer lag between 30 and 50 days, which is explored further in Appendix~\ref{ap:S4_CCF}.

\subsection{\textit{Spitzer} 1 - \textit{Spitzer} 2 Cross-Correlation Results}

The overall combined \textit{Spitzer} light curves were also cross correlated using the methods described to measure potential lags between the 3.6\,$\mu$m and 4.5\,$\mu$m emission regions, as displayed in Figure~\ref{fig:sp1_sp2_CCFs_all}, and in Appendix~\ref{ap:more_ccfs} for the individual seasons. The most consistent lag is measured with value between $\sim 0$--5 days, and is detected in each CCF, with a mean of {\color{black} $1.8 \pm 1.7$ days}. Additional lags are measured in the \mbox{M-ICCF} methods of the 2014 season with values of $\sim 47$ days and $\sim \pm 100$ days; however, as similar peaks are found in the \mbox{M-ICCF} ACFs at this time, the peak in the CCFs is likely a result of the shape of the light curves and not a delay between the 3.6\,$\mu$m and 4.5\,$\mu$m emission.

\begin{table*}
	\begin{minipage}{\textwidth}
        \centering
        \footnotesize{
        \caption{Descriptions of the parameters of the MCMC modelling that are used to find the best-fit dust transfer function and simulate the IR light curves. \label{tab:params}}        
        \begin{tabular}{C{0.3\textwidth}p{0.45\textwidth}C{0.15\textwidth}}
                \hline 
                Parameter & Definition &  Priors Range \\
                \hline 
                
                Radial power-law index, $\alpha$ & Power-law index which is used to describe the radial distribution of dust  &  $-5.5 \leq \alpha \leq -0.5$ \\
              
                Vertical scale height power-law index, $\beta$ & Power-law index which is used to describe the vertical distribution of the dust above the equator & $0.05 \leq \beta \leq 2.05$  \\
                
                Time lag (days), $\tau$ & The projected time delay between optical and IR light curves &  $5 \leq \tau \leq 30$\,days \\
               
                Inclination angle (degrees), $i$  & The angle between the system and line of sight of the observer &  $1 \leq i \leq 69$ \\
                
                Optical-IR amplitude conversion factor, $w_\textrm{eff}$ & The scaling factor of the amplitudes of the DTFs to account for uncertainties within the light curves, including different amounts of host-galaxy flux in the different observations and accretion-disk subtraction & $0.2 \leq w_\textrm{eff}\leq 11$ \\   
               
                Offset & Additional offset between relative light curves & $-0.2 \leq \textrm{Offset} \leq 0.2$ \\
                
                \hline 
                
                \centering Resampled relative optical light curve & As the observed optical light curves are not necessarily evenly sampled, they are interpolated and then treated as free parameters to find the best-fit optical light curve that, when convolved with the DTF, best fits the IR observations. & Interpolated optical light curve $\pm$ the 1$\sigma$ uncertainties at that point \\
                
            \hline
        \end{tabular}}
	\end{minipage}
\end{table*}

\section{Light-Curve Modelling}
\label{Sect:IRModelling}

The light curve of the reprocessed optical emission into the IR at time $t$ can be expressed as a convolution of the optical light curve with a dust transfer function (DTF) \citep{Peterson1993}, 

\begin{equation}
    \centering
    \label{eq:TF}
    f_\textrm{IR}^\textrm{rep}(t)~=~\int^{+\infty}_{-\infty} \Psi(\tau') \ f_\textrm{opt}(t - \tau') \ d\tau' \,.
\end{equation}
\noindent Here $f_\textrm{opt}(t-\tau')$ is the optical light curve at an earlier time $t - \tau'$, $\tau'$ is an arbitrary delay, $\Psi(\tau')$ is the DTF,  and $f_\textrm{IR}^\textrm{rep}(t)$ is the reprocessed IR light curve at time $t$:

\begin{equation}
    \label{eq:IRsubAD}
    f_\textrm{IR}^\textrm{rep}(t) = f_\textrm{IR}^\textrm{obs} (t) - f_\textrm{IR}^\textrm{disk} (t) \,, 
\end{equation}
\noindent where $f_\textrm{IR}^\textrm{obs} (t)$ is the total observed IR light curve at time $t$, and $f_\textrm{IR}^\textrm{disk} (t)$ is the contribution to the observed IR light curve from the accretion disk at time $t$, which is estimated from

\begin{equation}
    \label{eq:IRAD}
    f_\textrm{IR}^\textrm{disk} (t) = f_\textrm{opt} (t) \bigg(\frac{\nu_\textrm{IR}}{\nu_\textrm{opt}} \bigg)^{\alpha_\nu} \,,
\end{equation}
\noindent where $f_\textrm{opt}(t)$ is the optical flux at time $t$, $\nu_\textrm{IR}$ and $\nu_\textrm{opt}$ are respectively the effective frequencies of the IR and optical filters, and $\alpha_\nu$ is the power-law index which is set to the expected value of $\alpha_\nu=1/3$ in the standard accretion-disk model \citep{Shakura1973}.

The DTF describes how emission from the optical continuum is reprocessed into the observed IR emission. It depends on the geometry and structure of the dust emission region, and as such, can be used to provide further information on the inner regions of the AGN that cannot be spatially resolved. However, the DTF is not directly measurable, and finding a unique solution for it in Equation~\ref{eq:TF} would require high-quality data; thus, in this paper, it is estimated by first simulating the distribution of the dust in the inner regions of the AGN.

\subsection{Simulating Dust Transfer Functions}

Currently, the dust distribution in the inner regions of the AGN can only be spatially resolved using interferometry on relatively bright, nearby AGN. However, it can also be estimated using dust reverberation time delays, as dust at different locations correspond to different light-travel times to the observer, so the overall IR dust response to the optical emission is made up of a range of lags. 

For this paper, 10,000 dust clouds are randomly distributed as follows. Firstly, in the equatorial plane, the clouds are distributed following a radial power law with index $\alpha$, as given in Equation~\ref{eq:alpha} below, and are uniformly distributed with azimuthal angle ($\phi$) between 0 and 2$\pi$. Steep radial power-law indices (i.e., small values of $\alpha$) correspond to a compact object, while in shallow radial brightness distributions, the dust is extended as shown in Appendix~\ref{ap:dust_dists}. The heights of the dust clouds above the equator ($h$) are distributed following a vertical scale height power law with index $\beta$, as given by Equation~\ref{eq:beta} below, where values of $\beta = 0$ correspond to a flat disk, $\beta = 1$ is a flared disk of constant ratio of height to radius, and $\beta > 1$ follows an outflow-like distribution as shown in Appendix~\ref{ap:dust_dists}. The sizes of the dust clouds are assumed to be much smaller than the size of the overall dust distribution, so the dust clouds are treated as points in this model. 

\begin{equation}
    \label{eq:alpha}
     r \propto \bigg( \frac{r'}{r_\text{sub}} \bigg)^{\alpha}\,,
\end{equation}
where $r'$ is the distance of each dust cloud from the centre and $r_\text{sub}$ is the dust sublimation radius, and

\begin{equation}
    \label{eq:beta}
    h \propto r^{\beta}\,,  
\end{equation}
where $r$ is the radial position of each dust particle from Equation~\ref{eq:alpha}.

These optically thick dust clouds are assumed to be directly heated from the AGN; thus, only the side facing the AGN is illuminated while the far side is not. The observed emission therefore depends on the angle ($\psi$) that describes the dust cloud's position relative to the equatorial plane and relative to the observer, 

\begin{equation}
\label{eq:illum_ang}
    \text{cos} \ \psi = \text{cos} \ \theta  \ \text{cos} \phi \ \text{sin} \ i + \text{sin} \ \theta \ \text{cos} \ i \, ,
\end{equation}
\noindent where $\theta$ is the complement of the polar angle and $i$ is the inclination angle of the system. The inclination angle affects the delay maps as shown in Appendix~\ref{ap:dust_dists}, as when the disk is tilted (e.g., represented by an angle of $i = 69^\circ$) the far-side clouds have larger delays and the near-side clouds have smaller delays relative to the dust clouds in a face-on disk (i.e., $i=0^\circ$).

As only the side of the dust cloud that is facing the AGN is illuminated, the fraction of a dust cloud's illuminated surface that is then visible to the observer ($\kappa$) is estimated from

\begin{equation}
\label{eq:illum}
    \kappa = 0.5 \, (1-\text{cos} \ \psi)\, ,
\end{equation}
\noindent where an angle of $\psi = 180^\circ$ corresponds to the entire illuminated side of the dust cloud being visible to the observer, while for an angle of $\psi = 0^\circ$, the entire nonilluminated side is visible to the observer. An example of the illumination is displayed in Appendix~\ref{ap:dust_dists}. 

The dust clouds are assumed to radiate as blackbodies, and after accounting for the illumination effects, the DTF can be estimated from the combined emission response of each dust cloud (calculated using the Planck function, $B_\nu(T(r))$, for temperature $T(r)$ at radial distance $r$ given by Equation~\ref{eq:temp_dist}) to a delta-function input continuum pulse, 

\begin{equation}
\label{eq:temp_dist}
    T(r) = T_\text{sub} \bigg(\frac{r}{r_\text{sub}}\bigg)^{\frac{-2}{4+\gamma}} \, ,
\end{equation} 
\noindent where $T_\text{sub} = 1500$\,K is the sublimation temperature, and $\gamma$ is the dust IR opacity power-law index, which is set to 1.6 to correspond to standard interstellar dust material dust grains \citep{Barvainis1987}.

Using this method, a grid of DTFs was simulated for each IR wavelength, for a range of radial power-law indices \mbox{$-5.5 \leq \alpha \leq -0.5$}, a range of vertical scale height power-law indices \mbox{$0.05 \leq \beta \leq 2.05$}, and a range of inclination angles \mbox{$0^\circ \leq i \leq 69^\circ$}. The DTFs are calculated over (0--16)\,$\tau$, where $\tau$ corresponds to the projected light-travel time between the source of the optical and IR emission. Note that this model includes a number of simplifying assumptions in the distribution of dust and emission from the dust clouds that is observed, and excludes additional complexities that will likely effect the simulated dust transfer functions, which will be discussed in more detail in Section~\ref{sect:limitations}.

\subsection{MCMC Modelling of the DTFs and Light Curves}

To estimate the DTF that best describes the IR response of an AGN to the driving optical light curve, and as such further constrain properties of the inner regions of the AGN, Markov chain Monte Carlo (MCMC) modelling is used to fit models of the IR light curve to the IR observations. This modelling is performed on the relative light curves, $g_i(t) = f_i(t)/\langle f_i(t) \rangle -1 \ (i=\text{IR, opt})$, where $\langle f_i(t) \rangle$ is the mean flux of the light curves, $f_i(t)$, over the entire observational period being modelled. As most AGN (excluding blazars) will vary annually in the IR by $\sim 10$\% (e.g., \citealt{Lyu2017}), the uncertainty in the IR light curves is set to 1\% of the flux, to prevent the model from trying to overfit the shorter timescale variations (i.e., the nightly/weekly variations). 

An MCMC search is performed over the parameters described in Table~\ref{tab:params} to find the DTFs that result in reasonable fits of the simulated IR light curve to observations. These parameters include the radial power-law index, vertical scale height power-law index, inclination angle, and time lag that describe the parameter space of the DTF grids created in the previous section. They also include parameters such as the optical-IR amplitude conversion factor, $w_\textrm{eff}$, which is a scaling factor of the amplitudes of the transfer functions to account for uncertainties such as different amounts of host-galaxy contributions in the different light curves or the accretion-disk subtraction in the IR light curves, and an additional offset that could arise owing to using relative light curves. 

The driving optical light curve that is convolved with the best-fit DTF needs to be uniformly sampled; hence, resampled light curves were found by linearly interpolating the observations with a cadence of 4 days. While the mean of the interpolations could be used as the input optical light curve, they could underestimate the variability within the seasonal gaps, so instead the input optical light curves are treated as free model parameters with a prior range that is constrained by the data, and a simultaneous MCMC search is performed to also find the best optical light curve that when convolved with the DTF best fits the IR observations.

\subsection{Results of MCMC Light-Curve Modelling}
\label{Sect:IRModelling_results}

\subsubsection{Modelling the Entire Light Curves}

\begin{figure*}
    \begin{minipage}{\textwidth}
        \centering
        \includegraphics[width=\textwidth]{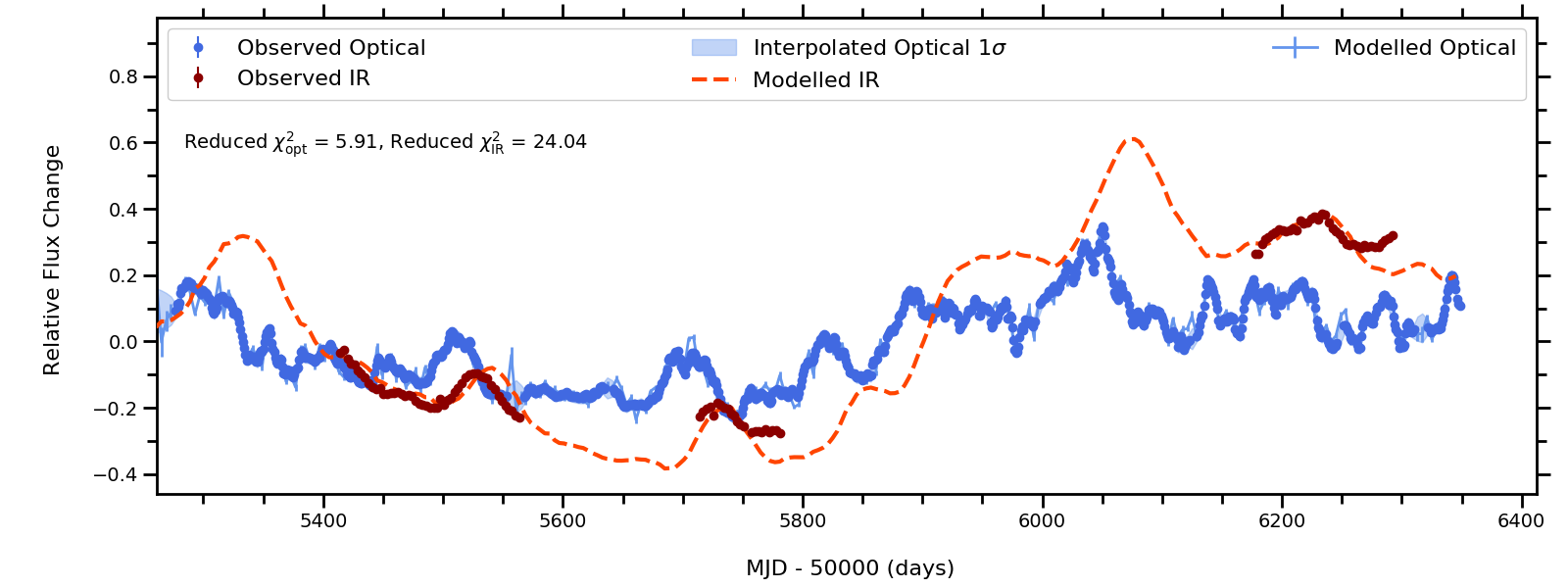}
        \caption{Simulated light curves of \textit{Kepler}-\textit{Spitzer} 1 for the entire overlapping observational periods, plotted with the parameters that corresponded to the highest posterior distribution, with values of $\alpha=-0.50$, $\beta=0.20$, $w_\textrm{eff}=2.67$, $\tau=15.94$\,days, and $i=54.08^\circ$. \label{fig:model_kep_sp1_all}}
    \end{minipage}
    \begin{minipage}{\textwidth}
        \includegraphics[width=\textwidth]{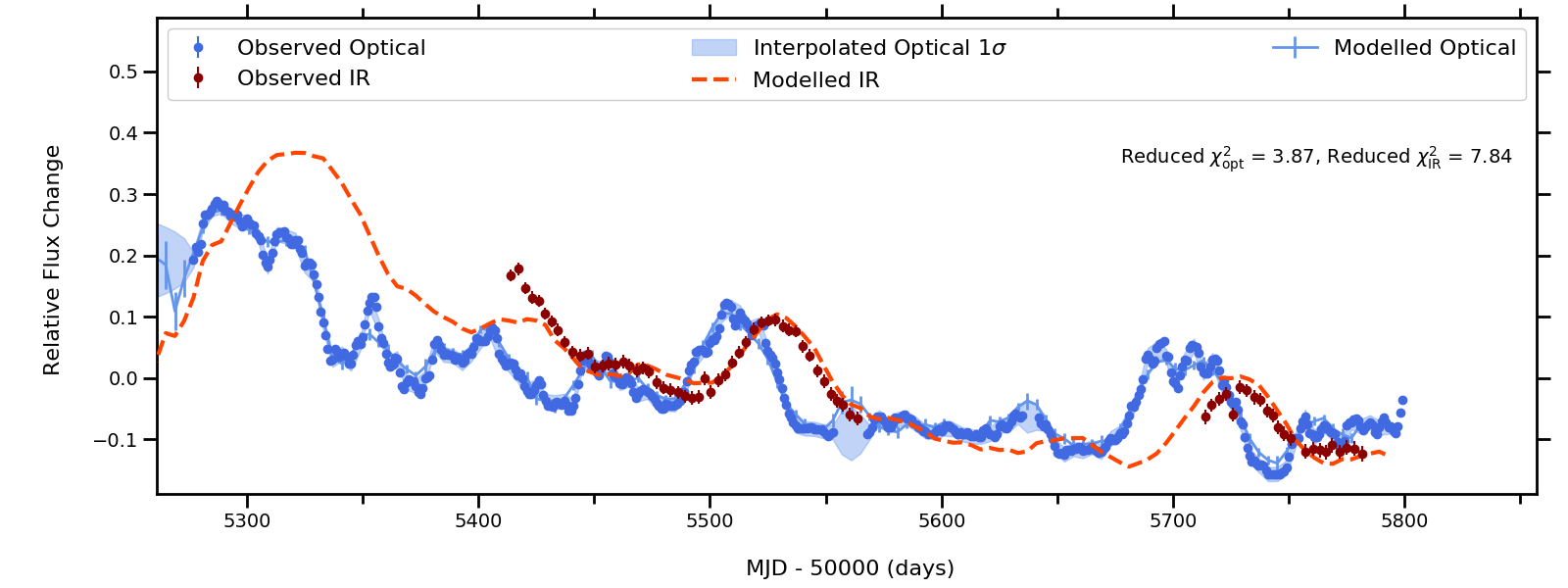}
        \caption{Simulated light curves of \textit{Kepler}-\textit{Spitzer} 1 for the observation seasons starting 2010--2011, plotted with the parameters that corresponded to the highest posterior distribution, with values of $\alpha=-0.56$, $\beta=0.51$, $w_\textrm{eff}=1.63$, $\tau=12.12$\,days, and $i=45.67^\circ$. \label{fig:model_kep_sp1_S1-2}}
        \vspace{0.3cm}
    \end{minipage}
    \begin{minipage}{\textwidth}
        \includegraphics[width=\textwidth]{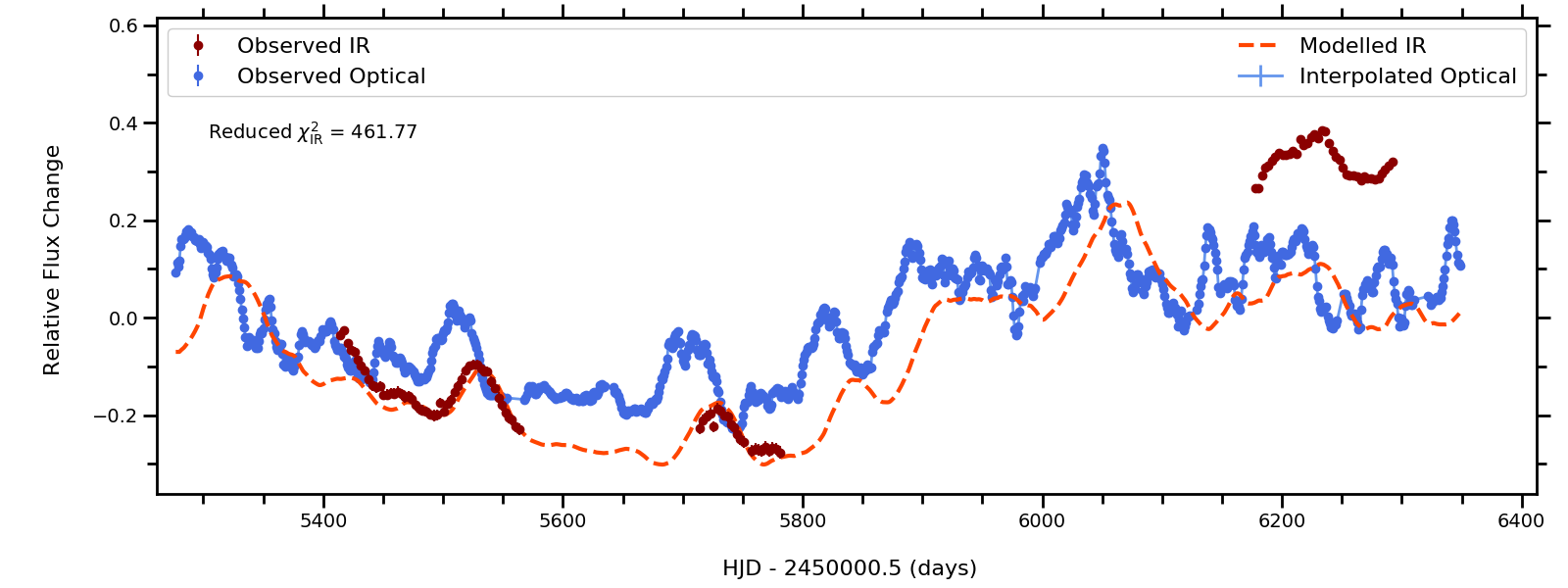}
        \caption{Simulated IR light curve of \textit{Kepler}-\textit{Spitzer} 1 for the entire overlapping observational periods, made using the linearly interpolated \textit{Kepler} light curve convolved with the DTF corresponding to Figure~\ref{fig:model_kep_sp1_S1-2}. \label{fig:model_kep_sp1_S1-2_extended}}
    \end{minipage}
\end{figure*}

\begin{table*}
    \captionof{table}{Mean output parameters of the MCMC modelling for each combination of optical (ground (gr) and \textit{Kepler} (kep)) and IR (\textit{Spitzer} 1 (sp1) and \textit{Spitzer} 2 (sp2)) light curves covering multiple observation seasons that have been separated to exclude the increase in flux between between 2011 and 2012. The uncertainties here represent the 1$\sigma$ standard deviations of the distributions for each parameter in each model. In this table, values of the offset have been multiplied by $10^{2}$.} \label{tab:entire_split_params}
        \begin{tabular}{C{0.09\textwidth}C{0.07\textwidth}C{0.08\textwidth}C{0.1\textwidth}C{0.08\textwidth}C{0.08\textwidth}C{0.08\textwidth}C{0.08\textwidth}C{0.06\textwidth}C{0.06\textwidth}}
        \hline
        
        Light-Curve Combination & Seasons Starting & Radial Power-Law Index & Vertical Scale Height Power-Law Index & Amplitude Conversion Factor & Lag (days) & Inclination (degrees) & Offset & $\chi_\textrm{opt}^2$ & $\chi_\textrm{IR}^2$ \\
        
        \hline
        

       {Gr-Sp1} & 2010--2011 &   
       $-0.52^{+0.01}_{-0.11}$ & $0.16^{+0.48}_{-0.10}$ & $2.66^{+0.16}_{-0.21}$ & $8.02^{+2.16}_{-0.55}$ & $49.71^{+13.81}_{-8.83}$ & $-0.74^{+0.55}_{-0.40}$  & 7.87 & 7.66 \\

        {Gr-Sp1} & 2012--2014 &  $-0.52^{+0.01}_{-0.03}$ &  $0.23^{+0.45}_{-0.18}$ & $1.65^{+0.07}_{-0.06}$ & $26.67^{+2.06}_{-1.71}$ & $47.76^{+20.42}_{-6.76}$ & $1.19^{+0.25}_{-0.23}$ & 5.99 & 8.74 \\
       
       
       \hline
       
       {Kep-Sp1} & 2010--2011 &  $-0.52^{+0.02}_{-0.05}$ & $0.08^{+0.51}_{-0.02}$ & $1.58^{+0.12}_{-0.10}$ & $10.83^{+1.05}_{-1.38}$ & $46.68^{+18.81}_{-11.28}$ & $0.47^{+0.38}_{-0.38}$ & 3.87 & 7.84 \\
       
      \hline

      {Gr-Sp2} & 2013--2014 & $-0.53^{+0.02}_{-0.51}$ & $0.08^{+0.53}_{-0.02}$ & $1.21^{+0.05}_{-0.04}$ & $28.54^{+0.44}_{-5.66}$ & $52.09^{+15.75}_{-7.70}$ & $0.95^{+0.33}_{-0.22}$ & 5.19 & 5.20 \\      

       
            \hline
    \end{tabular}
\end{table*}

\begin{figure*}
    \begin{minipage}{\textwidth}
        \subfloat[figure][Comparison between the mean output radial power-law indices. \label{fig:comp_alphas}]{\includegraphics[width=0.49\textwidth]{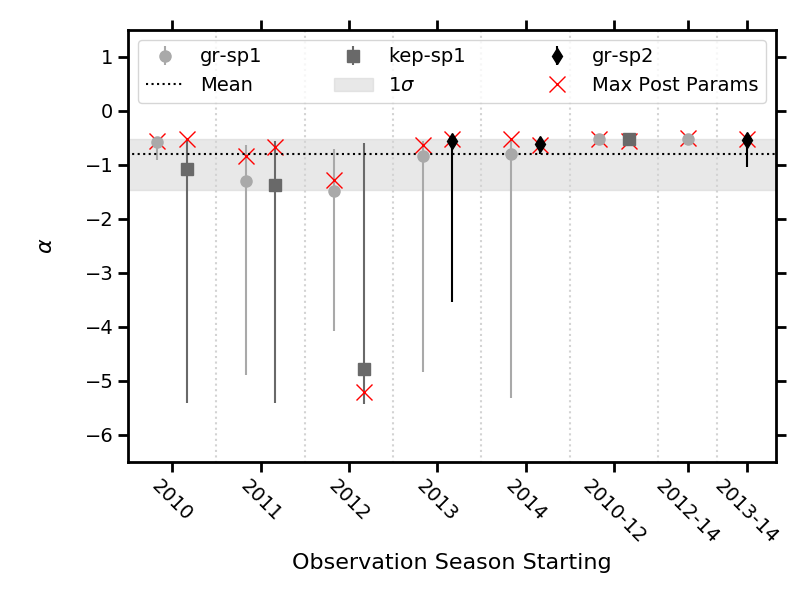}} 
        \hfill
        \subfloat[figure][Comparison between the mean output vertical scale height power-law indices. \label{fig:comp_betas}]{\includegraphics[width=0.49\textwidth]{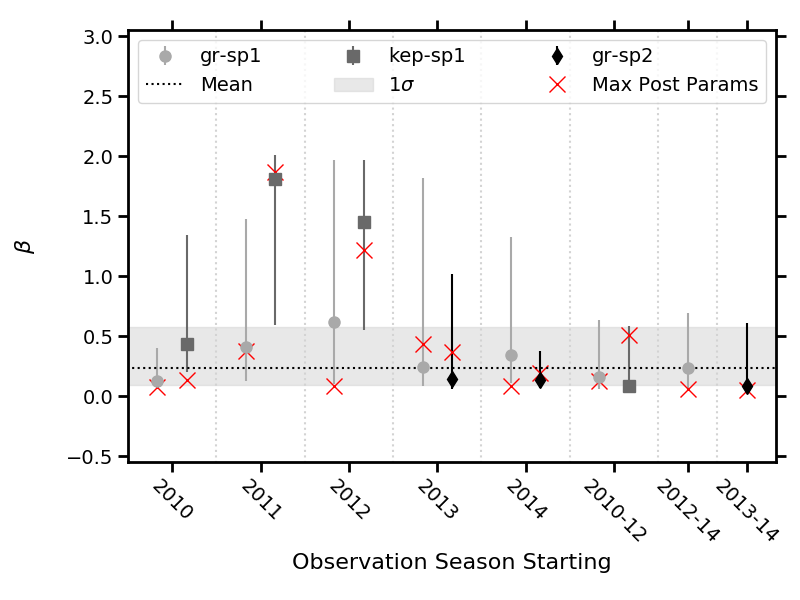}} 
        \\
        \subfloat[figure][Comparison between the mean time lags. \label{fig:comp_lags}]{\includegraphics[width=0.49\textwidth]{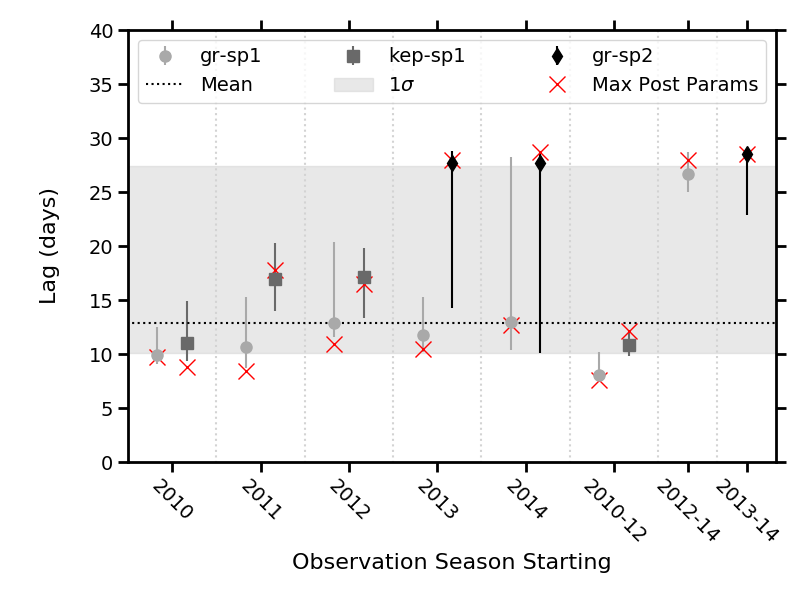}} 
        \hfill
        \subfloat[figure][Comparison between the mean output inclination angles. \label{fig:comp_incs}]{\includegraphics[width=0.49\textwidth]{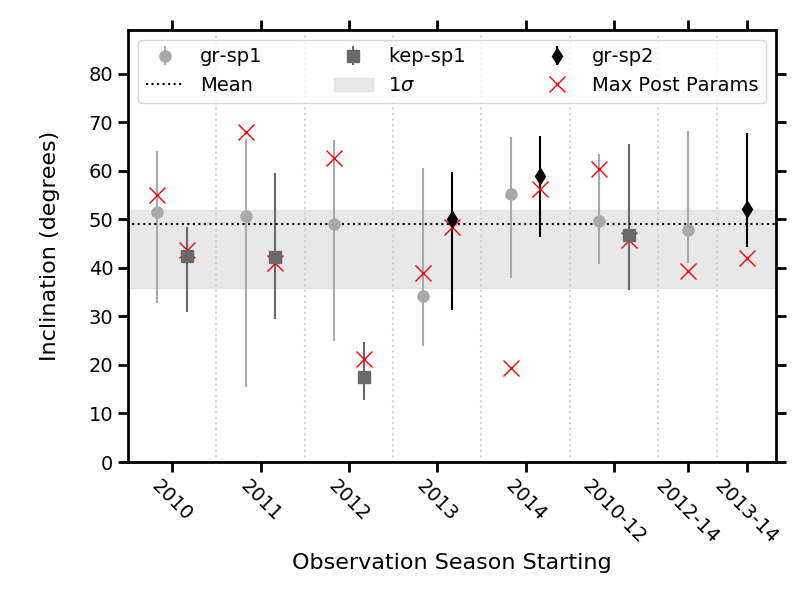}} 
        \\
        \caption{Comparison between the mean output parameters of the MCMC modelling of each combination of optical and IR light curves, for the individual observation seasons and multiseason light curves. The values corresponding to the maximum posterior distribution are also plotted in red.} \label{fig:comp_params}
    \end{minipage}
\end{figure*}

\begin{figure*}
    \begin{minipage}{\textwidth}
        \subfloat[figure][Distribution of dust clouds in $xy$.]{\includegraphics[width=0.33\textwidth]{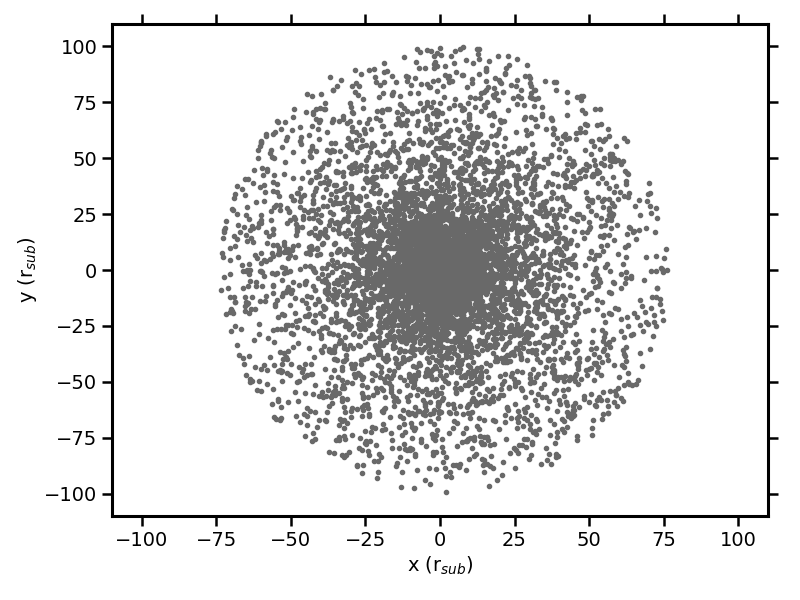}} 
        \hfill
        \subfloat[figure][Distribution of dust clouds in $yz$.]{\includegraphics[width=0.33\textwidth]{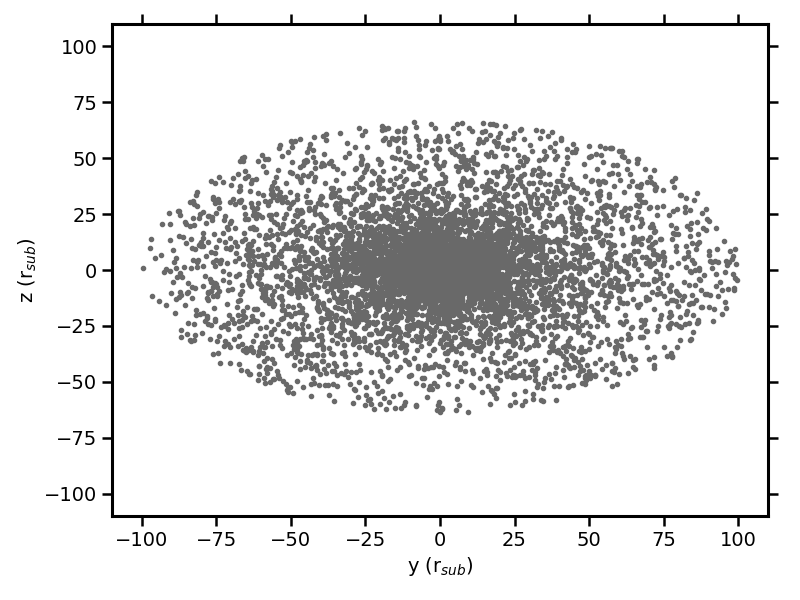}} 
        \hfill
        \subfloat[figure][Distribution of dust clouds in $xz$.]{\includegraphics[width=0.33\textwidth]{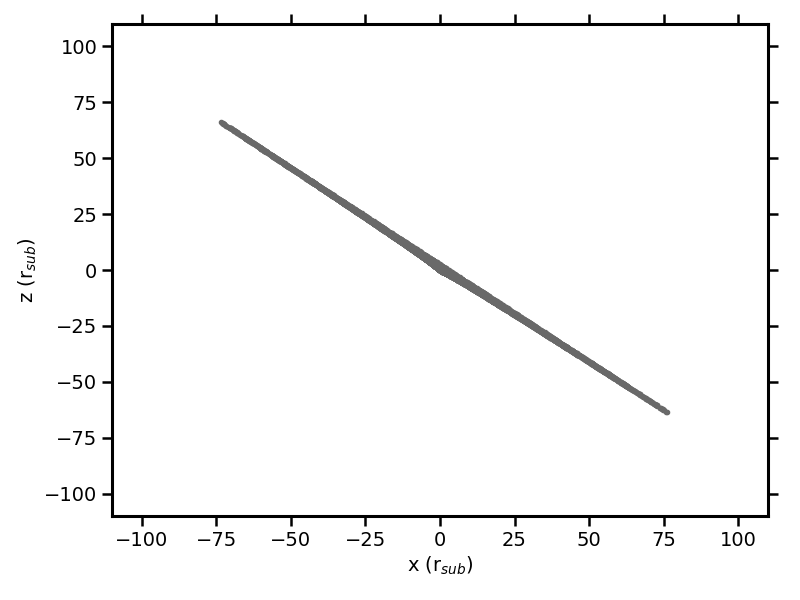}}
    \end{minipage}
    \begin{minipage}{\textwidth}
        \centering
        \subfloat[figure][Delay map for the distribution of dust clouds.]{\includegraphics[width=0.45\textwidth]{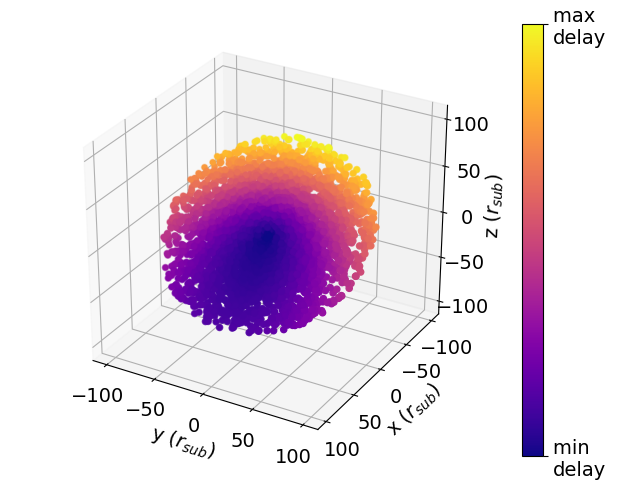}}
        \hspace{0.5cm}
        \subfloat[figure][Illumination map for the distribution of dust clouds.]{\includegraphics[width=0.45\textwidth]{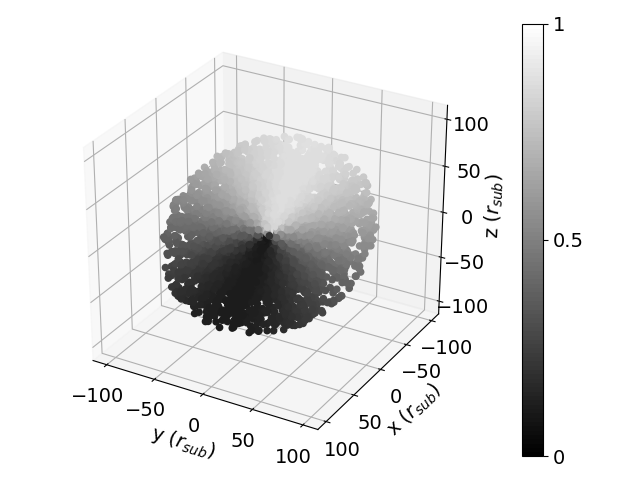}}
    \end{minipage}
    \caption{Distribution of 10,000 dust clouds corresponding to the mean parameters from the MCMC modelling of the Zw229-015 light curves, and the corresponding delay and illumination maps. \label{fig:zw229_dist}} 
\end{figure*}

Each combination of optical and IR light curve of \mbox{Zw229-015} was modelled using the methods described above to attempt to further constrain the inner regions of the AGN. The MCMC modelling was performed using 64 chains, each with a length of 50,000 iterations, which was sufficient for the parameters to converge. The range of priors for each parameter is listed in Table~\ref{tab:params}, where the time lag (for example) is constrained by the results of the DRM in Section~\ref{Sect:DustRM}. The quality of the fit is measured using the reduced $\chi^2$ to compare the simulated light curves to the observations. 

Initially, the entire overlapping light curves were modelled for each combination of optical and IR; for example, Figure~\ref{fig:model_kep_sp1_all} displays the model that corresponds to the maximum of the posterior distribution for the entire \textit{Kepler}-\textit{Spitzer}~1 light curves. It can be seen here that the overall variability trends of the simulated IR light curve on the order of several months to years fits the observations relatively well; however, the shape of the simulated IR light curve for the individual seasons deviates from the observations, specifically in the 2011 and 2012 seasons where the amplitude of variability appears to be overestimated. A dramatic increase in flux occurs between the 2011 and 2012 seasons, which is greater in the IR than optical and could affect the modelling, so to investigate this, only the 2010--2011 seasons were modelled in Figure~\ref{fig:model_kep_sp1_S1-2}. It can be seen that this simulated light curve follows the shape of the 2011 season substantially better than Figure~\ref{fig:model_kep_sp1_all}, both by eye and by the reduced $\chi_\text{IR}^2$, although the flux is now underestimated at the start of the 2010 season. Furthermore, the peak of the simulated light curve in Figure~\ref{fig:model_kep_sp1_S1-2} at $\sim$~HJD~55525 matches the observations better than Figure~\ref{fig:model_kep_sp1_all}. The means of the parameters that correspond to the best-fit DTFs in each of the 64 MCMC chains are compared in Table~\ref{tab:entire_params} for the \textit{Kepler}-\textit{Spitzer}~1 model over 2010--2011 and 2010--2012, and are shown to be consistent for the radial power-law index, vertical scale height power-law index, and inclination angle. However, the amplitude conversion factor is larger for the 2010--2012 simulated light curves, which explains why the variability of the 2011 season appears overestimated. Also, the time lag between optical and IR emission is {\color{black} $\sim 20$ days} in the 2010--2012 model compared to the {\color{black} $\sim 10$ days} found by only fitting the simulated light curves to the 2010--2011 seasons, which again could explain why the simulated peak at $\sim$~HJD~55525 is later than the observations in Figure~\ref{fig:model_kep_sp1_all}. 

Furthermore, in Figure~\ref{fig:model_kep_sp1_S1-2_extended}, the best-fit DTF found for Figure~\ref{fig:model_kep_sp1_S1-2} has been convolved with the entire observed \textit{Kepler} light curve. Here, the flux of the simulated light curve in the 2012 season is underestimated; however, the shape matches the observations well, which can be further seen in Figure~\ref{fig:mean_sub_models}. This suggests that the increase in flux between the 2011 and 2012 seasons is not well modelled by a single dust component; thus, to exclude any effect from the dramatic increase in flux going forward, the entire light curves are instead separated into the 2010--2011 seasons and the 2012--2014 seasons when modelling light curves that cover multiple seasons. The simulated ground-\textit{Spitzer}~1 and ground-\textit{Spitzer}~2 light curves are discussed in Appendix~\ref{ap:more_models} and displayed in Figures~\ref{fig:model_gr_sp1_all} and \ref{fig:model_gr_sp2_all}, respectively. 

Table~\ref{tab:entire_split_params} contains the corresponding mean and 1$\sigma$ uncertainties for each parameter in the MCMC modelling that describe the best-fit DTF in each combination of optical and IR light curves. It can be seen that the results for the radial power-law index and vertical scale height power-law index are consistent over the different combinations of light curves, as each model suggests a shallow radial power-law index with a mean of {\color{black} $\alpha = -0.52 \pm 0.01$} and a small vertical scale height power-law index with a mean of {\color{black} $\beta = 0.14 \pm 0.06$}. The inclination angles are less well constrained in each model than the other parameters, demonstrating large 1$\sigma$ uncertainties of $\sim 20^\circ$--$30^\circ$, but they return an overall mean of {\color{black} $i = 49.1^\circ \pm 2.1^\circ$}. Differences in the time lag can be seen depending on the seasons modelled, as the ground-\textit{Spitzer}~1 and \textit{Kepler}-\textit{Spitzer}~1 light curves over the 2010--2011 seasons find lags of $\sim 10$ days, while the ground-\textit{Spitzer}~1 and ground-\textit{Spitzer}~2 light curves over the 2012--2014 and 2013--2014 seasons (respectively) find longer time lags that tend toward the upper limit of the prior range, with values of $\tau \approx 25$-30 days. This gives an overall mean value of {\color{black} $\tau = 18.5 \pm 9.2$\,days}, but could also imply that the delay between light curves is increasing with time. It is worth noting that the small uncertainties returned for some of the parameters of these modelled IR light curves are likely a product of the simplified clumpy dust models used to simulate the dust transfer functions, and the exclusion of further complexities that can affect the dust transfer functions which are discussed further in Section~\ref{sect:limitations}. 

\subsubsection{Comparing with Models of the Individual Observation Seasons}

As the time lag increases when modelling the multiseason light curves using the later seasons, the individual observation season light curves were also modelled in Appendix~\ref{ap:individ_models} to see whether a change in the parameters was detected over time. The means of the output DTF parameters for these individual season models are compared with the output parameters from the simulated light curves covering multiple observational seasons in Figure~\ref{fig:comp_params}, for the radial power-law index, vertical scale height power-law index, time lag, and inclination angle. Comparisons of the output parameters for the optical-IR amplitude conversion factor, and the offset are given in Figure~\ref{fig:comp_params_extra}.

It can firstly be seen that modelling of the individual seasons is not able to constrain the parameters as well as the multiseason plots, specifically for the radial power-law index and the vertical scale height power-law index which often have uncertainties that cover nearly the entire prior range $-5.5 \leq \alpha \leq -0.5$ and $0.05 \leq \beta \leq 2.05$, respectively. This could be a result of the length of the observations, especially as some individual seasons only cover a range of $\sim 60$ days, which might not be long enough to properly constrain the parameters of the DTFs. Despite the large range in uncertainties, the overall mean is found to be {\color{black} $\alpha = -0.80^{+0.28}_{0.66}$} and {\color{black} $\beta = 0.23^{+0.35}_{-0.14}$}, consistent with the multiseason models.

Similarly, in Figure~\ref{fig:comp_incs}, the inclination angles found for the individual seasons often have a broad spread, with uncertainties covering $\sim 20^\circ$--$40^\circ$, showing that they are not very well constrained. Most combinations of optical and IR light curve are shown to be consistent amongst themselves over the different observation seasons, and the overall mean inclination angle is found to be {\color{black} $i = 49.08^{+2.89}_{-13.17}$ degrees}. However, some models deviate from this value; for example, the \textit{Kepler}-\textit{Spitzer} 1 light curves in the 2012 season find a value of $i \approx 20^\circ$, which again could be a result of the length of the individual seasons not being sufficient to properly constrain the parameters. It can also be seen that the value of the inclination angle that corresponds to the highest posterior distribution in the 2014 season of the ground-\textit{Spitzer}~1 light curve is outside of the 1$\sigma$ uncertainties, which could imply that for this season the MCMC modelling does not search the appropriate range of priors well.

Finally, in Figure~\ref{fig:comp_lags}, the time lags are found to be better constrained in most of the individual observation seasons than the other parameters, especially in the earlier observation seasons. In all individual observation seasons these lags are found with values between $\sim 10$ and 20 days for the ground-\textit{Spitzer}~1 and \textit{Kepler}-\textit{Spitzer}~1 light curves, but the ground-\textit{Spitzer}~2 light curves in the 2013 and 2014 seasons find a higher lag of $\sim 25$--30 days, which could be influencing the larger lag found in ground-\textit{Spitzer}~2 combined 2013--2014 light curves. However, the corresponding ground-\textit{Spitzer}~1 light curves in the individual seasons all find a shorter lag with a value of 10--15 days, although the 2014 season lag has large uncertainties, while the ground-\textit{Spitzer}~1 light curve over the combined 2012--2014 seasons finds the larger lag. Appendix~\ref{ap:individ_models} explores these larger uncertainties in the time lag in the later seasons and finds that the models actually return double-peaked distributions for the time lag with peaks at $\sim 10$ and 30 days. When plotting the models corresponding to the individual peaks, the model for the $\sim 10$ day lag is shown to better replicate the shape of the variability of the IR observations, but the larger lag has a lower reduced $\chi^2$ value. If the increase in time delay is then not a result of the variability in the individual seasons, the difference between the individual and multiple season models could alternatively still be impacted by modelling their long-term variability and the months-long gaps between observations. The overall mean time lag is found with a value of {\color{black} $\tau = 12.89^{+14.54}_{-2.83}$\,days}, or for each combination of optical and IR light curve, the means have values of {\color{black} $11.19^{+1.73}_{-2.97}$ days, $11.04^{+5.99}_{-0.29}$ days, and $27.66^{+0.47}_{-0.04}$ days} for the ground-\textit{Spitzer}~1, \textit{Kepler}-\textit{Spitzer}~1, and ground-\textit{Spitzer}~2 light curves, respectively.

Using the mean radial power-law index, vertical scale height power-law index, and inclination angle from modelling the light curves of Zw229-015, the distribution of dust clouds and the associated delay and illumination maps are plotted in Figure~\ref{fig:zw229_dist}.

\section{Discussion}
\label{Sect:Discuss}

In the previous sections, optical and IR light curves of \mbox{Zw229-015} were studied to further understand the inner regions of the AGN that are not spatially resolved.

\subsection{Dust Reverberation Mapping}

\begin{figure*}
    \subfloat[figure][Plot of the radius of the BLR against 5100\,\AA\,luminosity for AGN from \cite{Bentz2013} in the black squares and for Zw229-015 using the BLR-RM result from \cite{Barth2011} in red.  \label{fig:R-L_BLR} ]{\includegraphics[width=0.49\textwidth]{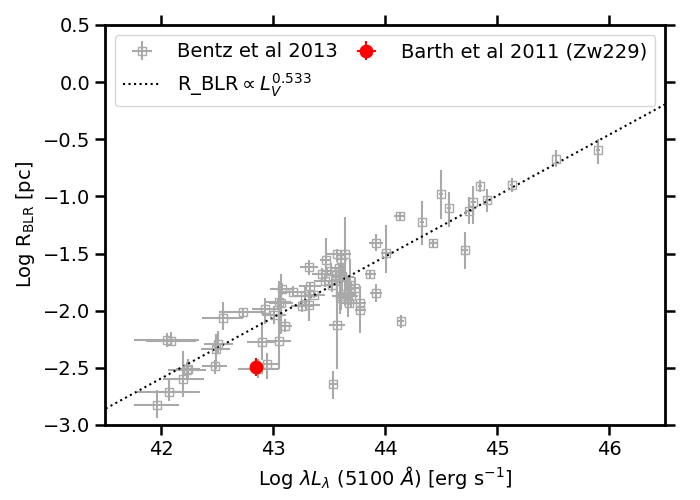}}
    \hfill
    \subfloat[figure][Plot of the radius of the dust emitting region against $V$-band luminosity for AGN from the literature in the black squares \citep{Koshida2014, Minezaki2019} and for Zw229-015 using the DRM result from this paper and from \cite{Mandal2020} in red.  \label{fig:R-L_dust} ]{\includegraphics[width=0.49\textwidth]{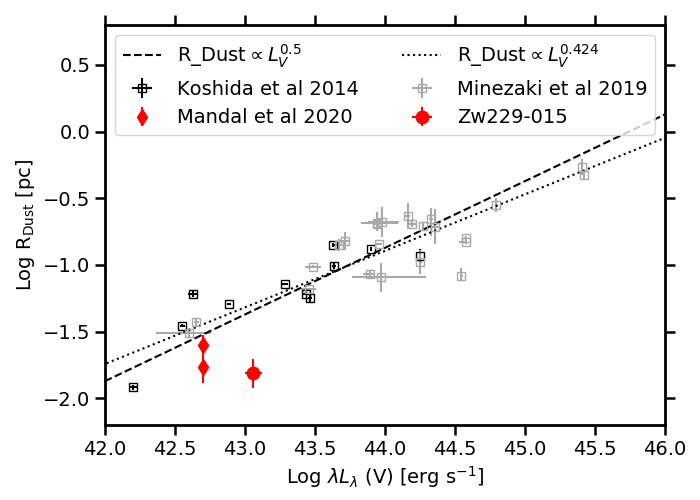}} 
    \caption{Radius-Luminosity relations using AGN from the literature compared to Zw229-015. \label{fig:R-L}}
\end{figure*}

The optical ground-based and \textit{Kepler} light curves were cross correlated with the IR \textit{Spitzer} channel~1 (3.6\,$\mu$m) and channel~2 (4.5\,$\mu$m) light curves to determine the possible lags between them, and their corresponding dust reverberation radii. The most consistent observed lag was detected between 5 and 30\,days in 70\% of the CCFs, with an overall mean of {\color{black}$\tau_\textit{obs} = 18.8 \pm 4.6$\,days}, which corresponds to a radius of {\color{black} $0.016 \pm 0.004$\,pc} (or in the rest frame of \mbox{Zw229-015}, {\color{black}$\tau_\textit{rest} = 18.3 \pm 4.5$\,days}, which corresponds to a radius of {\color{black} $0.015 \pm 0.004$\,pc}). Additional less-consistent lags were also detected in multiple seasons in the different combinations of optical--IR light curves, including a lag between 55 and 80\,days; however, a peak in the corresponding optical ACF in these seasons was detected at $\sim 60$\,days, implying that the lag is not between optical and IR light curves but is a result of aliasing in the light curves. This peak in the optical ACFs is further investigated as a possible periodicity in Appendix~\ref{ap:periodicty}; however, it was found to be a result of the overlapping light-curve regions used for cross correlation, as the entire \textit{Kepler} light curves for each season did not display this peak, nor did the 2013 ground-based light curve or the overall ground and \textit{Kepler} light curves.

The \textit{Spitzer} channel~1 and 2 light curves were also cross correlated with each other to look for possible lags between the emission from different IR wavelength ranges. The most consistent lag which was detected in all CCFs was found between 0 and 5\,days, with a mean of {\color{black}$\tau_\textit{obs} = 1.8 \pm 1.7$\,days} (= $\tau_\textit{rest}$). This $\sim 0$ lag implies that the 3.6\,$\mu$m and 4.5\,$\mu$m light is dominated by emission from the same dust component. This delay was further investigated in the ground-\textit{Spitzer}~1 and ground-\textit{Spitzer}~2 CCFs. In the 2013 season, a lag was only detected in the \mbox{RM-ICCF} method of ground-\textit{Spitzer}~2, but comparing this with the ground-\textit{Spitzer}~1 shows that the 4.5\,$\mu$m lag is longer by $\sim 13$\,days; however, this could be due to the broad nature of the CCFs and the corresponding ACFs, which is explored further in Appendix~\ref{ap:S4_CCF}. Furthermore, comparing the lags detected in the 2014 season of the ground-\textit{Spitzer}~1 and ground-\textit{Spitzer}~2 CCFs shows that they are consistent with each other in the individual methods within 1$\sigma$. 

\subsubsection{IR wavelength dependence on dust lags}

The mean observed lag is found to be consistent between each combination of optical and IR light curves, and also with the rest-frame delay of $20.36^{+5.82}_{-5.68}$\,days measured between the $V$ and $K_s$ bands using observations from 2017--2018 by \cite{Mandal2020}. 

\cite{Lyu2019} found the radius of dust reverberation mapping with longer IR wavelengths to be larger than those of smaller IR wavelengths, as they found a ratio between the $K$ (2.2\,$\mu$m), W1 (3.4\,$\mu$m), and W2 (4.5\,$\mu$m) dust reverberation radii of \mbox{$R_K:R_{W1}:R_{W2} = 0.6:1.0:1.2$} for 17 quasars in the $K$ and W1, and 67 quasars in the W1 and W2. Our results shows anomalous behaviour for Zw229-015 in comparison to the results reported in \cite{Lyu2019} however, as the lags between the optical and 3.6\,$\mu$m, and the optical and 4.5\,$\mu$m are shown to be consistent within the 1$\sigma$ uncertainties of the rest-frame dust reverberation lag measured between the $V$ and $K_s$ bands using observations from 2017--2018 by \cite{Mandal2020}. If we assume that the $V$ and $K_s$ reverberation lag remained approximately consistent in 2010--2018, the similarities between the $K_s$ band (2.15\,$\mu$m), 3.6\,$\mu$m, and 4.5\,$\mu$m reverberation lags indicate that in \mbox{Zw229-015} they are all tracing the same regions, which corresponds to the hottest dust at the inner radius of the dust emitting region. \cite{Honig2011} explain that the NIR emission (i.e., the $K_s$ band) predominantly originates in the peak of the blackbody emission of the hot dust in the inner region, while the predominant source of MIR emission (i.e., the 3.6\,$\mu$m and 4.5\,$\mu$m bands) can be either from the peak of the blackbody emission of cooler dust at larger distances from the central engine than the hot dust, or from the Rayleigh-Jeans tail of the hot-dust emission in the inner regions. If the object is compact, and the brightness distributions are therefore steep, the Rayleigh-Jeans tail of the hot-dust emission can dominate the MIR because of the lack of extended dust. This therefore suggests that in the previous MIR reverberation studies that found a larger lag between the optical and MIR than the optical and NIR (e.g., \citealt{Lyu2019}), the MIR emission is originating in the peak of the blackbody emission of the cooler dust, whereas for \mbox{Zw229-015}, where the lag between the optical and MIR is consistent with the lag between the optical and NIR, the 3.6\,$\mu$m and 4.5\,$\mu$m emission is dominated by the Rayleigh-Jeans tail of the hottest dust.

\subsubsection{Comparison with BLR lags}

The dust reverberation lag measured in this paper is found to be a factor of $\sim 4.7$ larger than the BLR reverberation lag measured by \cite{Barth2011}, consistent with the results found by \cite{Koshida2014}, in which the reverberation radius of broad emission lines is a factor of 4--5 smaller than dust reverberation radii. 

\cite{Czerny2011} proposed that some broad emission lines are formed in a failed radiatively accelerated dusty outflow (FRADO), and therefore predict the existence of a significant amount of dust within the BLR. This could be tested with the high-quality optical and IR data used in this paper, which allows for the detection of lags between light curves on timescales as short as $\sim 3$\,days, consistent with the BLR lag measured by \cite{Barth2011}. As a lag on such timescales is not measured in the high-cadence CCFs, it suggests that the presence of a significant amount of dust in the BLR is unlikely. 

\subsubsection{Reverberation Lag-Luminosity Relations}

Both the dust reverberation radii and BLR reverberation radii are expected to correlate with the optical luminosity of the AGN, with an approximate relationship of $r \propto L^{1/2}$. For example, Figure~\ref{fig:R-L_BLR} shows a plot of $R_\text{BLR}$ vs. optical 5100\,\AA\, luminosity for Zw229-015 and for other AGN taken from the literature \citep{Barth2011,Bentz2013}. For the AGN in \cite{Bentz2013}, the H$\beta$ BLR radius-luminosity relation was found to be $R_\text{BLR} \propto L^{0.533^{+0.035}_{-0.033}}$. Additionally, Figure~\ref{fig:R-L_dust} demonstrates the $R_\text{dust}$ vs. the $V$-band luminosity relationship using the DRM lag estimated for Zw229-015 in this paper and for other AGN taken from the literature \citep[e.g.,][]{Koshida2014, Minezaki2019}. \cite{Minezaki2019} fit the luminosity-dust radius relationship of the dust reverberation mapped AGN from their work and \cite{Koshida2014}, both forcing a fixed slope of 0.5 and allowing the slope to vary which resulted in a best fit to the data of $R_\text{dust} \propto L^{0.424 \pm 0.026}$. Furthermore, \cite{Gravity2020} estimated the hot dust radius vs. bolometric luminosity relation using NIR interferometric measurements, and found a best fit with a flatter slope of $R_\text{dust,int} \propto L_\text{bol}^{0.40 \pm 0.04}$. Alternatively, \cite{Lyu2019} perform a mid-IR reverberation mapping study of 87 quasars, and find radius-luminosity relationships with slopes of $R_\text{dust} \propto L_\text{bol}^{0.47 \pm 0.06}$ and $R_\text{dust} \propto L_\text{bol}^{0.45 \pm 0.05}$ in the W1 ($3.4 \mu$m) and W2 (4.5 $\mu$m) bands respectively, which are both consistent with the theorised relationship of $r \propto L^{1/2}$ within their 1$\sigma$ uncertainties. It is therefore currently unclear whether the deviation of the slope from the relationship of $r \propto L^{1/2}$ is an intrinsic effect: Selection biases could also affect the results as the brightest and most suitably variable objects are typically chosen as targets for both reverberation mapping and interferometric analysis. As it is easier to measure shorter lags in brighter objects, this could therefore lead to a biased selection on the short-lag side from the intrinsic dispersion in the correlation and, hence, flatten the observed slope. Indeed, comparing the distribution of AGN luminosities in Figures~\ref{fig:R-L_BLR} and \ref{fig:R-L_dust}, it can be seen that fewer dust lags of AGN with lower luminosity have been recorded, which means the dust plot could be biased by selection effects from selecting mostly more luminous AGN. Having fewer AGN dust lags for sources with luminosities $\lesssim 10^{43.5} \, \text{erg s}^{-1}$ could suggest that the entire distribution of AGN might not be sampled in this plot. This could be further implied as the dispersion of AGN from \cite{Koshida2014} and \cite{Minezaki2019} from the radius-luminosity relations at a luminosity of $\sim 10^{44} \, \text{erg s}^{-1}$ is greater than at a lower luminosity of $\sim 10^{42.5} \, \text{erg s}^{-1}$. Alternatively, selection biases could also result in steeper observed slopes than predicted. For example, AGN with lower continuum luminosities are likely to contain stronger broad and narrow emission lines compared to the continuum brightness \citep[the Baldwin effect;][]{Baldwin1989}. Due to the stronger emission line contributions, the continuum luminosities inferred from the broad-band imagining filters of these lower luminosity objects could be greater than would be predicted for their measured radii with the expected relationship of $r \propto L^{1/2}$. As this has a greater impact on the lower luminosity objects than the higher luminosity sources, it could therefore potentially lead to a steeper observed relationship between luminosity and radius.

As mentioned above, there are reasons to expect some intrinsic dispersion shown about the best fit trends in both Figure~\ref{fig:R-L_BLR} and \ref{fig:R-L_dust}. Several factors have been considered to explain the dispersion in the distributions. Firstly, the accretion rate of the AGN affects the slope of the $R_\text{BLR}$--$L$ relationship. \cite{Du2016} showed that AGN with higher dimensionless accretion rates and higher Eddington ratios have been linked to shorter H$\beta$ lags, which is theorised to be a result of self shadowing effects of slim accretion disks \citep{Li2010}. The self-shadowing region suppresses the ionizing flux seen by the BLR clouds \citep{Wang2014}, thereby reducing the ionisation parameter at a given distance compared to a normal accretion flow. This leads to a shorter time lag than implied by the luminosity based on the standard relation. Such self-shadowing is not restricted to the BLR, but would also affect dust temperature and, hence, reduce the sublimation radius. As a consequence, a high accretion rate AGN would have a shorter dust time lag. Secondly, the intrinsic scatter of the $R_\text{dust}$--$L$ relation could be due to the distribution of the dust, and whether it is illuminated isotropically or anisotropically by the central AGN \citep{Kawaguchi2010, Kawaguchi2011, Almeyda2017, Almeyda2020}. \cite{Kawaguchi2010} show that for an accretion disk emitting less radiation in the equatorial plane than towards the poles, the dust sublimation radius would be closer to the central black hole than if the emission was isotropic. Therefore, for AGN observed along angles closer to the pole, a higher luminosity would be inferred and the measured radius would be smaller compared to predictions using the this luminosity. Furthermore, most of the lags in Figure~\ref{fig:R-L_dust} are measured using a constant spectral index to correct the NIR fluxes for contributions from the accretion disk, though AGN often demonstrate spectral variability as the flux variations are not necessarily consistent across different wavelength ranges \citep[e.g.,][]{Kishimoto2008}. \cite{Mandal2020} therefore suggest that use of constant spectral index could affect the estimated lags. In their analysis with a constant spectral index, their estimated lag is found to lie closer to the predicted relation as shown in Figure~\ref{fig:R-L_dust}. Alternatively, a significant increase in brightness of the AGN is expected to result in a change in the dust sublimation radius as the dust in the innermost region is destroyed; however, the changing sublimation radius is found to follow the brightening with a delay of a few years. For example, \cite{Kishimoto2013} found that the radius at a given time was correlated not with the instantaneous flux but with the long-term average flux of the AGN in the previous $\sim 6$\,yr.

The results for Zw229-015, both from this paper and from literature, were compared to the $R_\text{dust}$--$L$ and $R_\text{BLR}$--$L$ relations. To put our dust lag measurements into context, we calculate a predicted lag using the correlation found by \cite{Minezaki2019}. We determine a mean luminosity of Zw229-015 using the Galactic extinction-corrected ground-based $V$-band light curves. The predicted time lag was found to be {\color{black}$\sim 60$\,days}, which is a factor of $\sim 3$ larger than the results measured in this paper. Using the theoretical $L^{1/2}$ relation leads to a similar offset. This deviation from the predictions of both the $R_\text{dust} \propto L^{1/2}$ and $R_\text{dust} \propto L^{0.424}$ relations can be seen in Figure~\ref{fig:R-L_dust}, in which $R_\text{dust}$ is plotted against the $V$-band luminosity using the DRM lag estimated for Zw229-015 in this paper and for other AGN taken from the literature \citep[e.g.,][]{Koshida2014, Minezaki2019}. Furthermore, the observed IR lag-luminosity relation of Zw229-015 from \cite{Mandal2020} is also plotted, using both a varying index of the power law for each epoch to estimate the contribution of emission from the accretion disk to the IR light curve, and using a fixed power-law index. Note that the luminosity of Zw229-015 in \cite{Mandal2020} was calculated from a modelled host-galaxy-flux-subtracted light curve, whereas our data point is still expected to have contributions from the host-galaxy flux. Additionally, the predicted BLR time-delay estimated using the relation from \cite{Bentz2013} for the optical 5100\,\AA\, luminosity of Zw229-015 is found to be {\color{black}$\sim 9$\,days}, which is a factor of $\sim 2.5$ larger than the delay measured by \cite{Barth2011}. Figure~\ref{fig:R-L_BLR} shows that the measured BLR radius of Zw229-015 is below the $R_\text{BLR} \propto L^{0.533}$ relation found by \cite{Bentz2013}; however, other objects are also shown to have a similar offsets from this relation, which has been suspected to be Eddington ratio dependent \citep[e.g.][]{Du2016}.

Both Figure~\ref{fig:R-L_dust} and \ref{fig:R-L_BLR} show that the Zw229-015 BLR and dust lags are consistently shorter than predicted by the canonical lag-luminosity relations. As previously mentioned, a recent change in intrinsic luminosity could cause this deviation if the sublimation radius did not adjust to this change yet. For that, we compare the fluxes used in this paper with photometric monitoring taken at earlier epochs. Zw229-015 was observed with the Catalina Real-time Transient Survey between 2005 and 2010 \citep[e.g.,][]{Drake2009} with the observed $V$ band photometry given in Appendix~\ref{ap:data}. The observed magnitudes did not change substantially over the 5 years prior to the data used in this paper. Using the observationally inferred typical delay time for structural changes \citep{Kishimoto2013}, we conclude that a substantial increase in brightness is not responsible for the short time lags measured. Two further potential reasons for a deviation from the relation we introduce above are Eddington ratio or anisotropy. Zw229-015 has been previously estimated to have a relatively low accretion rate \citep{Smith2018} and an inclination angle of $i = 32.9^{+6.1}_{-5.2}$ degrees \citep{Williams2018}. This inclination is quite typical for type 1 AGN and \citet{Du2016} associate offsets from the BLR lag-luminosity relation to high accretion rates, not low ones.

\subsection{Light-Curve Modelling}

The response of the IR dust emission to the optical variability was simulated and compared to observations for each combination of optical and IR light curve of \mbox{Zw229-015} in an attempt to further constrain the properties of the inner regions of the AGN, including the distribution of dust, time lag between light curves, and inclination angle of the system. MCMC modelling was used to find the best-fit dust transfer function (DTF) that, when convolved with a driving optical light curve, best replicated the variability of the IR emission. 

This modelling was initially performed on the entire overlapping light curves; however, Section~\ref{Sect:IRModelling_results} demonstrates that the increase in flux between the 2011 and 2012 seasons in the IR is not well replicated when modelling to fit the shape of the individual seasons in the light curves using a single dust component. The underestimation of the IR flux in the 2012 season could therefore imply that a secondary dust component is contributing to the overall emission, which has previously been seen in other AGN. For example, \cite{gravity2021} show that $\sim 5$\% of the total flux of NGC~3783 in the $K$ band is located 0.6\,pc away in projected distance from the central hot dust component, and they interpret this component as a cloud of dust and gas that is heated by the central AGN. To lessen the impact of the dramatic increase in flux on the results, the modelling was instead performed on the different combinations of optical and IR light curves in the 2010--2011 seasons and the 2012--2014 seasons separately. Furthermore, to examine whether the parameters corresponding to the best-fit DTFs were consistent over time, the modelling was also performed on the individual observation seasons. It was found that the parameters were less well constrained for the individual seasons compared to the multiseason plots, which could be due to the length of the observations of individual season light curves, but for the most part the results were consistent with the multiseason models. 

The radial power-law index was consistently found to be shallow in all of the models except the \textit{Kepler}-\textit{Spitzer}~1 2012 season, and had a mean value of {\color{black} $-0.80^{+0.28}_{-0.66}$}, which implies that the dust distribution is extended. This result contradicts the results from dust reverberation mapping which suggested that the dust should be compact, as the MIR and NIR lag the optical with similar delays, implying that they are tracing the same hot-dust region. An additional contribution to the total IR emission from another dust component located farther away from the central source could explain this, as the modelling could be influenced by the extended nature of the second component and therefore find the overall dust distribution to be extended. For example, multiple dust components were found by \cite{Lyu2021}, who analysed dust reverberation signals over a wavelength range of 1--40\,$\mu$m for NGC~4151, and detected lags corresponding to the sublimation radii of carbon and silicate dust, as well as a lag for cooler dust located farther from the central source.

A low value was consistently found for the vertical scale height power-law index in a majority of models, with a mean value of {\color{black} $\beta = 0.23^{+0.35}_{-0.14}$}, which indicates the dust is distributed in a relatively flat disk. \cite{Garcia-Burillo2021} showed that AGN with low Eddington ratios and/or luminosities are dominated by the disk and thought to display little to no polar dust emission as the radiative pressures are not enough to drive winds, which is observationally confirmed by \cite{Alonso-Herrero2021}. \mbox{Zw229-015} has an Eddington ratio of 0.125 and a bolometric luminosity of $\sim 10^{44}$\,erg\,s$^{-1}$ \citep{Smith2018}, which is therefore consistent with a small vertical scale height power-law index obtained from the modelling.

The time delay that corresponds to the light-travel time between the source of the optical and IR dust emission was consistently measured for a majority of models with values of $\sim 10$--20 days, and the overall mean time lag was found to be {\color{black} $12.89^{+14.54}_{-2.83}$\,days}, consistent with the result found with the DRM in Section~\ref{Sect:DustRM}. However, it was also found that all of the simulated ground-\textit{Spitzer}~2 light curves and the ground-\textit{Spitzer}~1 light curves that covered a combination of observation seasons in 2012--2014 detected larger time lags with values approaching the upper limit of the prior range of $\sim 25$\,days. The larger lags in the ground-\textit{Spitzer}~2 models could imply that the lag is different for the 3.6\,$\mu$m and 4.5\,$\mu$m light curves; however, for the individual seasons the uncertainties cover a broad range of lags. These large uncertainties are explored further in Appendix~\ref{ap:individ_models}, where they are shown to be due to a double-peaked distribution, with peaks at $\sim 10$ and 30\,days. A similar double-peaked distribution of lags is also found in the ground-\textit{Spitzer}~1 light curve in the 2013 season; however, the $\sim 10$\,day peak is stronger for this model. In Figure~\ref{fig:gr_sp2_S5_lag_distrib}, both best-fit models corresponding to each of the peaks in the lag distribution of the ground-\textit{Spitzer}~2 2014 season are plotted, and it is shown that the model corresponding to the $\sim 10$\,day lag follows the shape of the variability of the IR observations better, though the reduced $\chi^2$ is smaller for the $\sim 30$\,day lag model. The larger lag found in some of the individual season models, especially in the ground-\textit{Spitzer}~2 but also in the 2013 ground-\textit{Spitzer}~1 light curve, could therefore influence the larger lag found in the multiseason models in 2012--2014. Additionally, the modelling over these seasons could still be impacted by the long-term variability, or as a result of the large gaps between observations.

Finally, the mean inclination angle found by modelling the light curves was {\color{black} $49.08^{+2.89}_{-13.17}$} degrees, which is consistent with the inclination angle measured by \cite{Williams2018} of $i = 32.9^{+6.1}_{-5.2}$ degrees and \cite{Raimundo2020} of $i = 36.4^{+6.7}_{-6.4}$ degrees within 1$\sigma$.

\subsubsection{Limitations of the Dust Transfer Function Simulations used in Modelling the IR Light Curves in Response to the Optical Emission}
\label{sect:limitations}

In this paper, dust transfer functions are simulated using a simplified model of a clumpy dust distribution. Specifically, this model includes a distribution of dust clouds with both radial and vertical components. For further simplification, the dust clouds are assumed to be uniform in size, each with the same composition, and the effects of the geometrical distribution of dust, as well as orientation and illumination are considered when simulating the dust transfer functions. To create a more realistic model of the dust distribution however, further complex effects would need to be considered.

For example, it is assumed that the observed emission from the dust clouds comes from the surfaces directly illuminated from the central engine, and that the emission from each dust cloud is not attenuated by other dust clouds along the line of sight to the observer. In reality however, as the dust clouds are believed to be optically thick, some clouds could have the line of sight to the central source blocked by others (referred to as "cloud shadowing"). These clouds will therefore not be directly heated by the central engine, but instead will be heated by the diffuse radiation from nearby directly heated clouds. \cite{Almeyda2017} explored the effects of cloud shadowing in their model of the simulated dust emission response to optical emission, and found that cloud shadowing can dramatically effect the transfer functions at wavelengths $< 3.5 \mu$m. In addition, the diffuse radiation field should also heat the non-illuminated sides of the dust clouds, which would result in an increase in flux detected by the observer. Furthermore, the line of sight from a given cloud to the observer can be blocked by an intervening cloud (referred to as "cloud occultation"). \cite{Almeyda2020} showed that cloud occultation can impact the dust transfer functions, resulting in transfer functions that are more sharply peaked at shorter delays, and have shallower tail decays.

Additionally, in the model used in this paper, the optical continuum is assumed to emit isotropically, and therefore the flux absorbed by each cloud is dependent on the distance of the cloud from the AGN. {As mentioned previously, \cite{Kawaguchi2010} find that for an accretion disk that emits less radiation in the equatorial plane than towards the poles, the dust sublimation radius would be closer to the central black hole than if the emission was isotropic.} \cite{Almeyda2017} consider the effects of anisotropic illumination from the accretion disk and find that dust transfer functions for such illumination models typically peak at shorter lags and exhibit narrower peaks as a result of shorter travel times to the inner clouds that are located nearer to the central engine in the equatorial plane.  

Furthermore, a change in sublimation radius of the dust distribution as a result of a significant change in optical emission is not supported in the model used in this paper, though this behaviour is expected to occur in AGN in a realistic scenario \citep[e.g.][]{Kishimoto2013}. However, as discussed previously, for Zw229-015 the observed magnitudes did not change significantly over the 5 years prior to the data used in this paper.

\section{Summary}
\label{Sect:Concl}

We studied the optical and IR variability of Zw229-015, a Seyfert~1 galaxy at $z = 0.028$, using observations from optical ground-based telescopes and the \textit{Kepler} space telescope, and concurrent IR observations from the \textit{Spitzer Space Telescope} at 3.6\,$\mu$m and 4.5\,$\mu$m. The results are summarised below.

\begin{enumerate}
    \item We used multiple methods of cross correlation to measure dust reverberation lags. We found a mean rest-frame lag of $18.3 \pm 4.5$\,days for all combinations of optical and IR light curves, over the entire observation periods and the individual observation seasons. 
    \item No obvious difference was found between the DRM lags recovered using the \textit{Spitzer}~1 (3.6\,$\mu$m) or \textit{Spitzer}~2 (4.5\,$\mu$m) light curves;  furthermore, these lags were consistent with the result found between the $V$ and $K_s$ bands measured by \cite{Mandal2020}, which implies that the different IR wavelength ranges are dominated by dust within the same emission regions.
    \item Measured lags of Zw229-015 are consistently smaller than predictions from the lag-luminosity relations, for both IR dust lags and the BLR lag from \cite{Barth2011}. 
    \item By simulating a simplified, clumpy distribution of dust, the response of the IR dust emission to the optical variability was explored using MCMC modelling to further constrain parameters of the inner regions of the AGN that are not directly observable.
    \item An increase in flux between the 2011 and 2012 observation seasons, which is more extreme in the IR light curve than in the optical, was shown to not be modelled well by a single dust component, which could therefore imply that multiple dust components are contributing to the overall emission. 
    \item The MCMC modelling consistently found that the dust was distributed with a shallow radial power-law index and a relatively small vertical scale height power-law index, implying an extended, flat dust distribution.
    \item The modelling also estimated a mean inclination angle of {\color{black}$49^{+3}_{-13}$} degrees, consistent with the inclination angle found by \cite{Williams2018} and \cite{Raimundo2020} within 1$\sigma$ uncertainties, and also detected a mean projected time lag of {\color{black}$12.9^{+14.5}_{-2.8}$}\,days, consistent with the result found for the DRM.
\end{enumerate}

\section*{Acknowledgements}

E.G. and S.F.H. acknowledge support from the Horizon 2020 ERC Starting Grant \textit{DUST-IN-THE-WIND} (ERC-2015-StG-677117). E.G. is grateful to STFC for funding this PhD work. 
Research at UC Irvine has been supported by NSF grants AST-1412693 and AST-1907290.
A.V.F.'s group at UC Berkeley received financial assistance from the TABASGO Foundation, the Christopher R. Redlich Fund, the U.C. Berkeley Miller Institute for Basic Research in Science (where A.V.F. was a Miller Senior Fellow), and numerous individual donors. 

This paper includes data collected by the {\it Kepler} mission and obtained from the MAST data archive at the Space Telescope Science Institute (STScI). Funding for the {\it Kepler} mission is provided by the NASA Science Mission Directorate. STScI is operated by the Association of Universities for Research in Astronomy, Inc., under NASA contract NAS 5–26555.
This work is based in part on observations made with the \textit{Spitzer Space Telescope}, which was operated by the Jet Propulsion Laboratory, California Institute of Technology, under a contract with NASA. The research was partly carried out at the Jet Propulsion Laboratory, California Institute of Technology, under a contract with the National Aeronautics and Space Administration (80NM0018D0004).

This work makes use of observations from the Las Cumbres Observatory global telescope network.
KAIT and its ongoing operation were made possible by donations from Sun Microsystems, Inc., the Hewlett-Packard Company, AutoScope Corporation, Lick Observatory, the NSF, the University of California, the Sylvia \& Jim Katzman Foundation, and the TABASGO Foundation.
Research at Lick Observatory is partially supported by a generous gift from Google.

M.D.J. thanks the Department of Physics and Astronomy at Brigham Young University for continued support of research efforts at the BYU West Mountain Observatory. Research with the WMO 0.9\,m telescope was supported by NSF PREST grant AST–0618209 during the time the observations used in this investigation were secured.

\section*{Data Availability}

The data underlying this article are available in the article and in its online supplementary material.




\bibliographystyle{mnras}
\bibliography{bib} 



\appendix

\section{Photometric Data}
\label{ap:data}

A portion of the photometric data used throughout this paper for the ground-based optical and IR \textit{Spitzer}~1 and \textit{Spitzer}~2 are given in Table~\ref{tab:data}. These data are available in their entirety in the online journal. 

Additionally, Table~\ref{tab:CRTS_data} contains the mean photometric measurements of Zw229-015 from each epoch observed with the Catalina Real-time Transient Survey between 2005 and 2010. Note that this data likely contains a different amount of host galaxy flux contribution to the observations used in this paper.

\begin{table}
    \caption{Photometry of Zw229-015, with observation epoch in HJD--2,450,000.5 (days), and the corresponding magnitudes with their uncertainties.  \label{tab:data}   
}
    \subfloat[table][Example of the ground-based optical photometry, with the specific telescope listed. \label{tab:gr_opt_data} ]{
       \begin{tabular}{C{0.3\columnwidth}C{0.15\columnwidth}C{0.15\columnwidth}C{0.2\columnwidth}}
        \hline
        HJD--2,450,000.5 (days) & $V$ (mag) & Uncertainty in mag & Telescope\\
        \hline
        5352.30 & 15.980 & 0.0028 & WMO \\
        5357.37 & 16.037 & 0.0030 & WMO \\
        5362.34 & 16.054 & 0.0027 & WMO \\
        5363.42 & 16.063 & 0.0028 & WMO \\
        5364.31 & 16.034 & 0.0037 & WMO \\	
        \hline
    \end{tabular}}                    
    \\
    \subfloat[table][Example of the \textit{Kepler} optical photometry. \label{tab:kep_opt_data} ]{
       \begin{tabular}{C{0.35\columnwidth}C{0.3\columnwidth}C{0.25\columnwidth}}
        \hline
        HJD--2,450,000.5 (days) & \textit{Kepler} (mag) & Uncertainty in mag \\
        \hline
        5277 & 15.8822 & 0.0037 \\
        5278 & 15.8631 & 0.0011 \\
        5279 & 15.8700 & 0.0012 \\
        5280 & 15.8581 & 0.0026 \\
        5281 & 15.8297 & 0.0028 \\	
        \hline
    \end{tabular}}                    
    \\
    \subfloat[table][Example of the \textit{Spitzer} ch1 (3.6\,$\mu$m) photometry. \label{tab:sp1_data} ]{    \begin{tabular}{C{0.35\columnwidth}C{0.3\columnwidth}C{0.25\columnwidth}}
        \hline
        HJD--2,400,000.5 (days) & 3.6\,$\mu$m (mag) & Uncertainty in mag  \\
        \hline
        5414.03 & 11.9340 & 0.0027 \\
        5417.12 & 11.9250 & 0.0028 \\
        5420.37 & 11.9540 & 0.0027 \\
        5423.43 & 11.9690 & 0.0028 \\
        5426.14 & 11.9740 & 0.0027 \\
        \hline
    \end{tabular}}
    \\
    \subfloat[table][Example of the \textit{Spitzer} ch2 (4.5\,$\mu$m) photometry. \label{tab:sp2_data} ]{   \begin{tabular}{C{0.35\columnwidth}C{0.3\columnwidth}C{0.25\columnwidth}}
        \hline
        HJD--2,400,000.5 (days) & 4.5\,$\mu$m (mag) & Uncertainty in mag  \\
        \hline
        6460.23 & 10.8950 & 0.0017 \\
        6463.38 & 10.8900 & 0.0017  \\
        6466.34 & 10.8870 & 0.0017 \\
        6469.84 & 10.8910 & 0.0017 \\
        6472.77 & 10.8920 & 0.0017 \\
        \hline
    \end{tabular}}
\end{table}

\begin{table}
    \centering
    \caption{Mean photometry of Zw229-015 from the Catalina Real Time Transient Survey \citep{Drake2009}. \label{tab:CRTS_data} }
    \begin{tabular}{C{0.35\columnwidth}C{0.3\columnwidth}C{0.25\columnwidth}}
    \hline
    MJD--2,450,000.5 (days) & $V$ (mag) & Uncertainty in mag \\
    \hline
    3524.4 & 14.71 & 0.06 \\
    3669.1 & 14.65 & 0.06 \\
    4586.4 & 14.66 & 0.06 \\
    5104.2 & 14.74 & 0.06 \\
    5339.4 & 14.70 & 0.06 \\
    5366.4 & 14.69 & 0.06 \\
    5369.3 & 14.69 & 0.06 \\
    \hline
    \end{tabular}
\end{table}

\section{Reverberation Mapping}

\subsection{Detrending Multiseason Light Curves}
\label{ap:detrend}

\begin{figure*}
    \begin{minipage}{\textwidth}
        \centering
        \subfloat[figure][CCFs and ACFs of the entire ground (gr)-\textit{Spitzer}~1 (sp1) light curves]{
            \begin{minipage}{0.32\textwidth}
                \includegraphics[width=\textwidth]{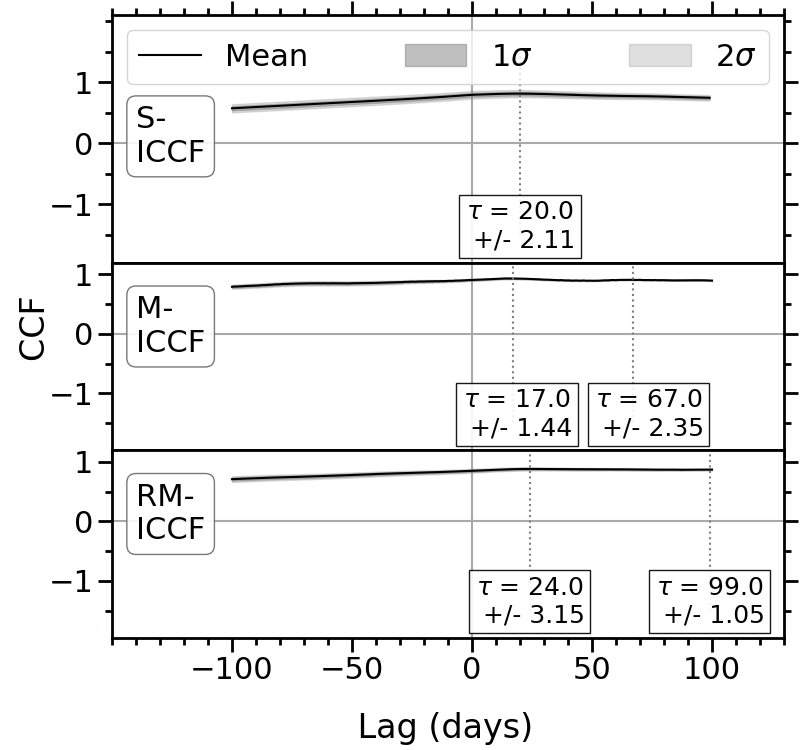} \\
                \includegraphics[width=\textwidth]{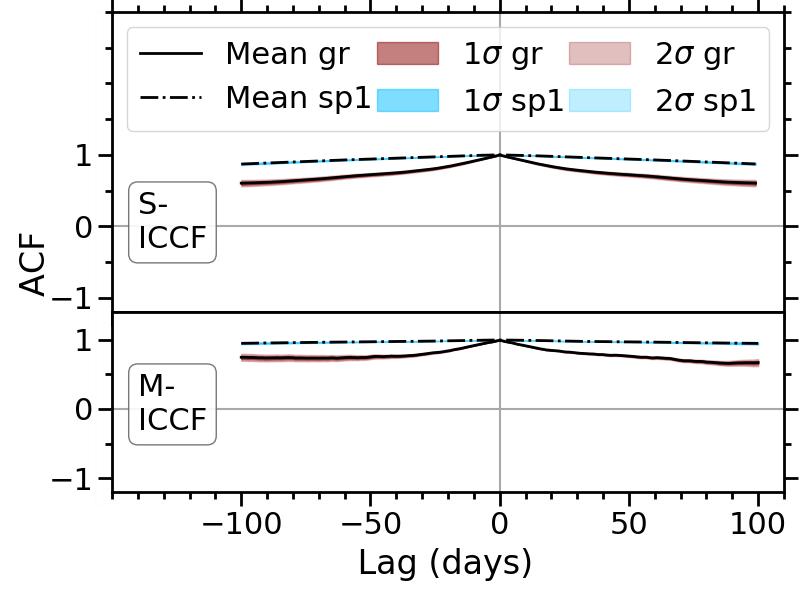}
        \end{minipage}}
        \hfill
        \subfloat[figure][CCFs and ACFs of the entire \textit{Kepler} (kep)-\textit{Spitzer}~1 (sp1) light curves]{
            \begin{minipage}{0.32\textwidth}
                \includegraphics[width=\textwidth]{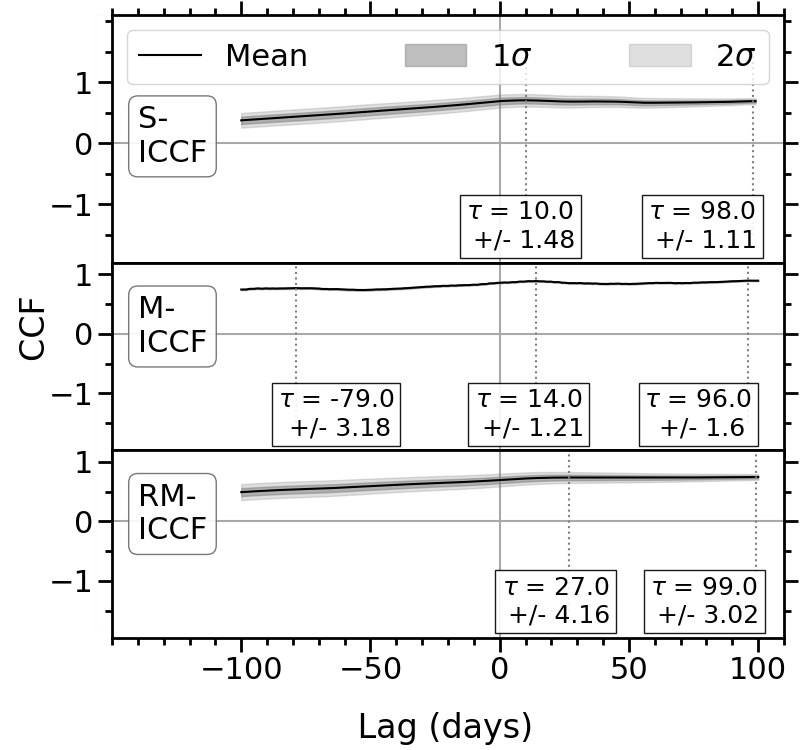} \\
                \includegraphics[width=\textwidth]{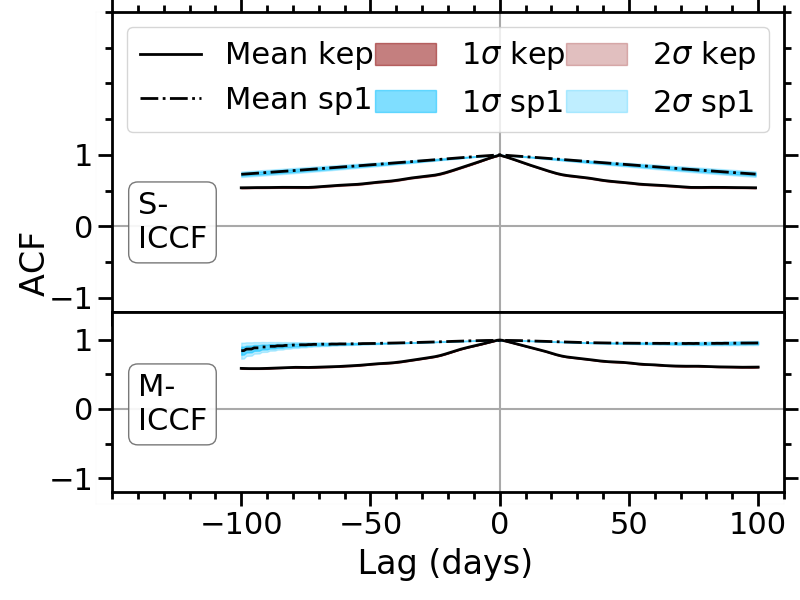}
            \end{minipage}}
        \hfill
        \subfloat[figure][CCFs and ACFs of the entire ground (gr)-\textit{Spitzer}~2 (sp2) light curves]{
            \begin{minipage}{0.32\textwidth}
                \includegraphics[width=\textwidth]{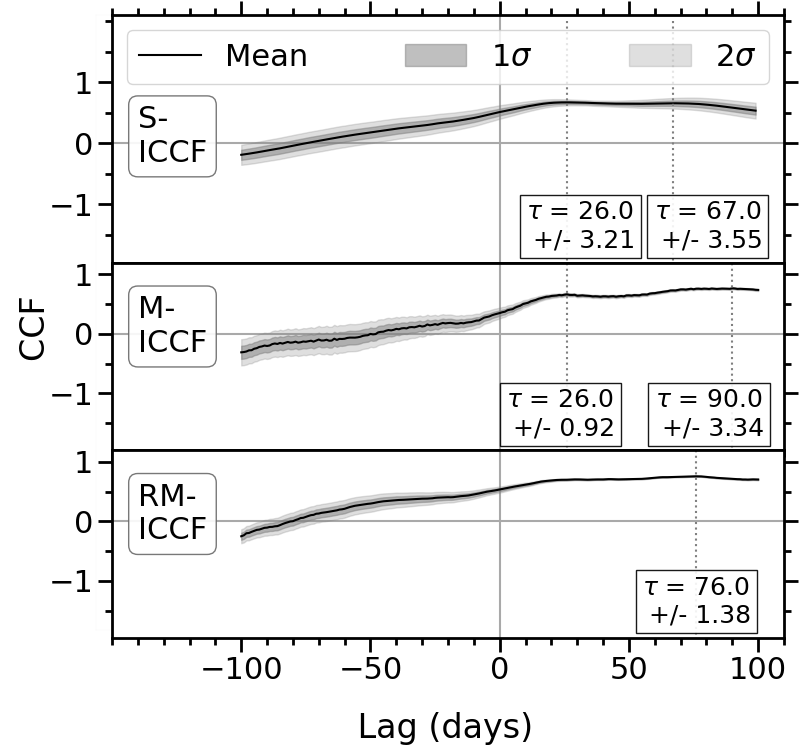} \\
                \includegraphics[width=\textwidth]{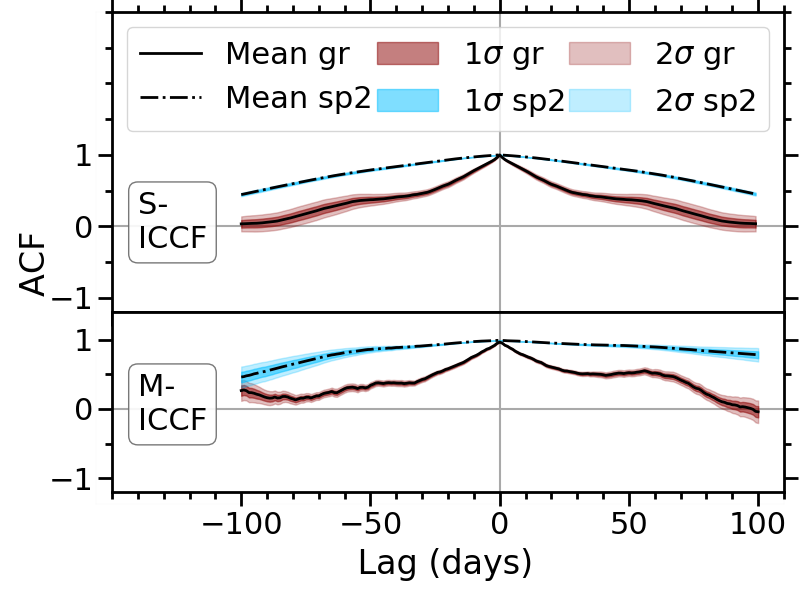}
            \end{minipage}}
        \captionof{figure}{CCF and ACF of each combination of the optical and IR light curves of \mbox{Zw229-015} over the entire overlapping observational periods, without subtraction of long-term variability. \label{fig:not_detrended}}
    \end{minipage}
    \vfill
    \begin{minipage}{\textwidth}
        \centering
        \subfloat[figure][Long-term variability detrended \textit{Kepler} light curve.]{\includegraphics[width=0.46\textwidth]{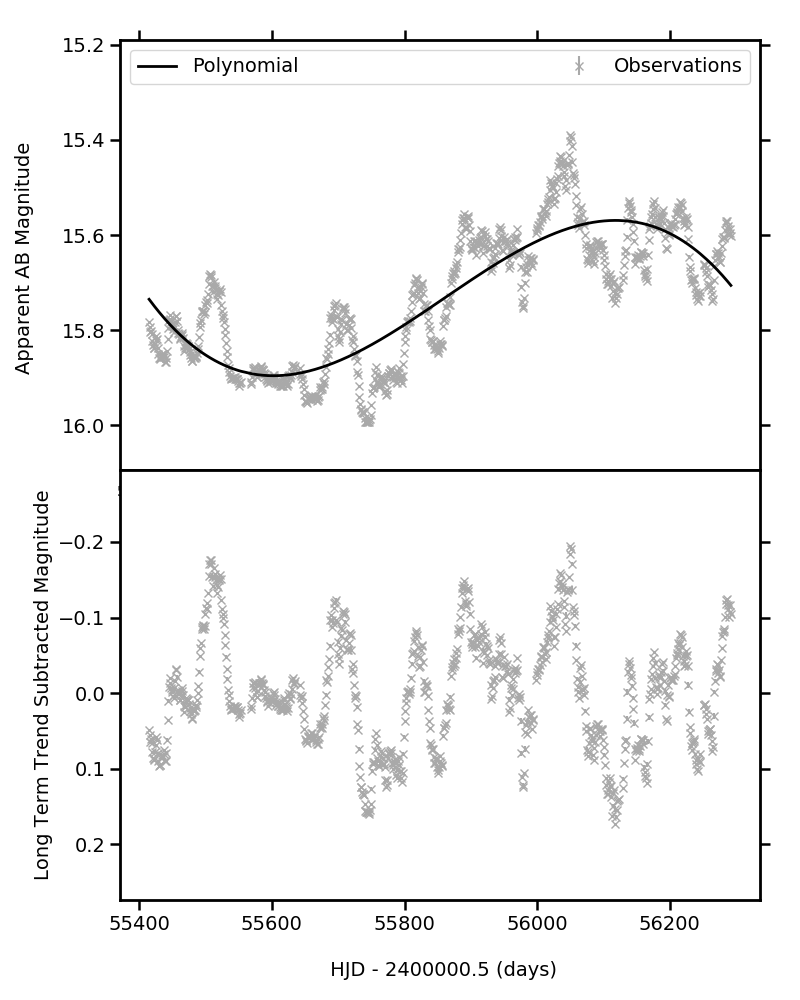}} 
        \hspace{0.3cm}
        \subfloat[figure][Long-term variability detrended \textit{Spitzer}~1 light curve.]{\includegraphics[width=0.46\textwidth]{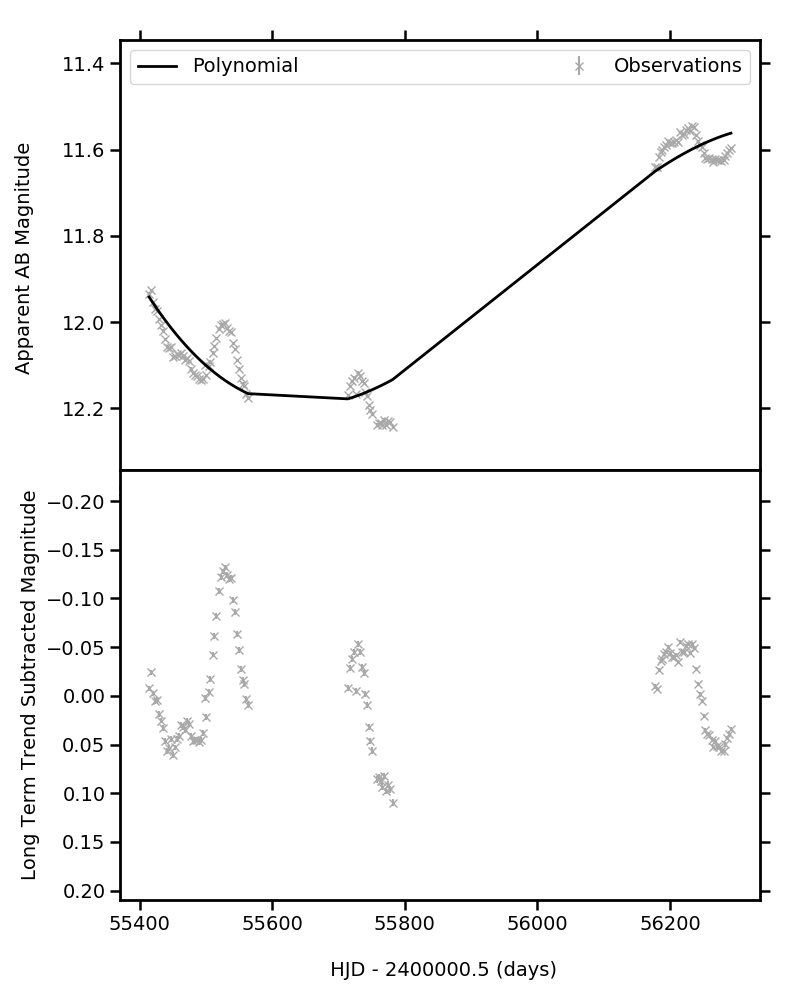}} 
        \caption{\textit{Upper panels:} Examples of the polynomials fitted to the light curves of Zw229-015 for the 2010--2012 observation seasons that demonstrated the long-term trends. \textit{Lower panels:} Examples of the light curves of Zw229-015 detrended of the long-term variations by subtracting the polynomial that was fitted to the data. \label{fig:polysub}}
    \end{minipage}
\end{figure*}

The long-term variations (i.e., the variability on timescales of months to years) could impact the CCFs as shown in Figure~\ref{fig:not_detrended}. For example, in the \textit{Spitzer}~1 light curve in Figure~\ref{fig:zw229lc}, the seasons starting 2012 onward are $\sim 0.5$\,mag brighter than the seasons starting 2010 and 2011, which can result in a CCF that is entirely positive. Therefore, to reduce the impact of the long-term variations, each light curve was detrended by fitting a third-order polynomial and subtracting it, as shown in Figure~\ref{fig:polysub}.

\subsection{Additional CCFs of the Individual Seasons}
\label{ap:more_ccfs}

\begin{figure*}
    \begin{minipage}{\textwidth}
        \centering
        \subfloat[figure][CCFs and ACFs of the 2011 ground and \textit{Spitzer}~1 light curves.]{
            \begin{minipage}{0.32\textwidth}
                \includegraphics[width=\textwidth]{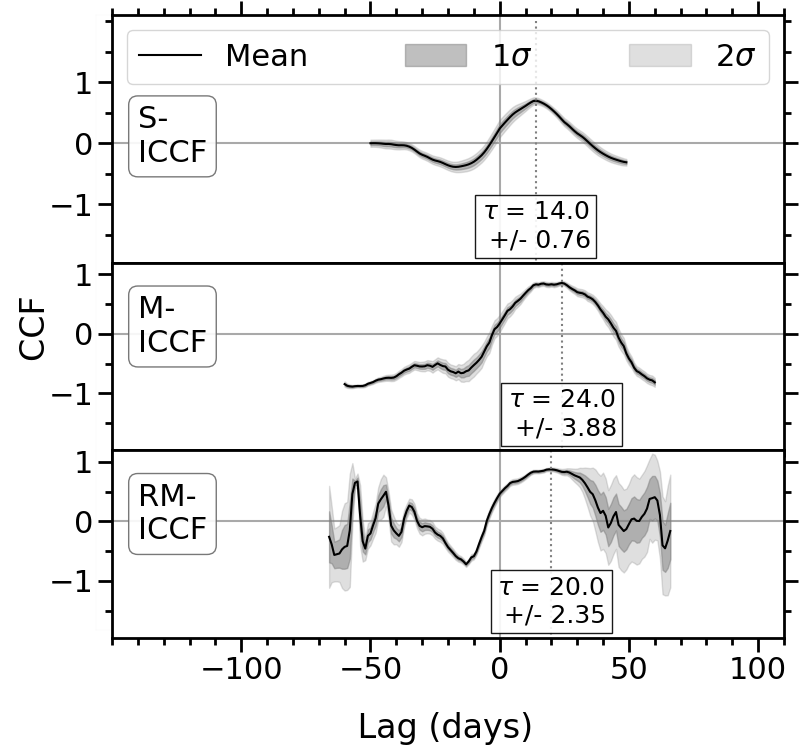} \\
                \includegraphics[width=\textwidth]{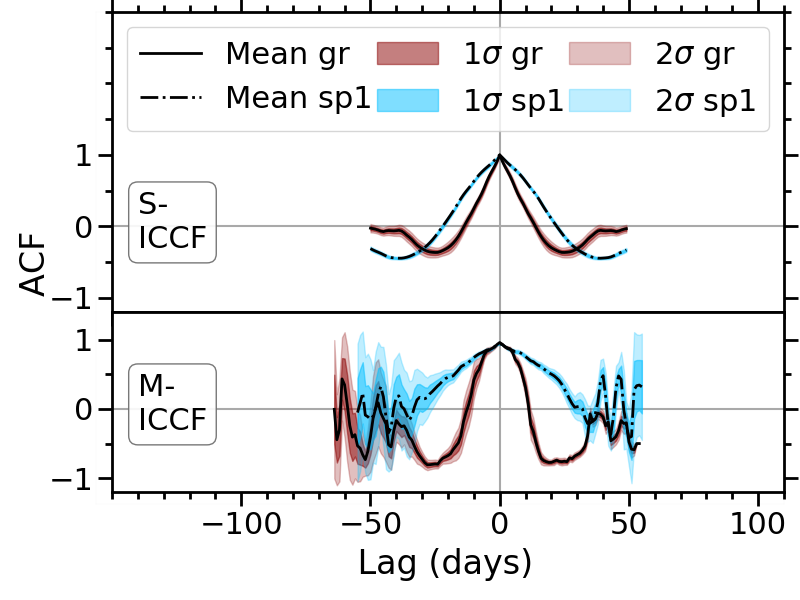}
            \end{minipage}}
        \hspace{2mm}
        \subfloat[figure][CCFs and ACFs of the 2012 ground and \textit{Spitzer}~1 light curves. \label{fig:gr_sp1_CCF_S3}]{
            \begin{minipage}{0.32\textwidth}
                \includegraphics[width=\textwidth]{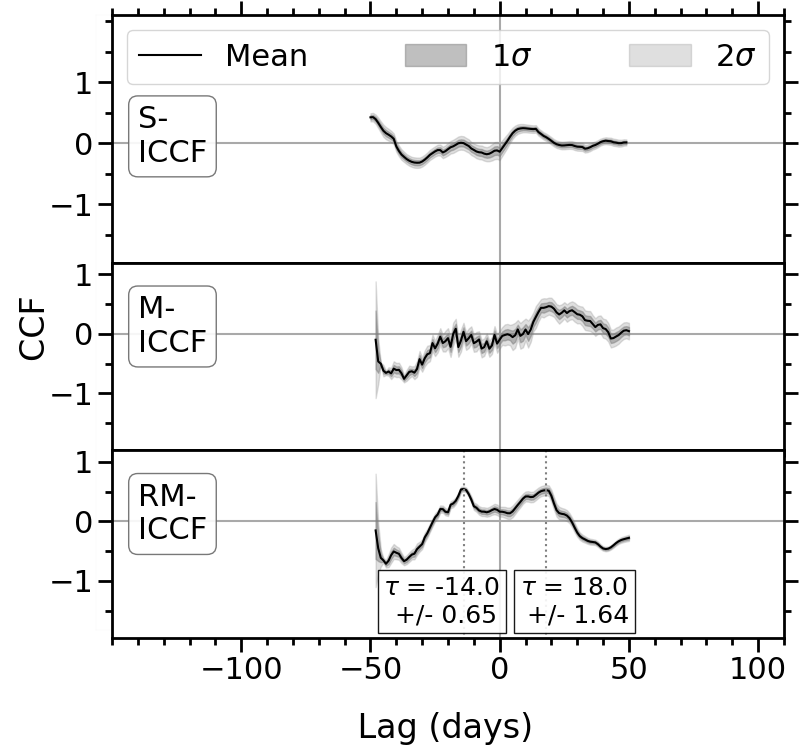} \\
                \includegraphics[width=\textwidth]{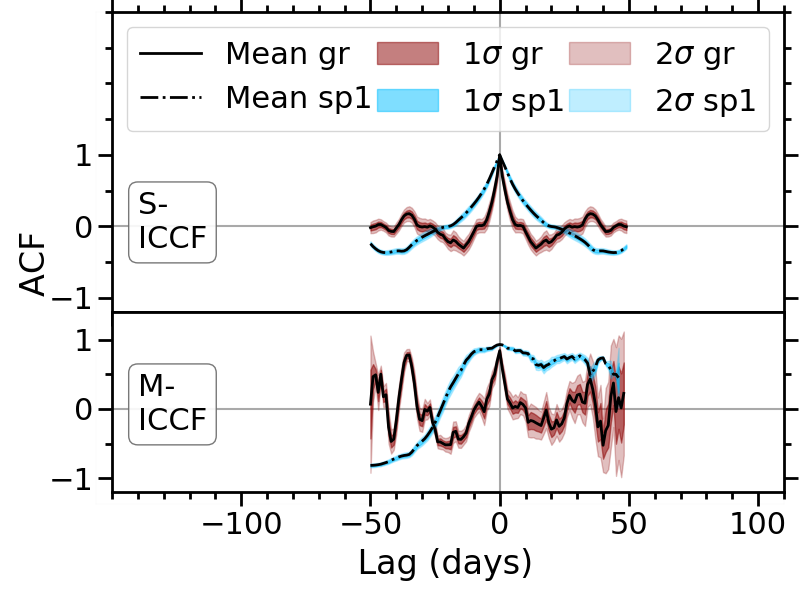}
             \end{minipage}}
        \caption{CCF and ACF of the ground optical (gr) and \textit{Spitzer}~1 (sp1) light curves for some of the individual observation seasons of \mbox{Zw229-015}, made using each method of CCF, where \mbox{M-ICCF} refers to interpolating the ground light curve and \mbox{RM-ICCF} refers to interpolating the \textit{Spitzer}~1 light curve. \label{fig:gr_sp1_CCFs_years_more}}
    \end{minipage}
    \vfill
    \begin{minipage}{\textwidth}
    \centering
        \subfloat[figure][CCFs and ACFs of the 2010 \textit{Kepler} and \textit{Spitzer}~1 light curves. \label{fig:kep_sp1_CCF_S1}]{
            \begin{minipage}{0.32\textwidth}
                \includegraphics[width=\textwidth]{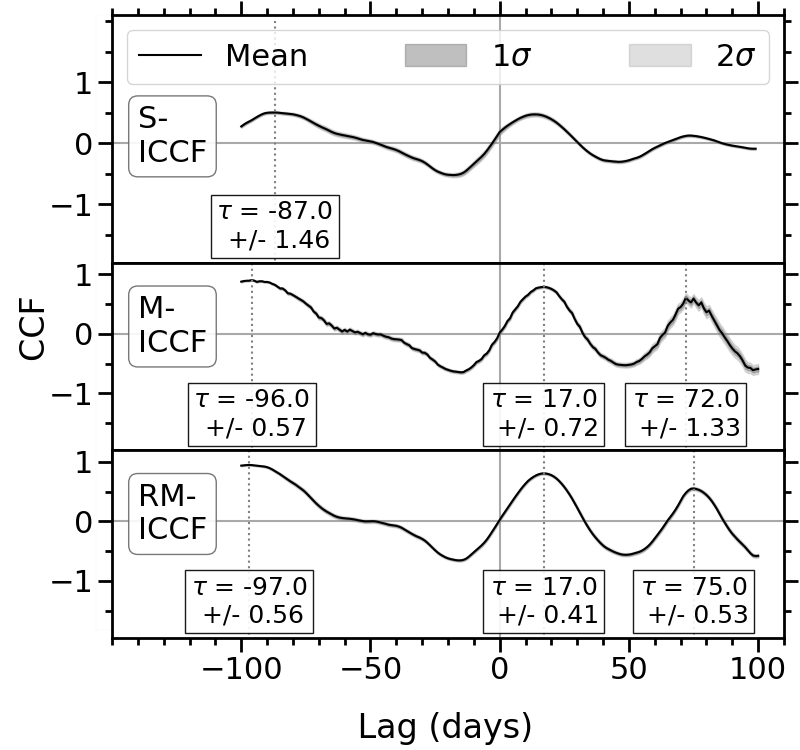} \\
                \includegraphics[width=\textwidth]{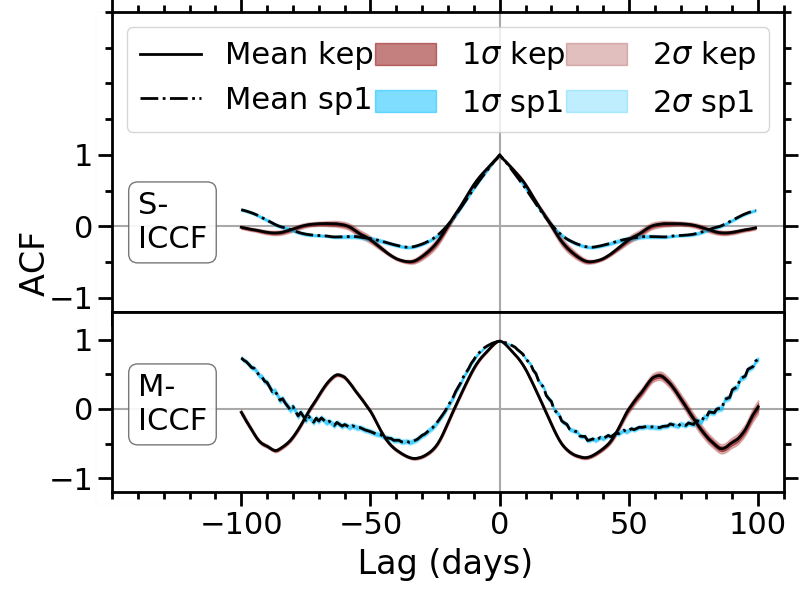}
            \end{minipage}}
        \hfill
        \subfloat[figure][CCFs and ACFs of the 2011 \textit{Kepler} and \textit{Spitzer} light curves. \label{fig:kep_sp1_CCF_S2}]{
            \begin{minipage}{0.32\textwidth}
                \includegraphics[width=\textwidth]{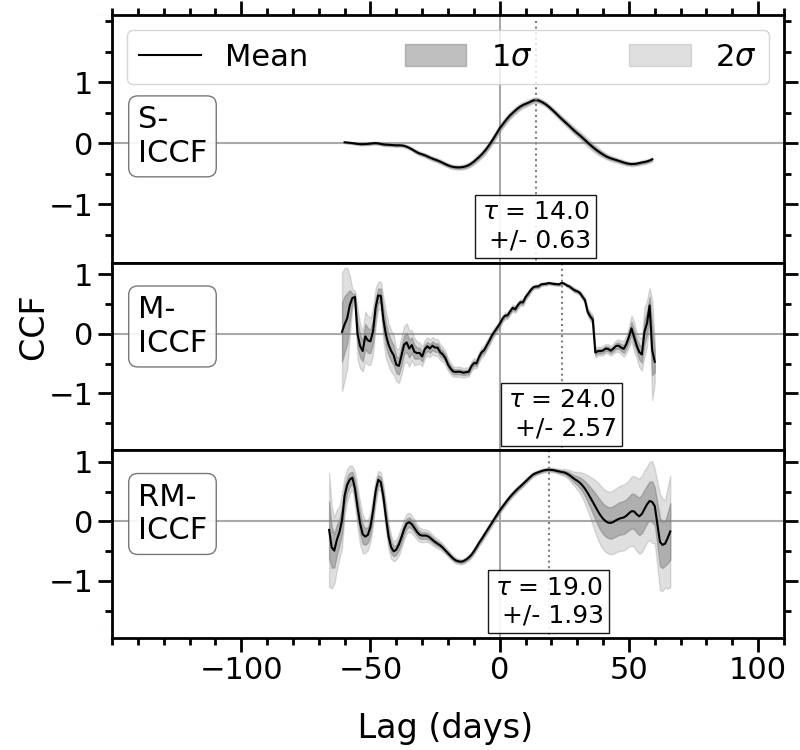} \\
                \includegraphics[width=\textwidth]{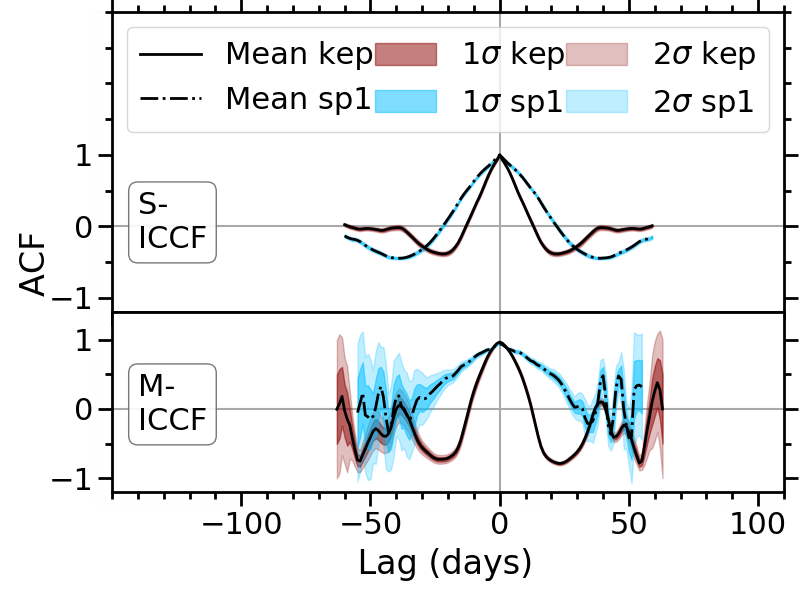}
            \end{minipage}}
        \hfill
        \subfloat[figure][CCFs and ACFs of the 2012 \textit{Kepler} and \textit{Spitzer}~1 light curves. \label{fig:kep_sp1_CCF_S3}]{
            \begin{minipage}{0.32\textwidth}
                \includegraphics[width=\textwidth]{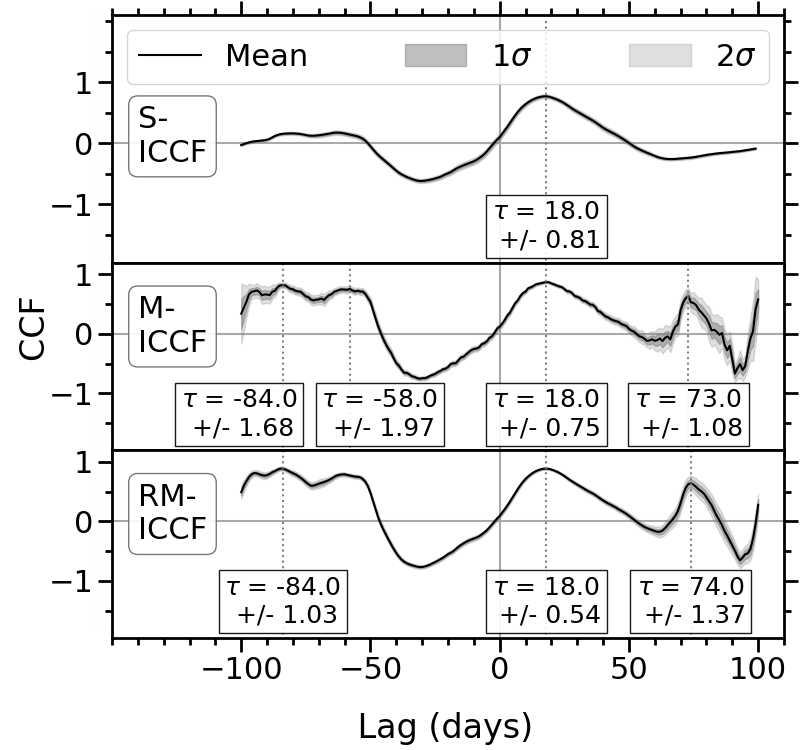} \\
                \includegraphics[width=\textwidth]{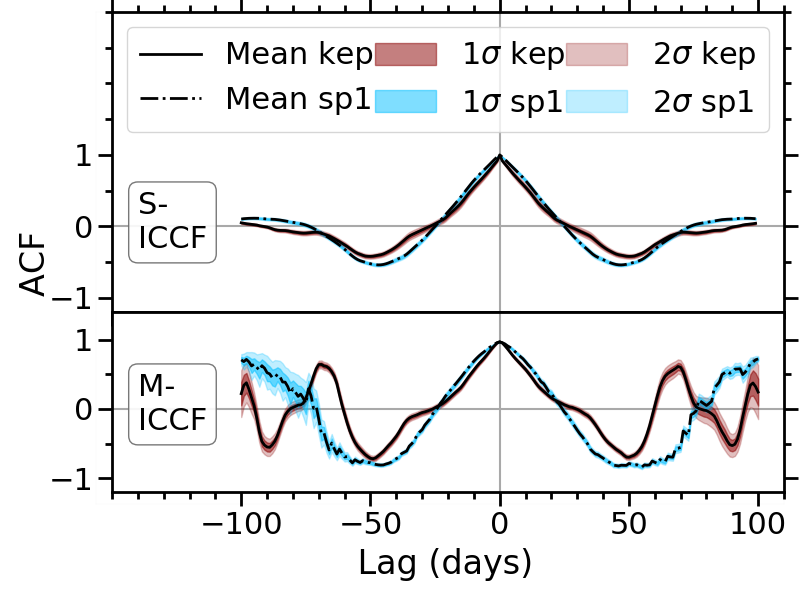}
            \end{minipage}}
        \caption{CCF and ACF of the \textit{Kepler} (kep) and \textit{Spitzer}~1 (sp1) light curves for the individual observation seasons of \mbox{Zw229-015}, made using each method of CCF, where \mbox{M-ICCF} refers to interpolating the \textit{Kepler} light curve and \mbox{RM-ICCF} refers to interpolating the \textit{Spitzer}~1 light curve. \label{fig:kep_sp1_CCFs_years_more}}
    \end{minipage}
\end{figure*}

\begin{figure*}
    \begin{minipage}{\textwidth}
    \centering
        \subfloat[figure][CCFs and ACFs of the 2013 ground and \textit{Spitzer}~2 light curves. \label{fig:gr_sp2_CCF_S4}]{
            \begin{minipage}{0.32\textwidth}
                \includegraphics[width=\textwidth]{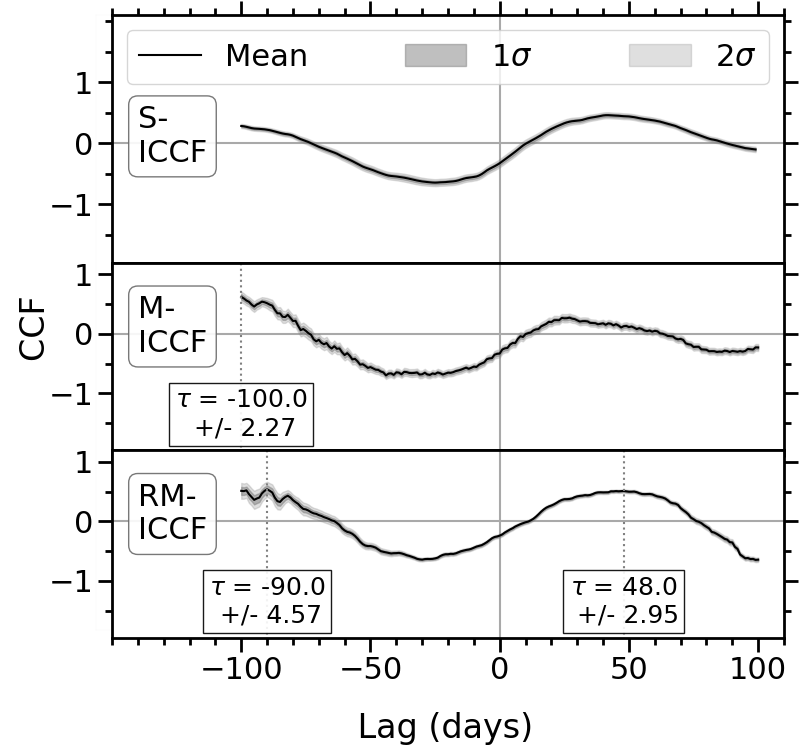}
                \\
                \includegraphics[width=\textwidth]{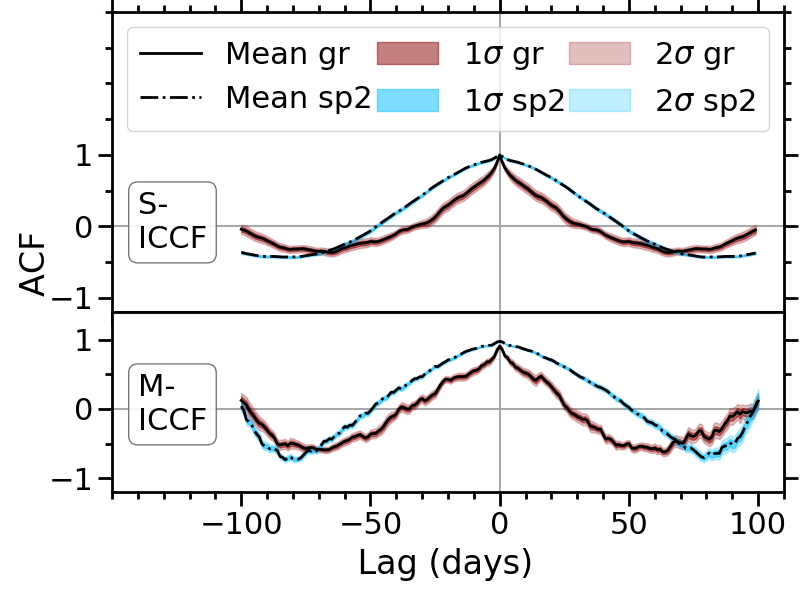}
            \end{minipage}} 
        \hspace{2mm}
        \subfloat[figure][CCFs and ACFs of the 2014 ground and \textit{Spitzer}~2 light curves. \label{fig:gr_sp2_CCF_S5}]{
            \begin{minipage}{0.32\textwidth}
                 \includegraphics[width=\textwidth]{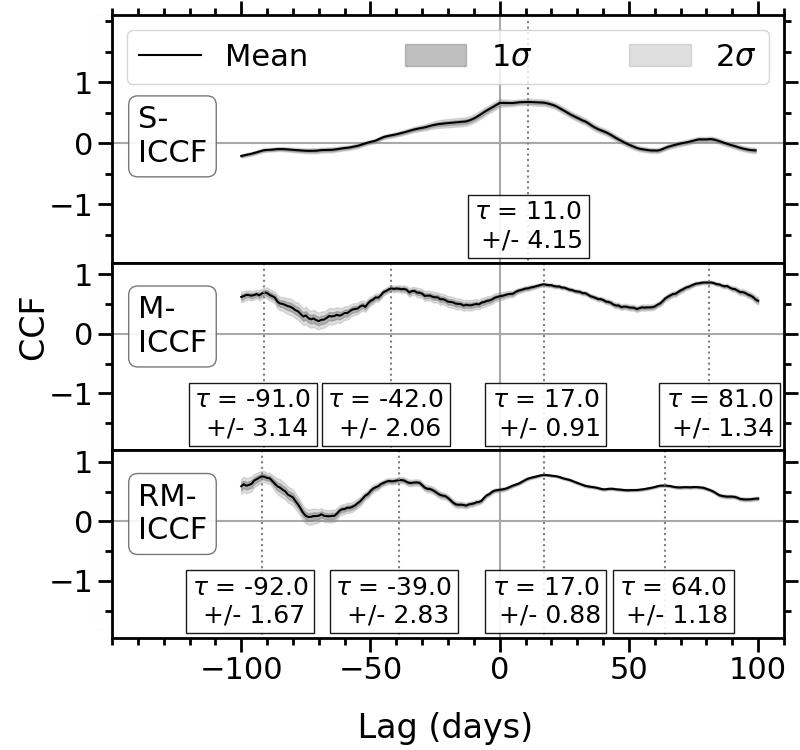} \\
                \includegraphics[width=\textwidth]{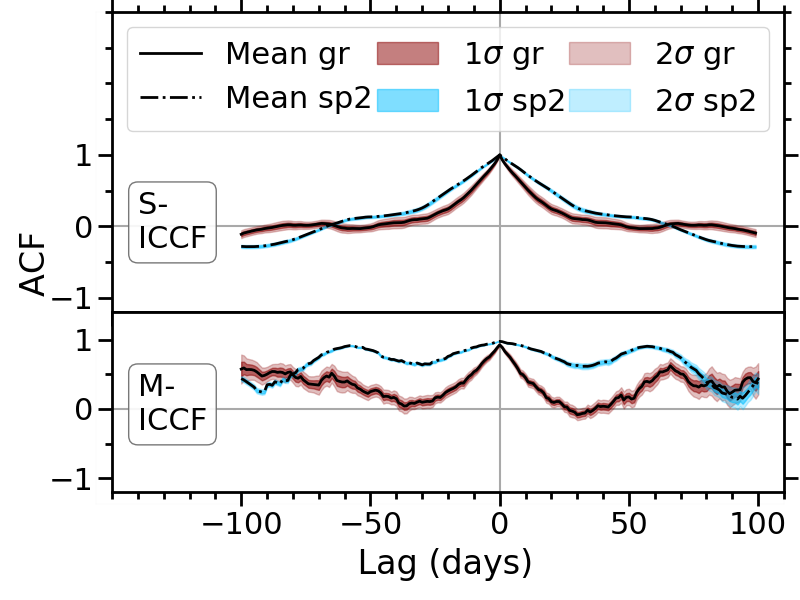}
            \end{minipage}}
        \caption{CCF and ACF of the ground optical (gr) and \textit{Spitzer}~2 (sp2) light curves for the individual observation seasons and the entire light curves of \mbox{Zw229-015}, made using each method of CCF, where \mbox{M-ICCF} refers to interpolating the ground light curve and \mbox{RM-ICCF} refers to interpolating the \textit{Spitzer}~2 light curve. \label{fig:gr_sp2_CCFs_years_more}}
    \end{minipage}
    \vfill
    \begin{minipage}{\textwidth}
        \centering
        \subfloat[figure][CCFs and ACFs of the 2013 \textit{Spitzer}~1 and \textit{Spitzer}~2 light curves. \label{fig:sp1_sp2_CCF_S4}]{
            \begin{minipage}{0.32\textwidth}
                \includegraphics[width=\textwidth]{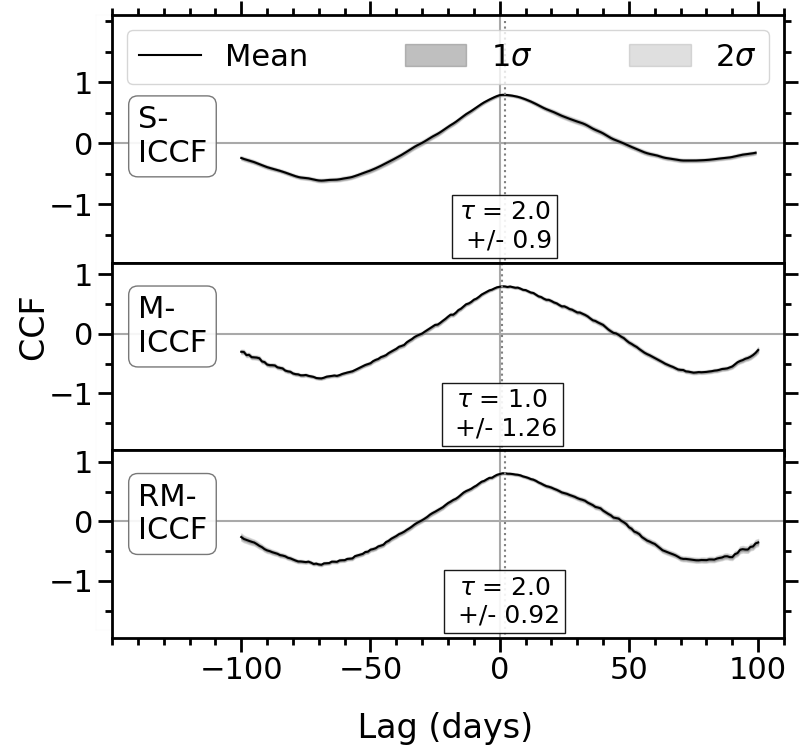} \\
                \includegraphics[width=\textwidth]{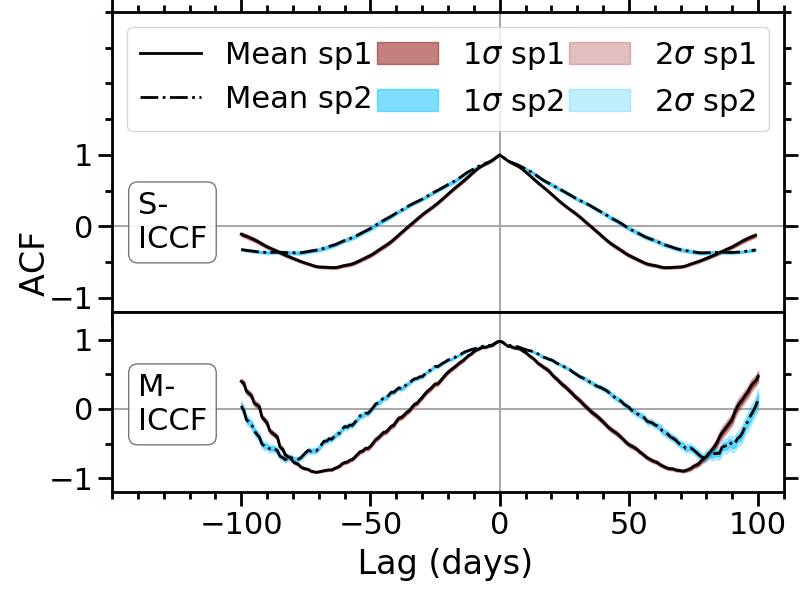}
            \end{minipage}}
        \hspace{2mm}
        \subfloat[figure][CCFs and ACFs of the 2014 \textit{Spitzer}~1 and \textit{Spitzer}~2 light curves. \label{fig:sp1_sp2_CCF_S5}]{
            \begin{minipage}{0.32\textwidth}
                \includegraphics[width=\textwidth]{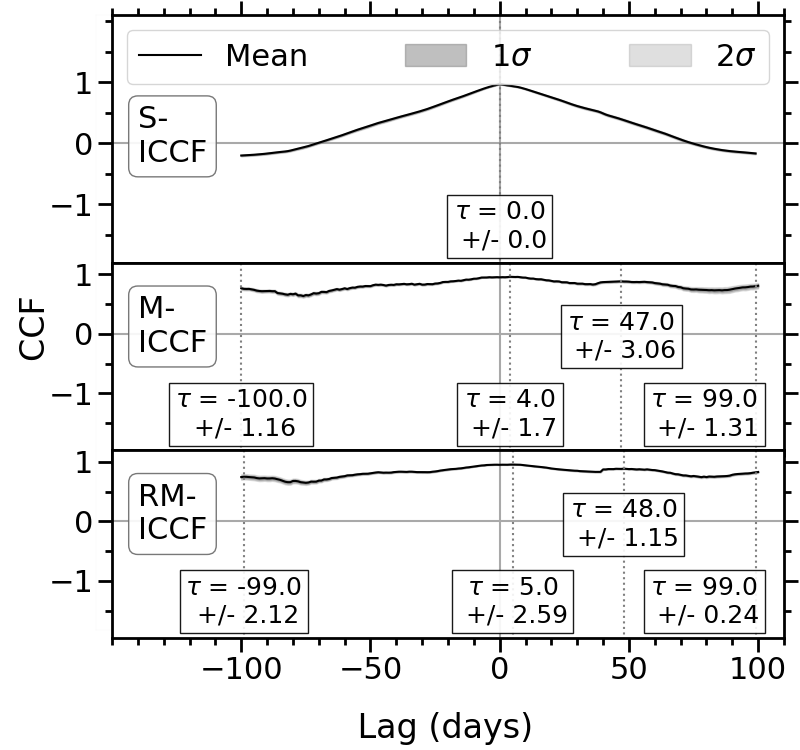} \\
                \includegraphics[width=\textwidth]{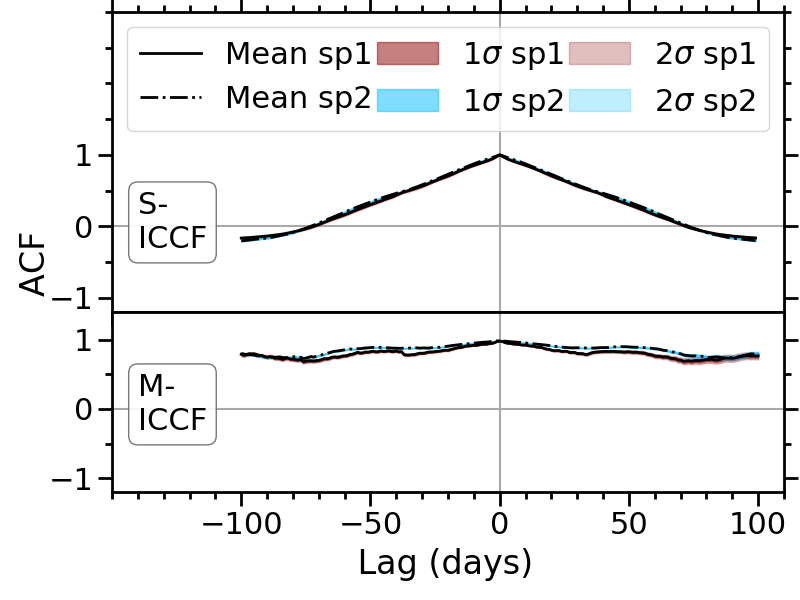}
            \end{minipage}}
        \vfill
        \caption{CCF and ACF of the \textit{Spitzer}~1 (sp1) and \textit{Spitzer}~2 (sp2) light curves for the individual observation seasons and the entire light curves of \mbox{Zw229-015}, made using each method of CCF, where \mbox{M-ICCF} refers to interpolating the \textit{Spitzer}~1 light curve and \mbox{RM-ICCF} refers to interpolating the \textit{Spitzer}~2 light curve. \label{fig:sp1_sp2_CCFs_years_more} \vspace{1cm}}
    \end{minipage}
\end{figure*}

The remaining CCFs and the corresponding ACFs of each combination of optical and IR light curve in each observation season are displayed in Figures~\ref{fig:gr_sp1_CCFs_years_more}, \ref{fig:kep_sp1_CCFs_years_more} and \ref{fig:gr_sp2_CCFs_years_more} for the ground-{\it Spitzer}~1, \textit{Kepler}-{\it Spitzer}~1, and ground-{\it Spitzer}~2 light curves, respectively, for each method of cross correlation as described in Section~\ref{Sect:DustRM}. Similarly, Figure~\ref{fig:sp1_sp2_CCFs_years_more} displays the \textit{Spitzer}~1-{\it Spitzer}~2 CCFs of the individual seasons. Unless otherwise stated, each CCF is tested with a lag range of $\pm 100$\,days owing to the length of the light curves in the individual seasons. Potential lags are measured from the peaks of the CCFs and are considered positive detections if the values are $> 0.5$ as discussed in Section~\ref{sect:CCFResults}, and are labelled on the plots. 

Figure \ref{fig:gr_sp1_CCFs_years_more} contains the CCFs of the ground-{\it Spitzer}~1 light curves in the 2011 and 2012 seasons, where the CCFs are only tested with lags between $\pm 60$\,days and $\pm 50$\,days for the 2011 and 2012 seasons, respectively, owing to the length of overlapping regions of the light curves. Each method of CCF for the season starting in 2011 detects a lag at $\sim 20$\,days, as does the RM-ICCF method of the 2012 season; however, the S-ICCF and M-ICCF methods in the 2012 season do not detect lags, which is likely due to the sampling of the light curves. 

Figure~\ref{fig:kep_sp1_CCFs_years_more} contains the \textit{Kepler}-{\it Spitzer}~1 CCFs for the individual seasons. The 2011 CCFs are only tested with a lag range of $\pm 60$\,days owing to the length of the \textit{Spitzer}~1 light curve in that season. Each method of CCF in the 2011 and 2012 seasons detects the lag at $\sim 20$\,days, as do the M-ICCF and RM-ICCF methods of the 2010 season. The S-ICCF method of the 2010 season does not detect a lag at $\sim 20$\,days, but it does display a peak at that time with a value $< 0.5$, which could imply that the interpolations are diluting the correlations. Additional lags are measured at $\sim 75$\,days in the M-ICCF and RM-ICCF methods of the 2010 and 2012 seasons, which could be due to aliasing of the light curves as peaks at $\sim 50$--60\,days are found in the corresponding optical ACFs. Finally, lags at $\sim -90$\,days are found in some of the CCF methods in the 2010 and 2012 seasons, but as described in Section~\ref{Sect:DustRM}, these are deemed unlikely as they occur only because of a small number of overlapping data points. 

Figure~\ref{fig:gr_sp2_CCFs_years_more} contains the ground-{\it Spitzer}~2 CCFs for the seasons starting 2013 and 2014. The 2014 season detects a lag at $\sim 15$\,days in each method of CCF, but also detects possible lags at additional intervals of $\sim 60$\,days which could be due to aliasing in the light curves. The 2013 season only detects a possible positive lag in the RM-ICCF method with a value of $ 48.00 \pm 2.95$\,days, which is much larger than the positive lags detected below 50\,days in the other seasons and other combinations of optical and IR light curves. This is further explored in Appendix~\ref{ap:S4_CCF}. The M-ICCF and RM-ICCF also detect a possible lag at $\sim -95$\,days; however, as this occurs owing to a small number of overlapping data points in the light curves, this is deemed unlikely.

Finally, Figure~\ref{fig:sp1_sp2_CCFs_years_more} contains the \textit{Spitzer}~1-{\it Spitzer}~2 CCFs for the seasons starting in 2013 and 2014. Each method of CCF in the 2013 and 2014 season detects a lag at $\sim 2$\,days, but the M-ICCF and RM-ICCF methods of the 2014 season also detect lags at intervals of $\sim 45$--50\,days which could be due to aliasing in the light curves.

\begin{figure*}
    \begin{minipage}{\textwidth}
    \centering
        \includegraphics[width=0.33\columnwidth]{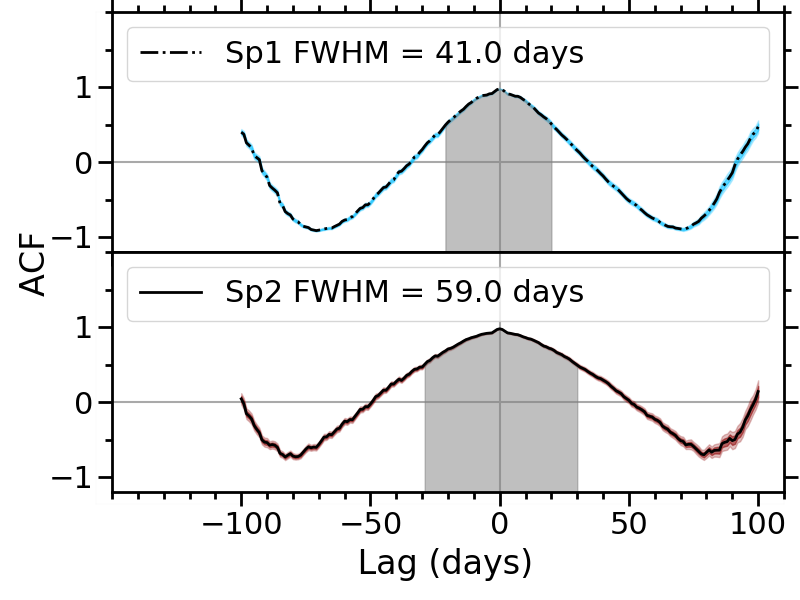}
        \caption{\mbox{M-ICCF} ACFs of the 2013 \textit{Spitzer}~1 and \textit{Spitzer}~2 light curves, with their corresponding FWHMs labelled. \label{fig:sp1_sp2_ACF_S4_FWHM}}
    \end{minipage}
    \vfill
    \begin{minipage}{\textwidth}
        \centering
        \subfloat[figure][ACFs of the 2010 season \textit{Kepler} light curves.
        \label{fig:alt_kepS1_ACF}]{\includegraphics[width=0.32\textwidth]{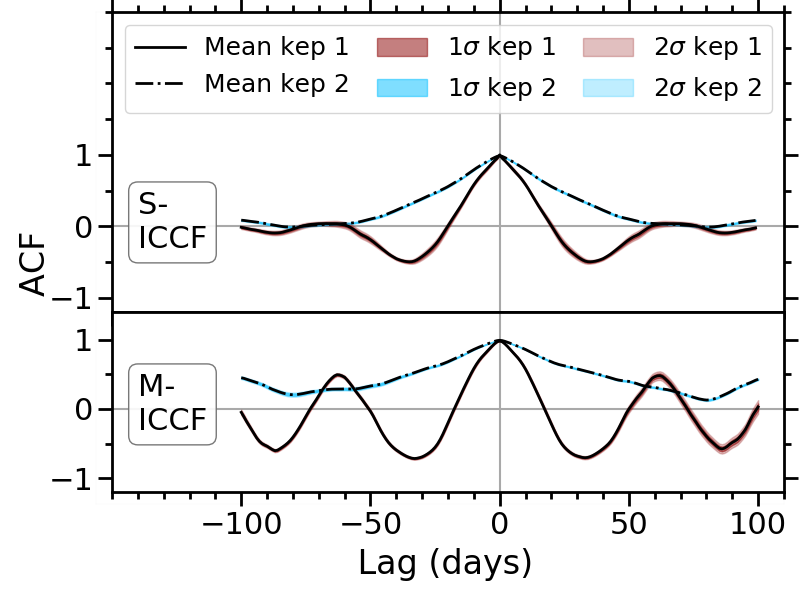}}
        \hspace{2mm}
        \subfloat[figure][ACFs of the 2012 season \textit{Kepler} light curves. \label{fig:alt_kepS3_ACF}]{\includegraphics[width=0.32\textwidth]{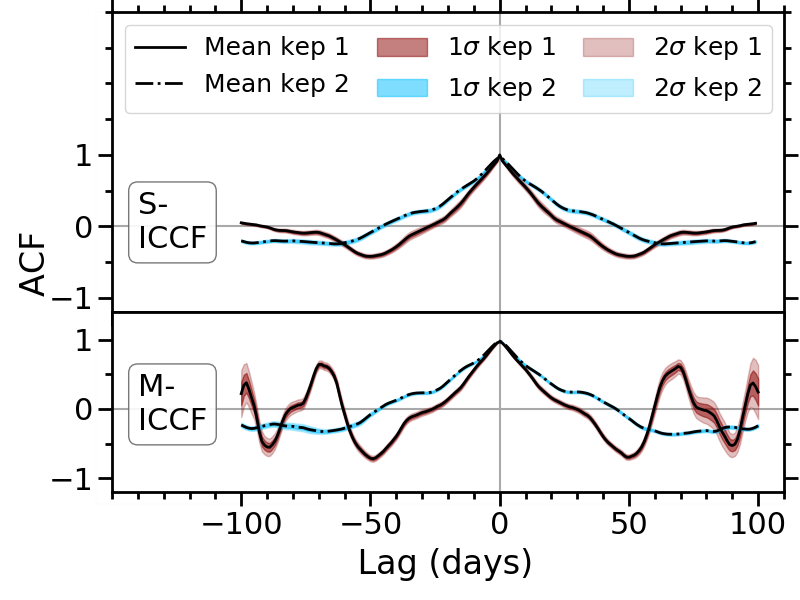}}
        \caption{ACFs of the entire seasons \textit{Kepler} light curves (kep 2) compared to the ACFs of the portion of \textit{Kepler} light curves that overlap with the IR observations (kep 1), in both the \mbox{S-ICCF} and \mbox{M-ICCF} methods. \label{fig:alt_kep_ACF}}
    \end{minipage}
\end{figure*}

\subsubsection{2013 CCFs and ACFs}
\label{ap:S4_CCF}

The CCFs in the 2013 season are broader than in other seasons, as shown in Figures~\ref{fig:gr_sp1_CCF_S4} and \ref{fig:gr_sp2_CCF_S4}, which is mirrored in the breadth of the ACFs, with the \textit{Spitzer}~2 \mbox{M-ICCF} ACF covering the widest range. While the lags detected in the ground-{\it Spitzer}~1 2013 CCFs are consistent with the other seasons with values between \mbox{$\sim 5$ and 30\,days}, the 2013 season of the ground-{\it Spitzer}~2 CCF detects a possible lag at \mbox{$48.00 \pm 2.95$}\,days in the \mbox{RM-ICCF} method, which is much larger than the lags detected below 50\,days in the other seasons. The increase in possible detected lag between the \textit{Spitzer}~1 and 2 \mbox{M-ICCF} CCFs could be a result of the breadth of the 2013 season \textit{Spitzer}~2 \mbox{M-ICCF} ACF.

To further explore these detected lags, the full width at half-maximum (FWHM) of the corresponding \mbox{M-ICCF} ACFs were measured, as shown in Figure~\ref{fig:sp1_sp2_ACF_S4_FWHM}. The ground-\textit{Spitzer}~1 \mbox{RM-ICCF} detected a lag at $\tau = 31.00 \pm 2.53$\,days, whereas the ground-\textit{Spitzer}~2 \mbox{RM-ICCF} detected a much larger lag at $\tau = 48.00 \pm 2.95$\,days, but both CCFs showed a broad peak around this lag, which is reflected in the ACFs. Figure~\ref{fig:sp1_sp2_ACF_S4_FWHM} shows that the FWHM of the peak in the \textit{Spitzer}~1 ACF is 41\,days, but in the \textit{Spitzer}~2 ACF the FWHM is much larger with a value of 60\,days, which could therefore cause the lag in the CCF to be larger in ground-\textit{Spitzer}~2.

\subsection{Potential Periodicity in the Optical Light Curves}
\label{ap:periodicty}

In Section~\ref{sect:CCFResults}, a peak in the optical ACFs at $\sim 55$--70\,days is observed in multiple seasons, which could imply a periodicity within the light curves. This peak appears predominantly in the \mbox{M-ICCF} ACFs, in all optical ACFs that are tested with lags of $\pm 100$\,days, except in the 2013 season. In these ACFs, the correlation of the light curve with itself is only tested for the range of light curve that overlaps with the IR observations ($\sim 50$--150\,days per season), so to further examine this potential periodicity, the entire \textit{Kepler} light curves were analysed for each season and overall.   

Figure~\ref{fig:alt_kep_ACF} shows the ACF of the entire \textit{Kepler} light curves in the 2010 and 2012 seasons, respectively, compared to the ACFs of the portion of the light curve that overlaps with the IR observations. The entire-season light curve ACFs do not demonstrate a peak at $\sim 60$\,days as is shown in the limited ACFs. Furthermore, the peak is not present in the entire \textit{Kepler} ACF in Figure~\ref{fig:kep_sp1_CCF_all}, or in the ground ACFs in the 2013 season in Figure~\ref{fig:gr_sp1_CCF_S4}, or in the entire ground ACFs in Figure~\ref{fig:gr_sp1_CCF_all}. The possible $\sim 60$\,day periodicity is therefore not observed consistently across all optical ACFs of all lengths, which implies that $\sim 60$\,days is not a characteristic timescale of \mbox{Zw229-015}, but a consequence of the overlapping observation epochs.

\newpage

\section{Light-Curve Modelling}

\subsection{Distribution of Dust Clouds and the Corresponding Dust Transfer Functions}
\label{ap:dust_dists}

\begin{figure*}
    \begin{minipage}{\textwidth}
        \subfloat[figure][Radial distribution of dust clouds with $\alpha = -0.5$. \label{fig:alpha-0.5}]{\includegraphics[width=0.32\textwidth]{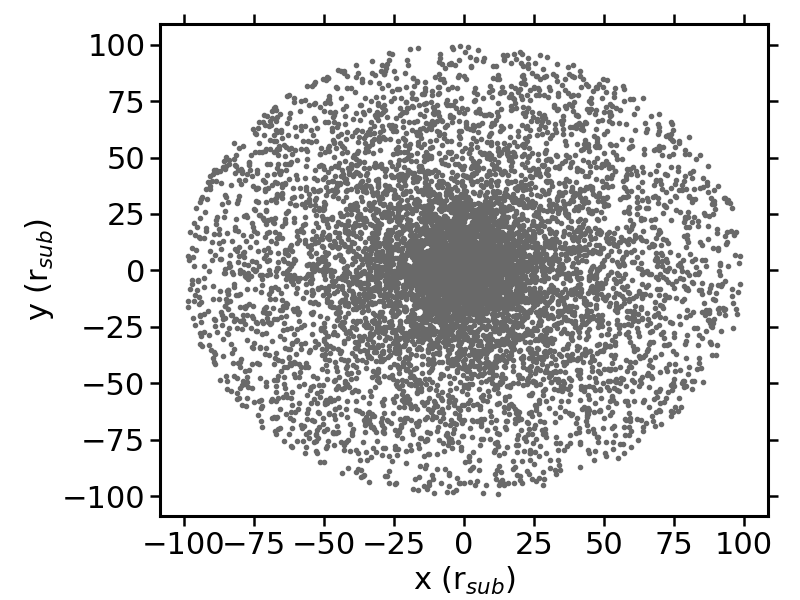}} 
        \hfill
        \subfloat[figure][Radial distribution of dust clouds with $\alpha = -3.0$.]{\includegraphics[width=0.32\textwidth]{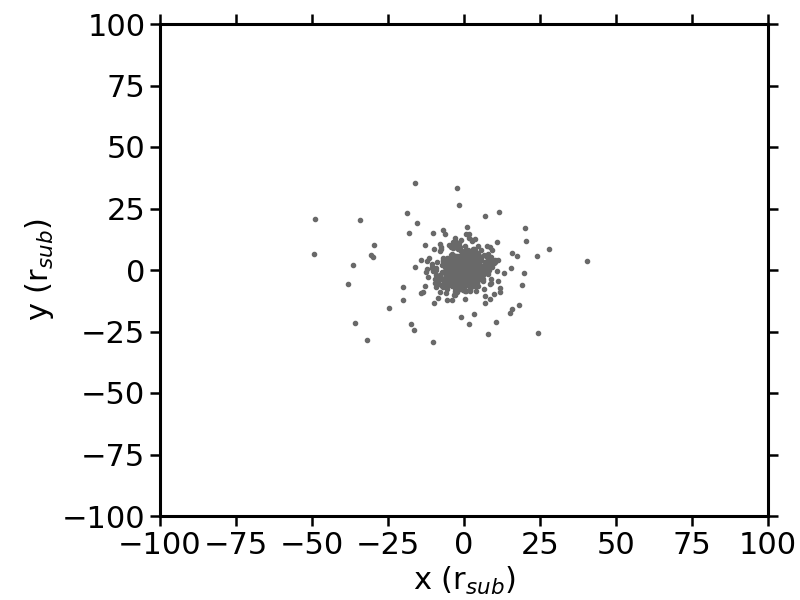}} 
        \hfill
        \subfloat[figure][Radial distribution of dust clouds with $\alpha = -5.5$. \label{fig:alpha-5.5}]{\includegraphics[width=0.32\textwidth]{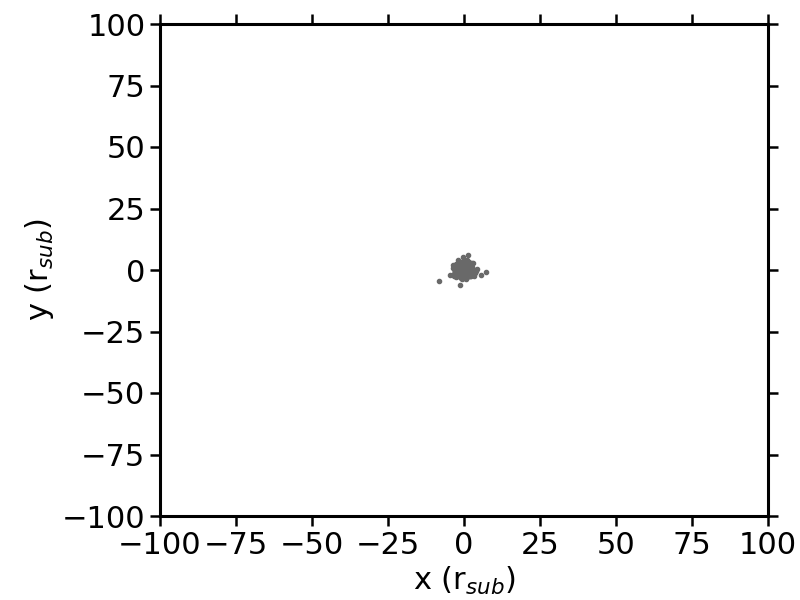}}
        \caption{Examples of the radial distribution of 10,000 dust clouds used to simulate dust transfer functions with different values of $\alpha$. In this figure, the inclination angle is set to $i=0^\circ$  and the vertical scale height power-law index is set to $\beta=0.05$. \label{fig:rad_dist}}      
    \end{minipage}
    \hfill
    \begin{minipage}{\textwidth}
        \subfloat[figure][Vertical distribution of dust clouds with $\beta = 0.05$. \label{fig:beta0.05}]{\includegraphics[width=0.32\textwidth]{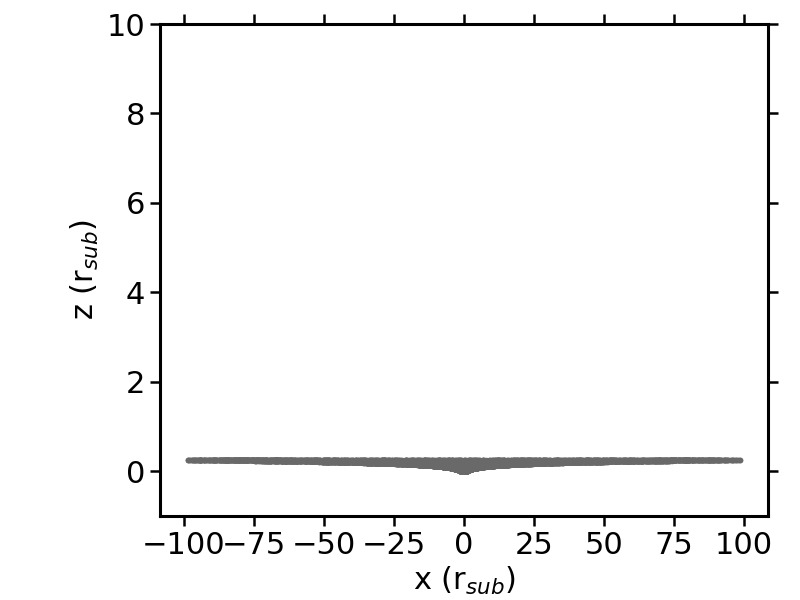}} 
        \hfill
        \subfloat[figure][Vertical distribution of dust clouds with $\beta = 1.05$. \label{fig:beta1.05}]{\includegraphics[width=0.32\textwidth]{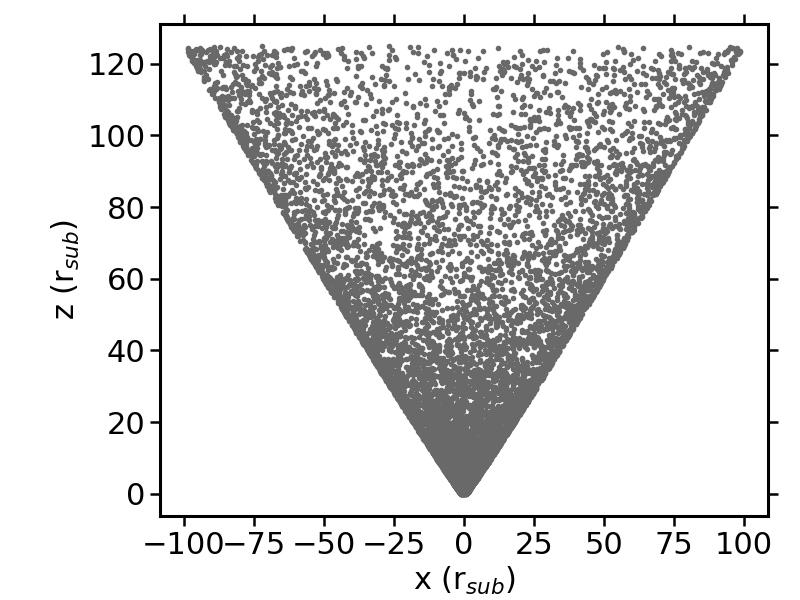}}
        \hfill
        \subfloat[figure][Vertical distribution of dust clouds with $\beta = 2.05$. \label{fig:beta2.05}]{\includegraphics[width=0.32\textwidth]{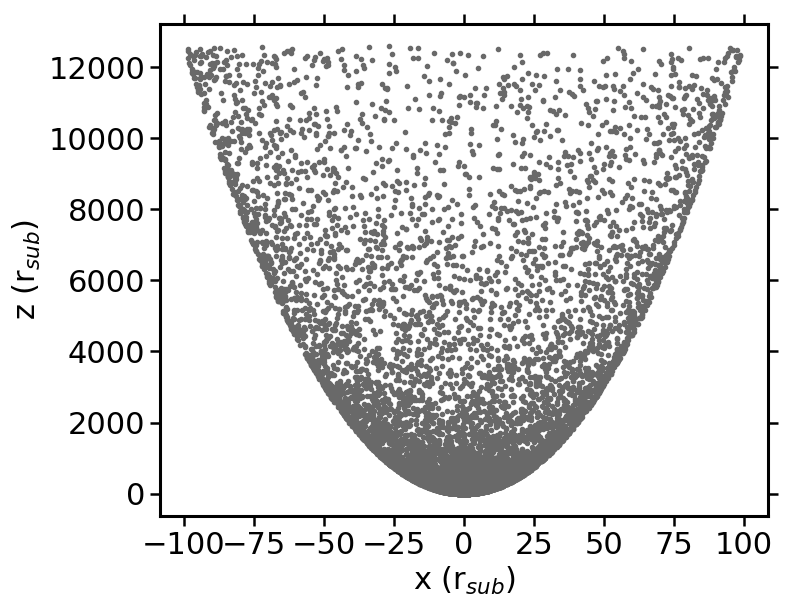}}
        \caption{Examples of the vertical distribution of 10,000 dust clouds used to simulate dust transfer functions with different values of $\beta$. In this figure, the inclination angle is set to $i=0^\circ$ degrees and the radial power-law index is set to $\alpha = -0.5$. \label{fig:vert_dist}}  
    \end{minipage}
    \begin{minipage}{\textwidth}
        \subfloat[figure][Change in DTF with different values of radial power-law index, $\alpha$. \label{fig:TFs_alpha}]{\includegraphics[width=0.49\textwidth]{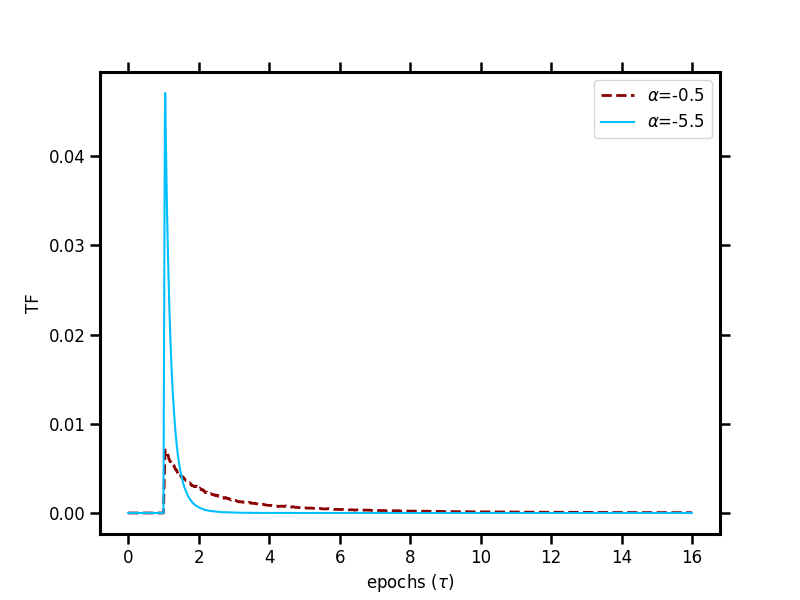}} 
        \hfill
        \subfloat[figure][Change in DTF with different values of vertical. scale height power-law index, $\beta$. \label{fig:TFs_beta}]{\includegraphics[width=0.49\textwidth]{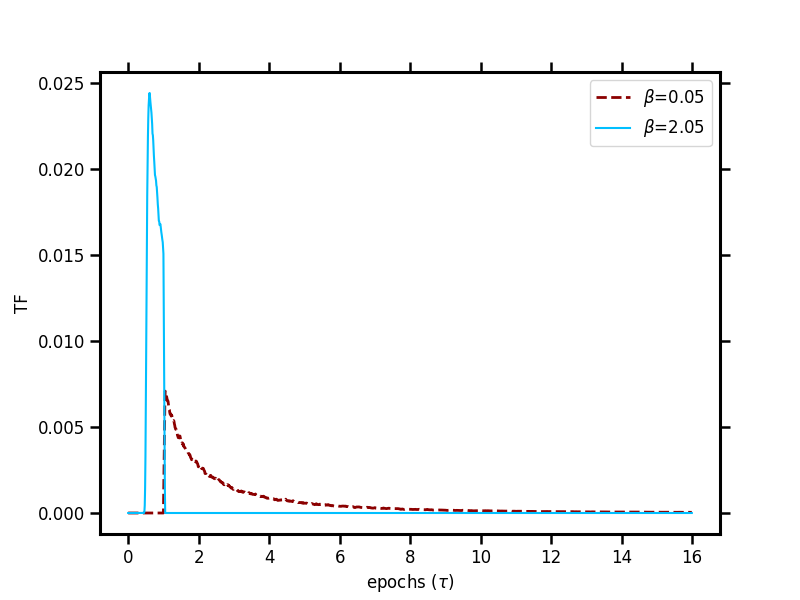}} 
        \caption{Examples of how the dust transfer functions change depending on the parameters used to model them. Unless otherwise specified in the legends of the individual plots, these DTFs correspond to values of $\alpha=-0.5$, $\beta=0.05$, and $i=0^\circ$. \label{fig:TFs_ab}} 
    \end{minipage}
\end{figure*}

The distribution of the dust affects the DTF as shown in Figure~\ref{fig:TFs_ab}. For example, Figure~\ref{fig:TFs_alpha} illustrates how the radial brightness distribution affects the DTF, as the DTF corresponding to a compact object (e.g., $\alpha = -5.5$, shown in Figure~\ref{fig:alpha-5.5}) contains a stronger peak that tails off faster than in the extended object (e.g., $\alpha = -0.5$ shown in Figure~\ref{fig:alpha-0.5}), as the emission comes from nearer to the centre of the source. Furthermore, the DTF also changes depending on the height of the dust clouds above the equator as shown in Figure~\ref{fig:TFs_beta}; for an AGN with an inclination angle of $0^\circ$, the clouds with $\beta=2.05$ (e.g., with the distribution shown in Figure~\ref{fig:beta2.05}) are closer to the observer than the clouds with $\beta=0.05$ (e.g., with the distribution shown in Figure~\ref{fig:beta0.05}), and therefore the DTF peaks earlier. 

\begin{figure*}
    \begin{minipage}{\textwidth}
        \centering
        \subfloat[figure][Delay map for i = 0 degrees.]{\includegraphics[width=0.45\textwidth]{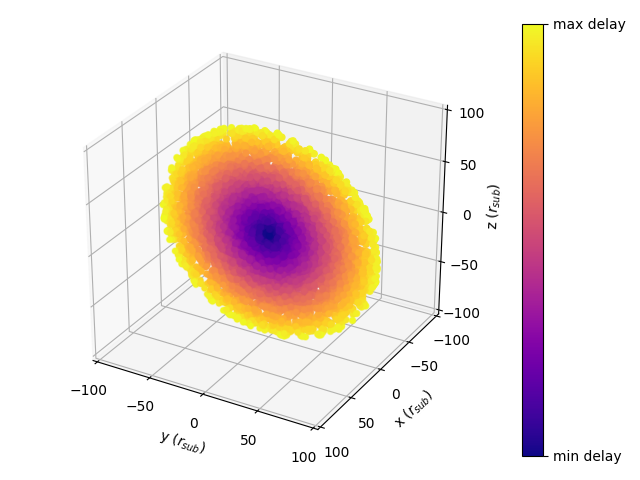}}
        \hspace{0.5cm}
        \subfloat[figure][Delay map for i = 69 degrees.]{\includegraphics[width=0.45\textwidth]{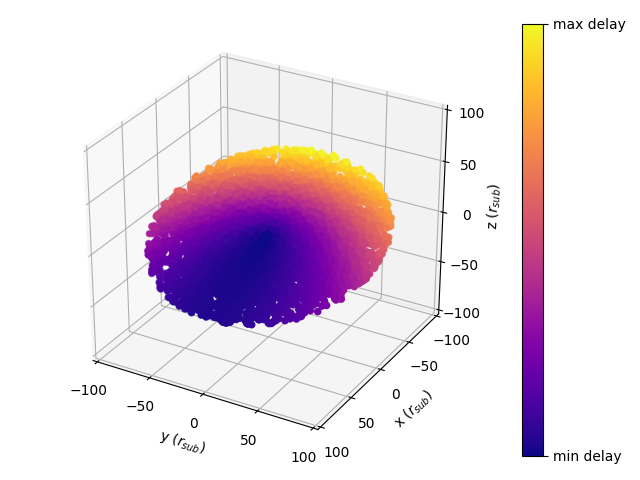}}
        \caption{Maps of the delay to each corresponding dust cloud depending on inclination angle and location of the cloud, where the line of sight of the observer is down the $y$ axis. 1000 dust clouds have been plotted in this figure, with the radial power-law index set to $\alpha=-0.5$ and the vertical scale height power-law index set to $\beta=0.05$. \label{fig:del_map}}    
    \end{minipage}
    \vfill
    \begin{minipage}{\textwidth}
        \centering
        \subfloat[figure][Illumination map for $i = 0^\circ$.]{\includegraphics[width=0.45\textwidth]{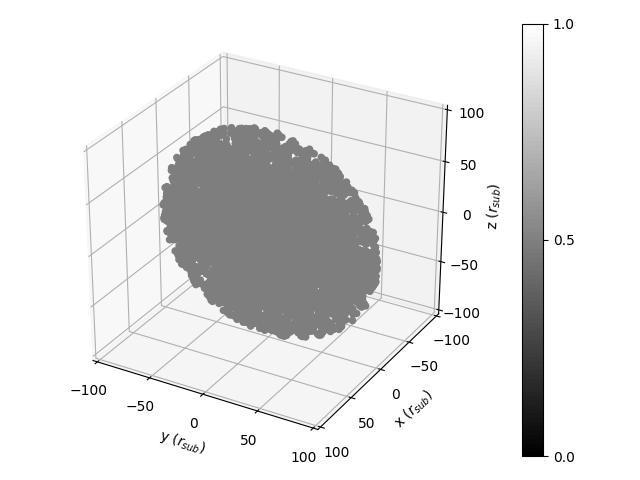}}
        \hspace{0.5cm}
        \subfloat[figure][Illumination map for $i = 69^\circ$.]{\includegraphics[width=0.45\textwidth]{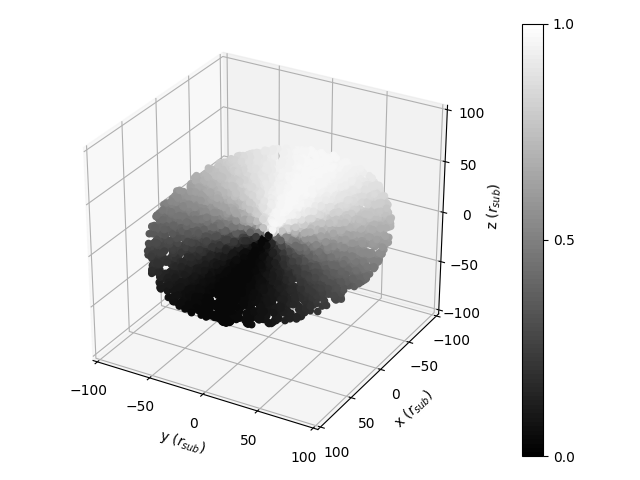}}
        \caption{Maps of the illumination of each corresponding dust cloud depending on inclination angle and location of the cloud, where the line of sight of the observer is down the $y$ axis. 1000 dust clouds have been plotted in this figure, with the radial power-law index set to $\alpha=-0.5$ and the vertical scale height power-law index set to $\beta=0.05$. \label{fig:illum_map}}    
    \end{minipage}
    \vfill
    \begin{minipage}{\textwidth}
        \centering
        \subfloat[figure][Change in DTF with different values of inclination angle ($i$) before the illumination effects are considered. \label{fig:TFs_inc}]{\includegraphics[width=0.42\textwidth]{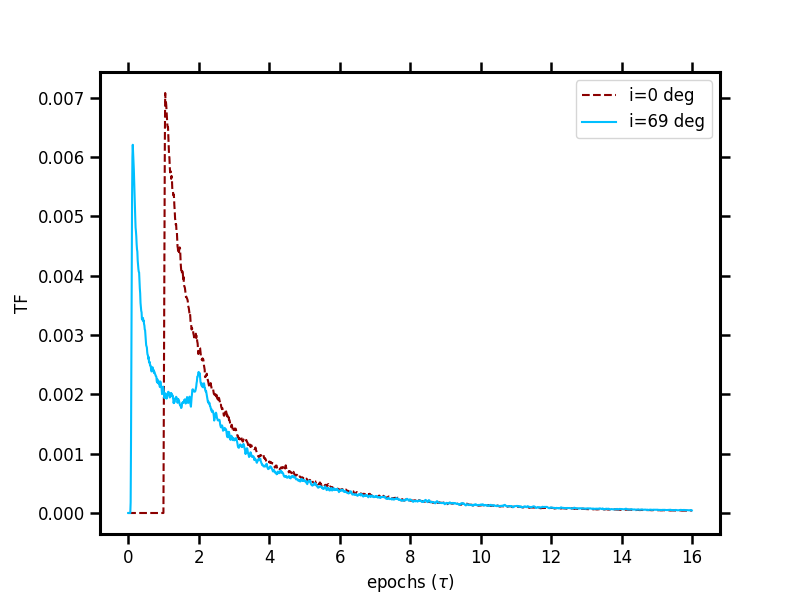}} 
        \hspace{0.5cm}
        \subfloat[figure][Change in DTF with different values of inclination angle ($i$) when illumination effects are considered \label{fig:TFs_inc_illum}]{\includegraphics[width=0.42\textwidth]{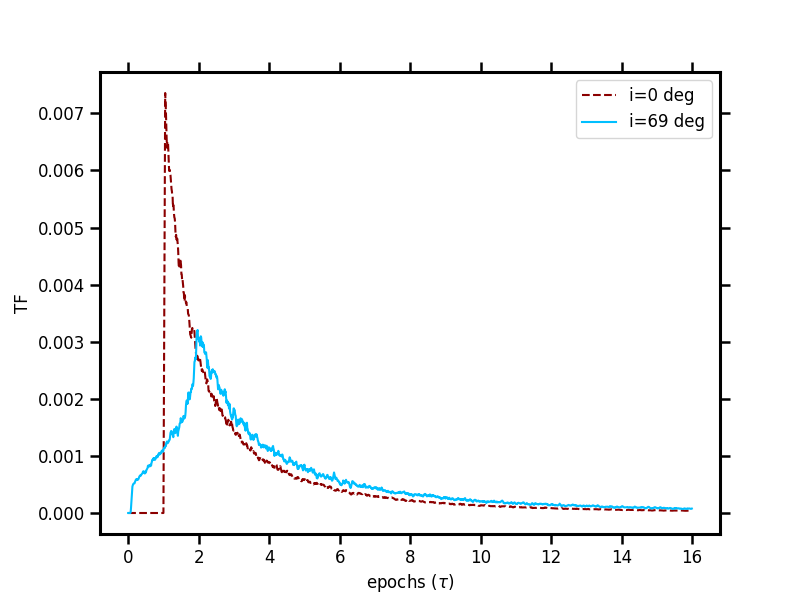}}   
        \caption{Examples of how the dust transfer functions change depending on the parameters used to model them. These DTFs correspond to values of $\alpha=-0.5$ and $\beta=0.05$. \label{fig:TFs}} 
        \vspace{2cm}
    \end{minipage}
\end{figure*}

Before the illumination is considered, the inclination angle can be seen to affect the DTFs as shown in Figure~\ref{fig:TFs_inc}, as for a tilted disk the DTF peaks at shorter lags and develops a tail toward larger lags compared to the DTF corresponding to a face-on disk. This can also be seen in the delay maps in Figure~\ref{fig:del_map}. Figure~\ref{fig:illum_map} shows how the illumination then affects the clouds at different locations for different inclination angles. For an inclination angle of $0^\circ$ the clouds are all similarly partially illuminated, and therefore the shape of the DTF remains the same, but for an inclination angle of $69^\circ$ the surface of the near-side dust clouds that are visible to the observer are mostly not illuminated, while the far-side clouds are mostly illuminated. This therefore changes the DTF, as shown in Figure~\ref{fig:TFs_inc_illum}; the first peak that is visible in Figure~\ref{fig:TFs_inc} for an inclination angle of $i=69^\circ$, which corresponds to the emission from the near-side clouds, is no longer present, while the peak corresponding to the far-side dust clouds remains.

\begin{figure*}
    \centering
    \subfloat[figure][Simulated light curves of Figure~\ref{fig:model_kep_sp1_all} mean-subtracted and compared to the observations. \label{fig:mean_sub_S1-3} ]{\includegraphics[width=\textwidth]{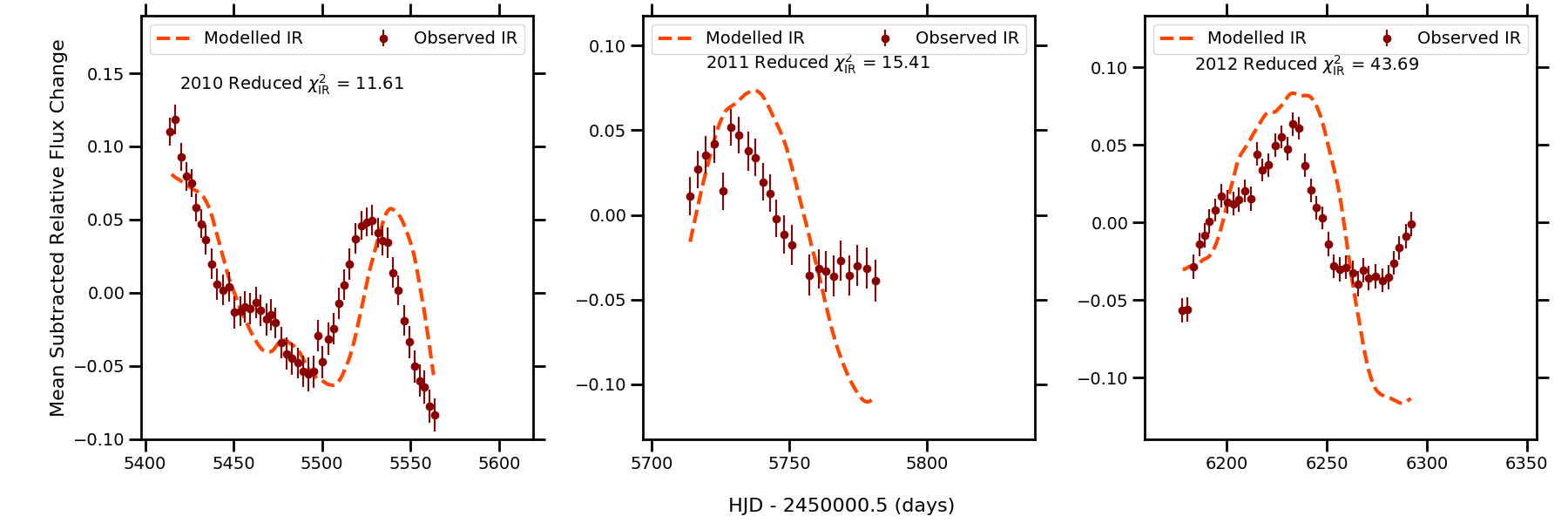}} 
    \vfill
    \subfloat[figure][Simulated light curves of Figure~\ref{fig:model_kep_sp1_S1-2_extended} mean-subtracted and compared to the observations. \label{fig:mean_sub_S1-2_extend} ]{\includegraphics[width=\textwidth]{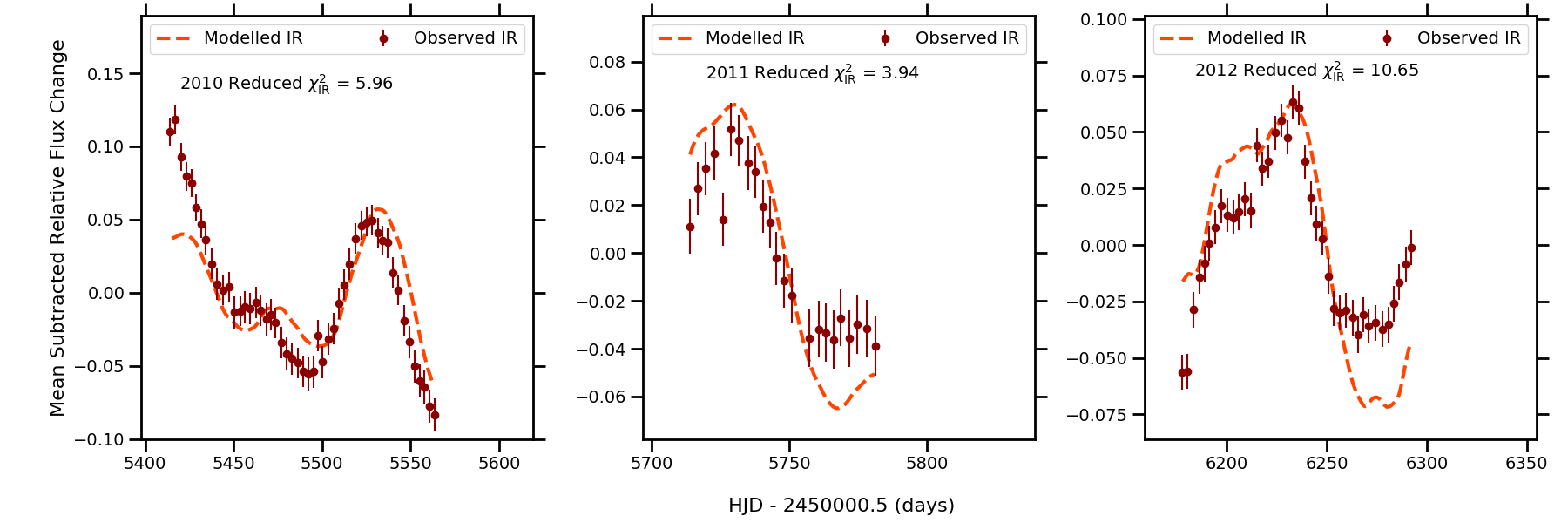}}
   \caption{Mean-subtracted simulated \textit{Spitzer}~1 light curves compared to the mean-subtracted observations in the individual observation seasons, to show the quality of fit independent of the offset. \label{fig:mean_sub_models}}
\end{figure*}

\subsection{Poor fitting model owing to dramatic increase in flux between 2011 and 2012}
\label{ap:longterm_models}

In Section~\ref{Sect:IRModelling_results}, it was shown that the dramatic increase in flux between the 2011 and 2012 seasons was not well-modelled by a single dust component; when it tries to simulate the entire light curve in Figure~\ref{fig:model_kep_sp1_all}, it does not match the shape of the individual seasons well. This is further demonstrated here in Figure~\ref{fig:mean_sub_S1-3}, which shows the mean-subtracted simulated and observed light curves in the individual seasons. When the model made for the 2010--2011 seasons only was extended to include the 2012 season, the shape of the individual seasons better match the observations, as shown for the mean-subtracted light curves in each season in Figure~\ref{fig:mean_sub_S1-2_extend}, both by eye and by the values of the reduced $\chi^2$.

\subsection{Additional Modelling of the Optical and IR Light Curves}
\label{ap:more_models}

\begin{figure*}
    \begin{minipage}{\textwidth}
        \subfloat[figure][Simulated light curve of ground-{\it Spitzer}~1 for the entire overlapping light curves, plotted with the parameters that corresponded to the highest posterior distribution, with values of $\alpha=-0.51$, $\beta=0.05$, $\tau=23.05$\,days, and $i=47.85^\circ$. \label{fig:model_gr_sp1_S1-5}]{\includegraphics[width=\textwidth]{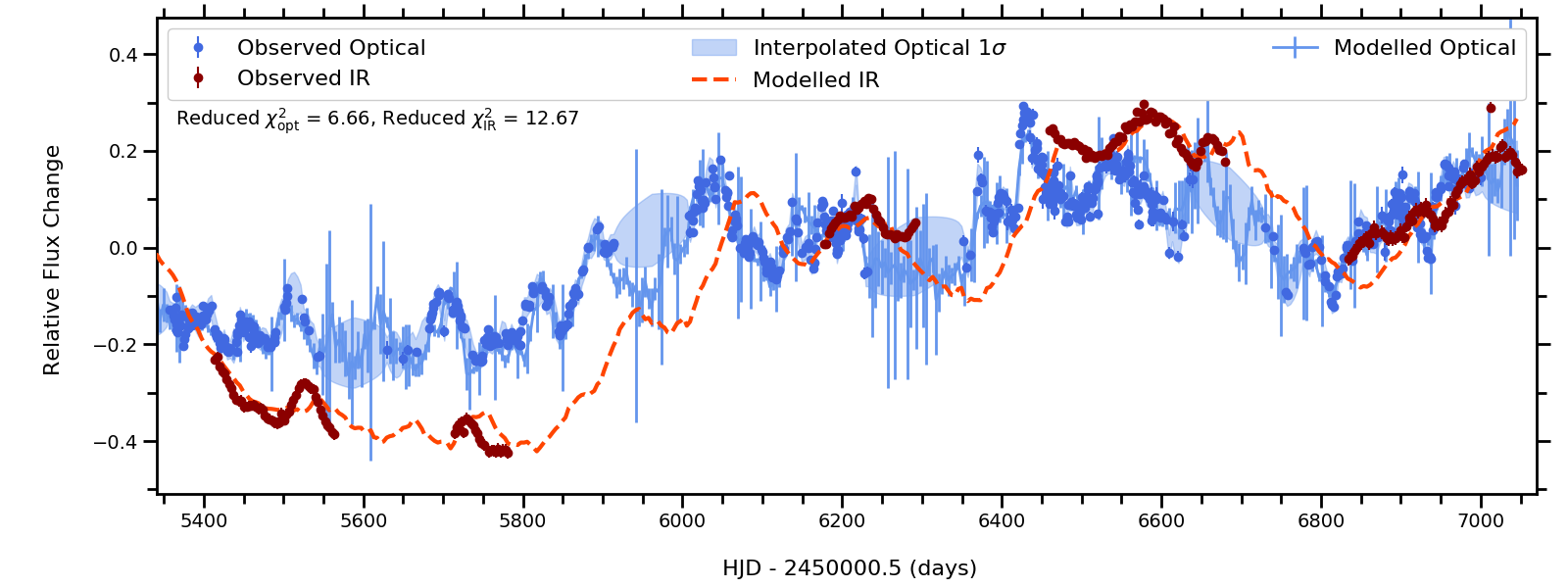}}
        \\
        \subfloat[figure][Simulated light curves for ground-{\it Spitzer}~1 in the 2010--2011 season, plotted with $\alpha=-0.51$, $\beta=0.13$, $\tau=7.62$\,days, and $i=60.32^\circ$. \label{fig:model_gr_sp1_S1-2}]{\includegraphics[width=\textwidth]{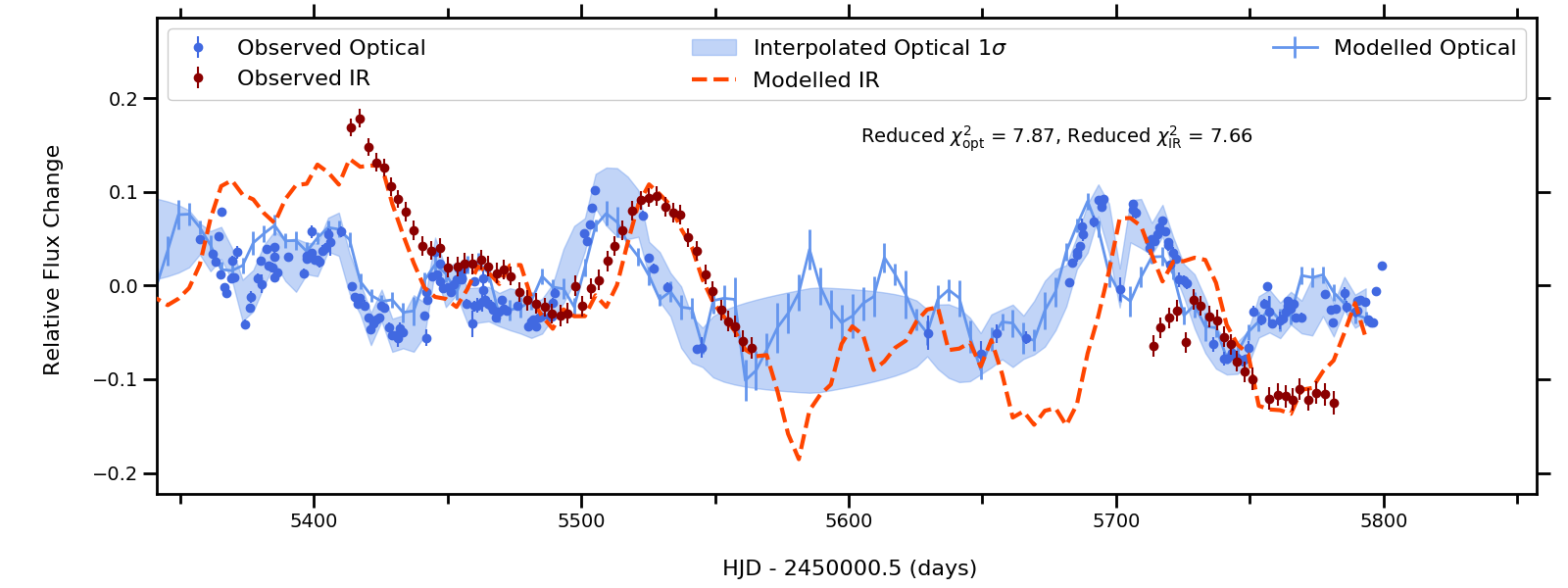}}
        \\
        \subfloat[figure][Simulated light curves for ground-{\it Spitzer}~1 in the 2012--2014 season, plotted with $\alpha=-0.51$, $\beta=0.06$, $\tau=27.92$\,days, and $i=39.34^\circ$. \label{fig:model_gr_sp1_S3-5}]{\includegraphics[width=\textwidth]{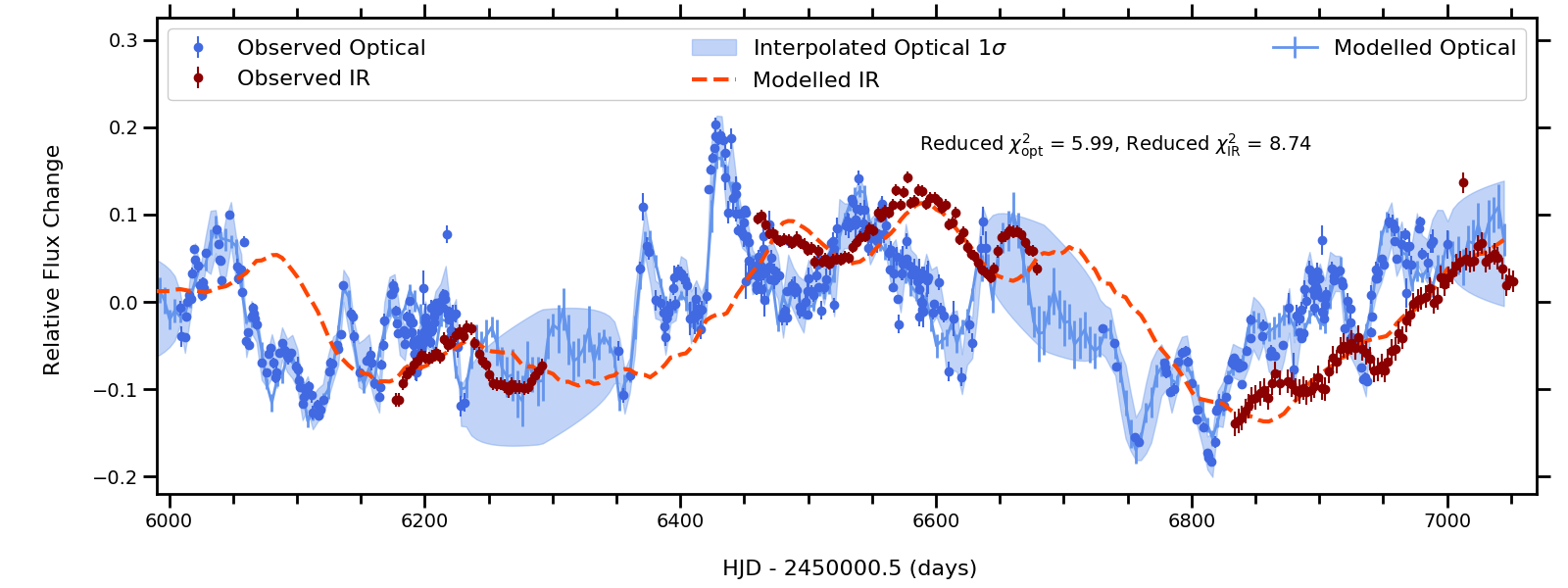}}
        \caption{Simulated light curves of the ground-{\it Spitzer}~1 observations, for the entire overlapping light curves and the light curves separated into the 2010--2011 seasons and 2012--2014 seasons, plotted with the parameters listed that corresponded to the highest posterior distribution. \label{fig:model_gr_sp1_all}}
        \vspace{0.5cm}
    \end{minipage}
\end{figure*}

\begin{figure*}
    \begin{minipage}{\textwidth}        
        \includegraphics[width=\textwidth]{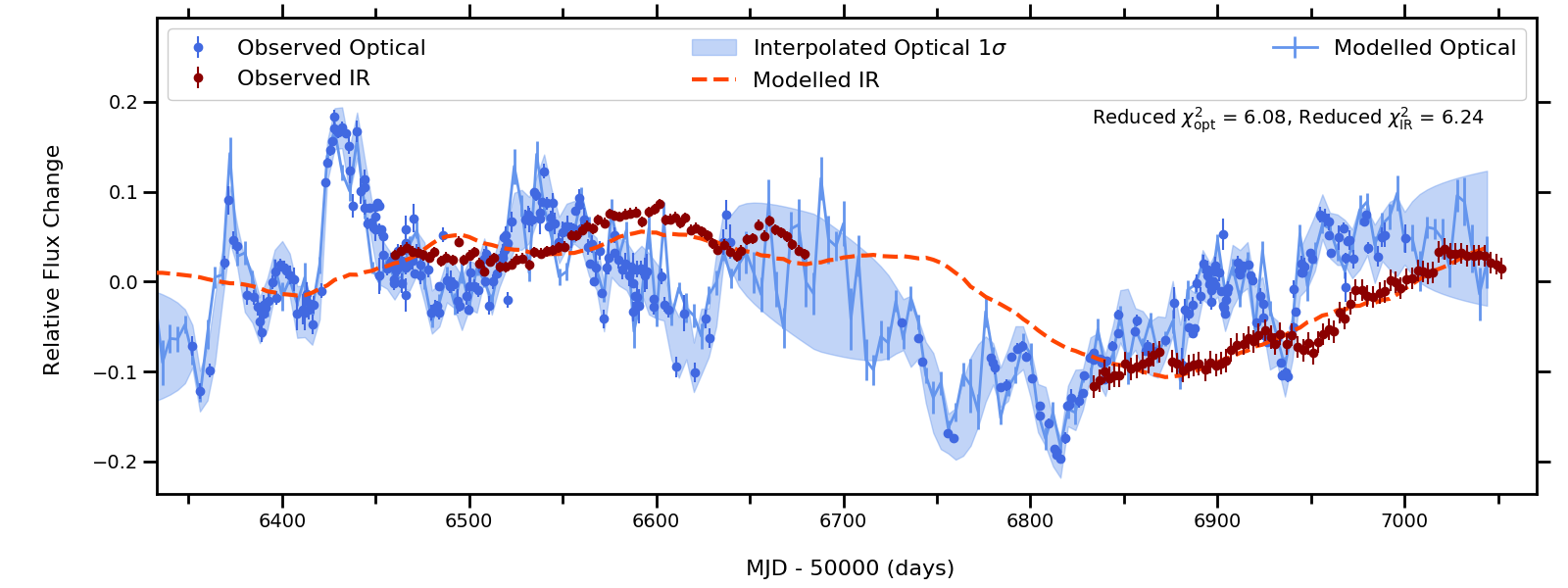}
        \caption{Simulated light curve of ground-{\it Spitzer}~2 for the observation seasons starting 2013--2014, plotted with the parameters that corresponded to the highest posterior distribution, with values of $\alpha=-0.52$, $\beta=0.05$, $\tau=28.53$\,days, and $i=41.94^\circ$. \label{fig:model_gr_sp2_all}}
        \vspace{0.3cm}
    \end{minipage}
    \begin{minipage}{\textwidth}        
        \includegraphics[width=\textwidth]{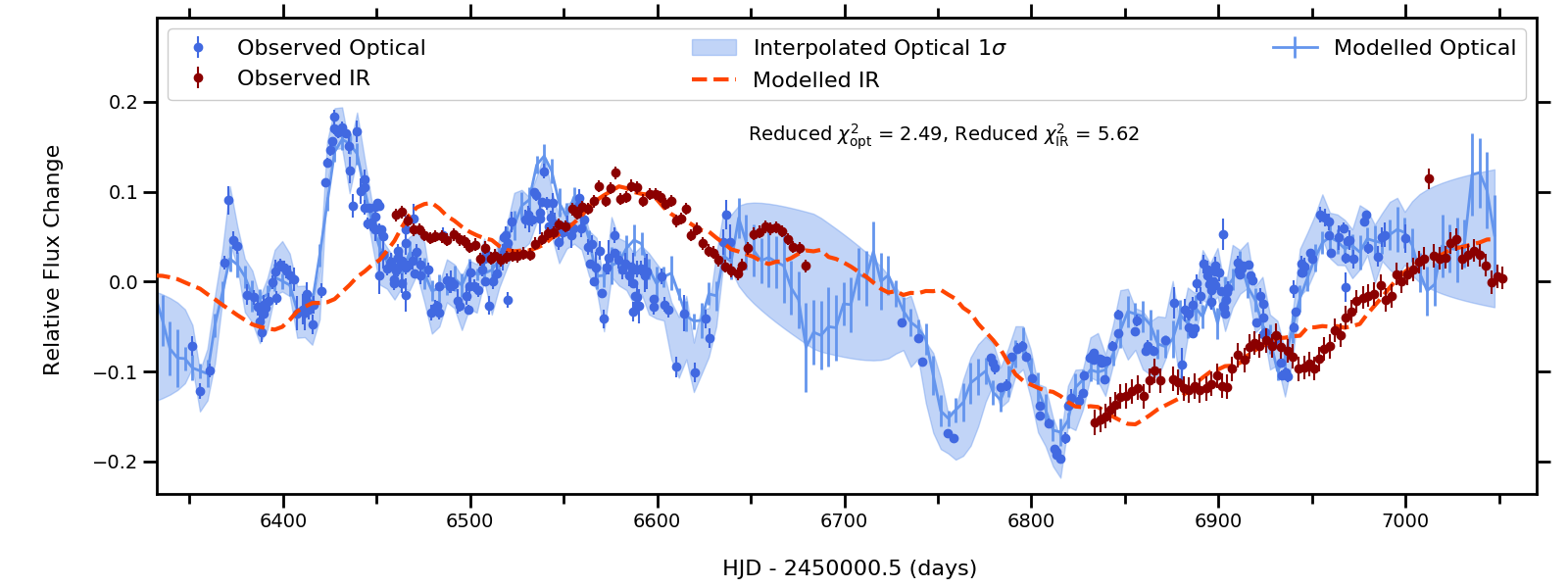}
        \caption{Simulated light curve of ground-{\it Spitzer}~1 for the observation seasons starting 2013--2014, plotted with the parameters that corresponded to the highest posterior distribution, with values of $\alpha=-0.51$, $\beta=0.06$, $\tau=19.66$\,days, and $i=61.02^\circ$. \label{fig:model_gr_sp1_S4-5}}
        \vspace{0.3cm}
    \end{minipage}
    \begin{minipage}{\textwidth}
        \captionof{table}{Mean output parameters for the MCMC modelling of the entire light curves over different combinations of observation seasons. \label{tab:entire_params}}
            \begin{tabular}{C{0.09\textwidth}C{0.07\textwidth}C{0.08\textwidth}C{0.1\textwidth}C{0.08\textwidth}C{0.08\textwidth}C{0.08\textwidth}C{0.08\textwidth}C{0.06\textwidth}C{0.06\textwidth}}
            \hline
            
            Light-Curve Combination & Seasons Starting & Radial Power-Law Index & Vertical Scale Height Power-Law Index & Amplitude Conversion Factor & Lag (days) & Inclination (degrees) & Offset ($\times \ 10^{-2}$) & $\chi_\textit{opt}^2$ & $\chi_\textit{IR}^2$ \\
            
            \hline


            {Gr-Sp1} & 2010--2014 & $-0.53^{+0.03}_{-0.25}$ &  $0.06^{+0.16}_{-0.01}$ & $1.92^{+0.09}_{-0.04}$ & $25.45^{+2.50}_{-2.35}$ & $45.54^{+3.45}_{-6.50}$ & $-0.91^{+1.50}_{-0.43}$ & 6.66 & 12.67 \\

            \hline
           
            {Kep-Sp1} & 2010--2012 &  $-0.52^{+0.02}_{-0.05}$ & $0.43^{+0.18}_{-0.35}$ & $2.64^{+0.05}_{-0.05}$ & $20.42^{+3.34}_{-2.77}$ & $44.42^{+9.71}_{-2.48}$ & $4.50^{+0.44}_{-0.52}$ & 5.91 & 24.04 \\
           
            \hline
           

           {Gr-Sp1} & 2010--2011 &   
           $-0.52^{+0.01}_{-0.11}$ & $0.16^{+0.48}_{-0.10}$ & $2.66^{+0.16}_{-0.21}$ & $8.02^{+2.16}_{-0.55}$ & $49.71^{+13.81}_{-8.83}$ & $-0.74^{+0.55}_{-0.40}$  & 7.87 & 7.66 \\
       
            {Kep-Sp1} & 2010--2011 &  $-0.52^{+0.02}_{-0.05}$ & $0.08^{+0.51}_{-0.02}$ & $1.58^{+0.12}_{-0.10}$ & $10.83^{+1.05}_{-1.38}$ & $46.68^{+18.81}_{-11.28}$ & $0.47^{+0.38}_{-0.38}$ & 3.87 & 7.84 \\
       
            \hline       
       
            {Gr-Sp1} & 2012--2014 &  $-0.52^{+0.01}_{-0.03}$ &  $0.23^{+0.45}_{-0.18}$ & $1.65^{+0.07}_{-0.06}$ & $26.67^{+2.06}_{-1.71}$ & $47.76^{+20.42}_{-6.76}$ & $1.19^{+0.25}_{-0.23}$ & 5.99 & 8.74 \\


            \hline    
            

          {Gr-Sp1} & 2013--2014 & $-0.54^{+0.03}_{-0.29}$ & $0.11^{+0.56}_{-0.05}$ & $1.38^{+0.07}_{-0.06}$ & $27.40^{+0.99}_{-3.87}$ & $42.18^{+23.81}_{-3.30}$ & $0.70^{+0.34}_{-0.43}$ & 2.49 & 5.62 \\


           {Gr-Sp2} & 2013--2014 & $-0.53^{+0.02}_{-0.51}$ & $0.08^{+0.53}_{-0.02}$ & $1.21^{+0.05}_{-0.04}$ & $28.54^{+0.44}_{-5.66}$ & $52.09^{+15.75}_{-7.70}$ & $0.95^{+0.33}_{-0.22}$ & 5.19 & 5.20 \\
           
           \hline       
           
        \end{tabular}
        \vspace{2cm}
    \end{minipage}
\end{figure*}

\subsubsection{Modelling Multiple Observation Seasons}

Figures~\ref{fig:model_gr_sp1_all} and \ref{fig:model_gr_sp2_all} display the simulated light curves made using the MCMC model described in Section~\ref{Sect:IRModelling} for the entire ground-{\it Spitzer}~1 and ground-{\it Spitzer}~2 light curves, respectively. The ground-{\it Spitzer}~1 light curves were divided into the 2010--2011 seasons and the 2012--2014 seasons as the dramatic increase in flux between 2011 and 2012 was shown to be poorly fit by a single dust component model in Sections~\ref{Sect:IRModelling_results} and Appendix~\ref{ap:longterm_models}.

Figure~\ref{fig:model_gr_sp1_S1-2} shows the model matching the highest posterior distribution of the 2010--2011 seasons of the ground-{\it Spitzer}~1 light curves. Overall, the shape of the simulated IR light curve is shown to follow the variability of the observed light curve quite well; however, the amplitude of the 2011 season still appears to be overestimated here, and furthermore, the start of the 2010 season is underestimated in the simulated light curve.

Figure~\ref{fig:model_gr_sp1_S3-5} shows the model of the ground-{\it Spitzer}~1 light curve in the seasons starting 2012--2014. The overall variability is once again modelled quite well by the simulated light curve; however, there are portions that deviate from the observed light curve. For example, the end of the 2012 season is overestimated, and the peak at the end of the 2013 season is not visible in the simulated light curve; however, both of these correspond to periods where there are very few, if any, observations in the optical and therefore the IR is simulated here based on the optical interpolations. Additionally, there is a dip in the 2014 season IR observed light curve that is not replicated by the simulated light curve. 

Figure~\ref{fig:model_gr_sp2_all} contains the simulated and observed ground-{\it Spitzer}~2 light curves for the observation seasons starting in 2013--2014. Like Figure~\ref{fig:model_gr_sp1_S3-5}, the simulated light curve does not replicate the peak in the IR observations at the end of the 2013 season, or the dip in the 2014 season, and it also slightly overestimates the flux between HJD~56450 and 56500 and underestimates the flux between HJD~56550 and 56625, making the variability in 2013 flatter than the observations. As a comparison, the ground-{\it Spitzer}~1 light curves in the seasons starting 2013--2014 were also modelled in Figure~\ref{fig:model_gr_sp1_S4-5}.

\subsubsection{Modelling the Individual Observation Seasons}
\label{ap:individ_models}

\begin{figure*}
    \centering
    \subfloat[figure][Simulated light curves for ground-{\it Spitzer}~1 in the 2010 season, plotted with $\alpha=-0.55$, $\beta=0.07$, $\tau=9.74$\,days, and $i=54.97^\circ$. ]{\includegraphics[width=0.49\textwidth]{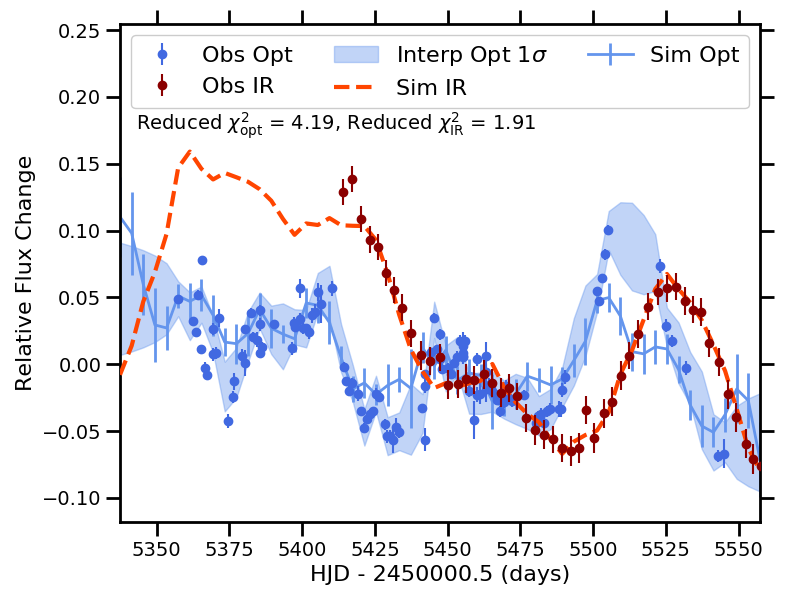}} 
    \hfill
    \subfloat[figure][{Simulated light curves for \textit{Kepler}-{\it Spitzer}~1 in the 2010 season, plotted with $\alpha=-0.52$, $\beta=0.13$, $\tau=8.82$\,days, and $i=43.76^\circ$. }]{\includegraphics[width=0.49\textwidth]{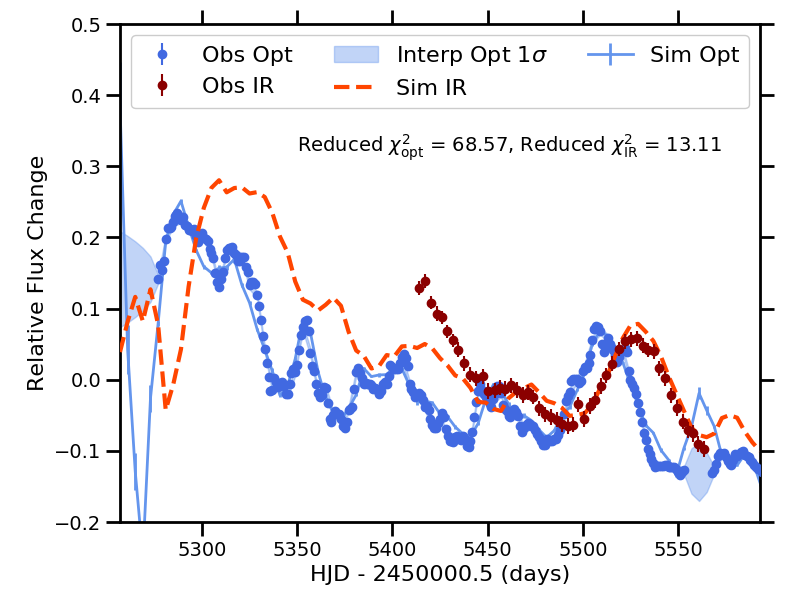}} 
    \\
    
    \subfloat[figure][Simulated light curves for ground-{\it Spitzer}~1 in the 2011 season, plotted with $\alpha=-0.83$, $\beta=0.37$, $\tau=8.43$\,days, and $i=68.00^\circ$. ]{\includegraphics[width=0.49\textwidth]{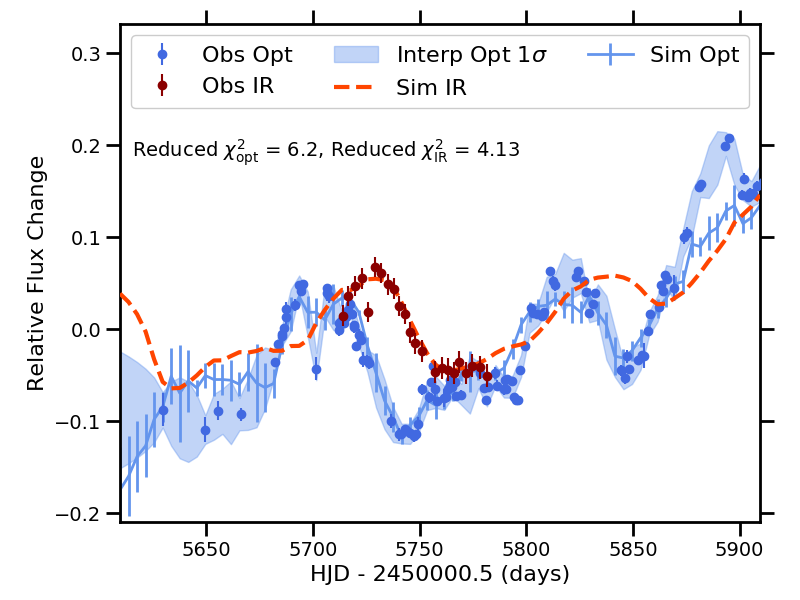}} 
    \hfill
    \subfloat[figure][Simulated light curves for \textit{Kepler}-{\it Spitzer}~1 in the 2011 season, plotted with  $\alpha=-0.66$, $\beta=1.87$, $\tau=17.77$\,days, and $i=40.98^\circ$. ]{\includegraphics[width=0.49\textwidth]{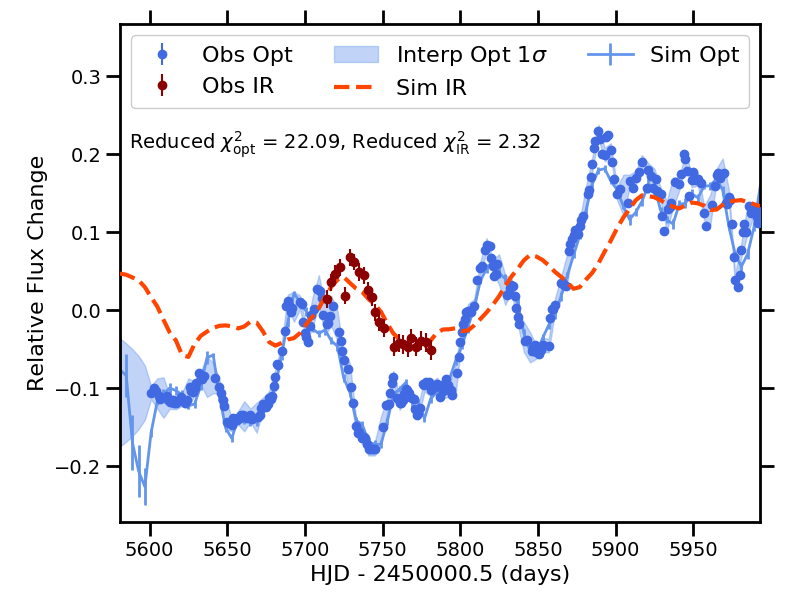}} 
    \\
    
    \subfloat[figure][Simulated light curves for ground-{\it Spitzer}~1 in the 2012 season, plotted with $\alpha=-1.28$, $\beta=0.08$, $\tau=10.92$\,days, and $i=62.64^\circ$. ]{\includegraphics[width=0.49\textwidth]{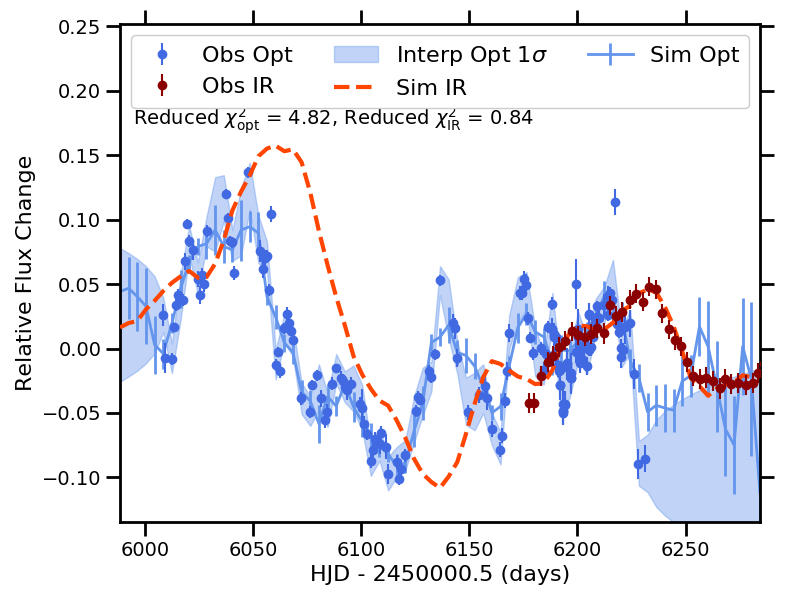}} 
    \hfill
    \subfloat[figure][Simulated light curves for \textit{Kepler}-{\it Spitzer}~1 in the 2012 season, plotted with $\alpha=-5.21$, $\beta=1.22$, $\tau=16.45$\,days, $i=21.22^\circ$. ]{\includegraphics[width=0.49\textwidth]{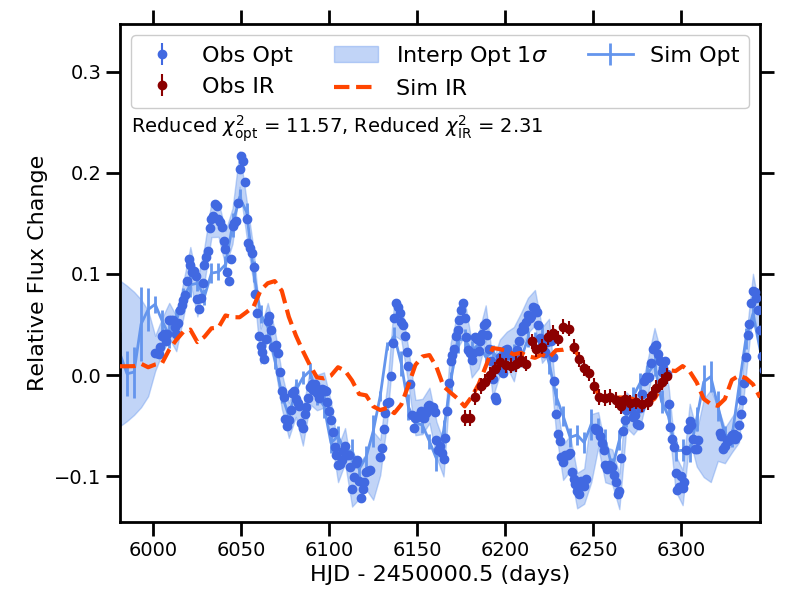}} 
    \\
    \caption{Simulated optical and IR light curves for the individual observation seasons, plotted with the specified values corresponding to the maximum posterior model. \label{fig:individ_seasons}}
\end{figure*}

\begin{figure*}
    \ContinuedFloat
    \begin{minipage}{\textwidth}
        \subfloat[figure][Simulated light curves for ground-{\it Spitzer}~1 in the 2013 season, plotted with $\alpha=-0.62$, $\beta=0.43$, $\tau=10.44$\,days, and $i=38.89^\circ$.]{\includegraphics[width=0.49\textwidth]{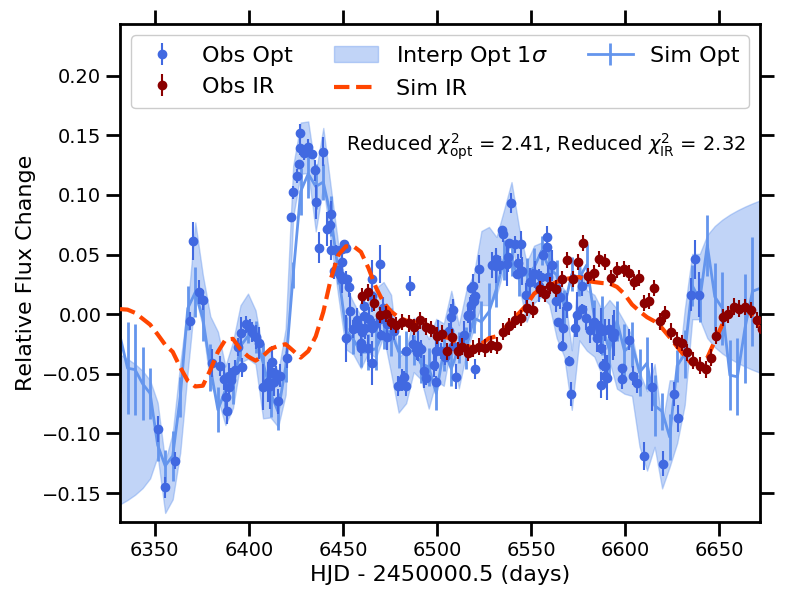}} 
        \hfill
        \subfloat[figure][Simulated light curves for ground-{\it Spitzer}~2 in the 2013 season, plotted with $\alpha=-0.51$, $\beta=0.37$, $\tau=27.94$\,days, and $i=48.41^\circ$. ]{\includegraphics[width=0.49\textwidth]{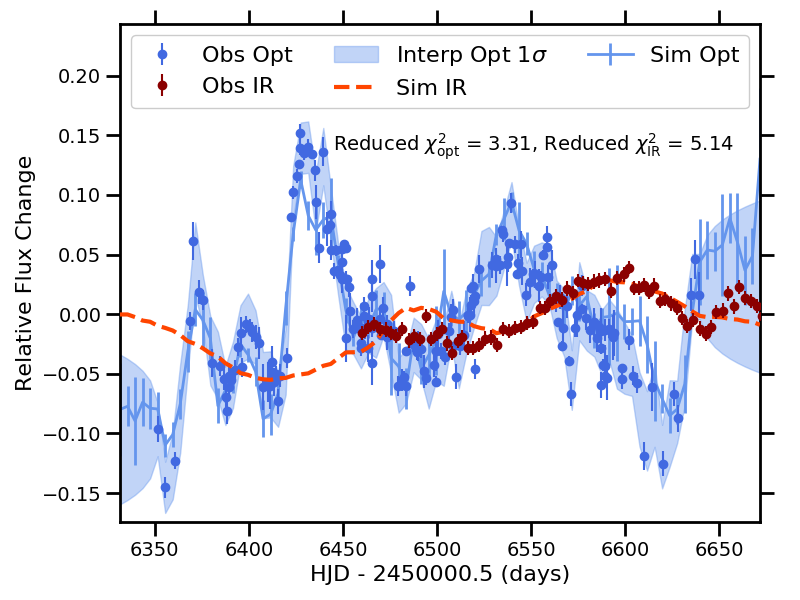}} 
        \\
        
        \subfloat[figure][Simulated light curves for ground-{\it Spitzer}~1 in the 2014 season, plotted with $\alpha=-0.51$, $\beta=0.09$, $\tau=12.70$\,days, and $i=19.45^\circ$.]{\includegraphics[width=0.49\textwidth]{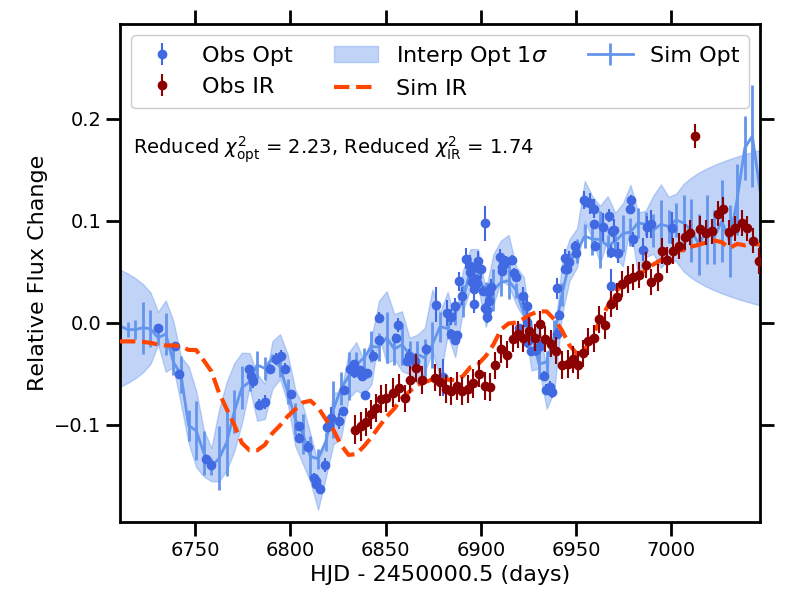}} 
        \hfill
        \subfloat[figure][Simulated light curves for ground-{\it Spitzer}~2 in the 2014 season, plotted with $\alpha=-0.62$, $\beta=0.19$, $\tau=28.73$\,days, and $i=56.22^\circ$.]{\includegraphics[width=0.49\textwidth]{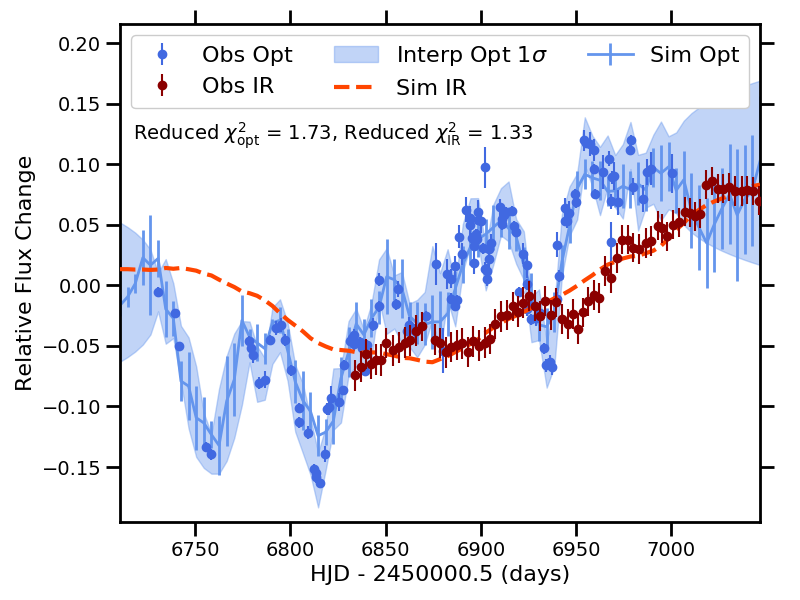}} 
        \\
        \caption{Continued.}
    \end{minipage}
    \vspace{0.3cm}
    \begin{minipage}{\textwidth}
        \captionof{table}{Mean output parameters corresponding to the best-fit DTFs found by MCMC modelling the individual observation seasons for each combination of optical and IR light curves, and the quality of their fits described by reduced $\chi^2$.  In this table, values of the offset have been multiplied by $10^{2}$. \label{tab:season_params}}
        \begin{tabular}{C{0.09\textwidth}C{0.07\textwidth}C{0.08\textwidth}C{0.1\textwidth}C{0.08\textwidth}C{0.08\textwidth}C{0.08\textwidth}C{0.08\textwidth}C{0.06\textwidth}C{0.06\textwidth}}
                \hline
                Light-Curve Combination & Season Starting & Radial Power-Law Index & Vertical Scale Height Power-Law Index & Amplitude Conversion Factor & Lag (days) & Inclination (degrees) & Offset & $\chi_\textit{opt}^2$ & $\chi_\textit{IR}^2$  \\
                \hline
                

               {Gr-Sp1} & 2010 & $-0.58^{+0.06}_{-0.33}$ &  0.13$^{+0.28}_{-0.06}$ & 3.23$^{+0.46}_{-0.50}$ & 9.90$^{+2.59}_{-0.86}$ & 51.57$^{+12.52}_{-18.91}$ & 1.17$^{+0.76}_{-0.88}$ & 4.19 & 1.91 \\
               
               
               {Kep-Sp1} & 2010 & $-1.07^{+0.55}_{-4.33}$ & 0.43$^{+0.92}_{-0.23}$ & 1.30$^{+0.29}_{-0.20}$ & 11.04$^{+3.89}_{-1.71}$ & 42.50$^{+5.91}_{-11.50}$ & 3.49$^{+0.47}_{-0.68}$ & 68.57 & 13.11 \\   
               
               \hline
               
               {Gr-Sp1} & 2011 &  $-1.29^{+0.65}_{-3.61}$ & 0.41$^{+1.06}_{-0.29}$ & 0.91$^{+0.23}_{-0.14}$ & 10.64$^{+4.62}_{-1.97}$ & 50.64$^{+35.21}_{-15.96}$ & 2.85$^{+0.86}_{-0.66}$ & 6.20 & 4.13 \\

    
               {Kep-Sp1} & 2011 & $-1.37^{+0.82}_{-4.04}$ & 1.81$^{+0.20}_{-1.22}$ & 0.68$^{+0.08}_{-0.08}$ & 16.96$^{+3.36}_{-2.99}$ & 42.26$^{+17.28}_{-12.81}$ & 5.19$^{+0.85}_{-0.67}$ & 22.09 & 2.33 \\
               
               \hline
    
               {Gr-Sp1} & 2012 &  $-1.48^{+0.79}_{-2.60}$ & 0.62$^{+1.34}_{-0.52}$ & 1.30$^{+0.36}_{-0.22}$ & 12.89$^{+7.52}_{-1.34}$ & 49.08$^{+17.21}_{-24.12}$ & 1.54$^{+0.47}_{-0.39}$ & 4.82 & 0.84 \\
    
    
               {Kep-Sp1} & 2012 &  $-4.77^{+4.17}_{-0.66}$ &  1.45$^{+0.52}_{-0.89}$ & 0.55$^{+0.09}_{-0.06}$ & 17.15$^{+2.68}_{-3.82}$ & 17.46$^{+7.27}_{-4.70}$ & 0.93$^{+0.28}_{-0.26}$ & 11.57 & 2.31 \\
               
               \hline

                {Gr-Sp1} & 2013  & $-0.84^{+0.28}_{-4.00}$ & 0.24$^{+1.57}_{-0.15}$ & 0.74$^{+0.12}_{-0.12}$ & 11.75$^{+3.55}_{-1.54}$ & 34.11$^{+26.54}_{-10.29}$ & 0.61$^{+0.21}_{-0.23}$ & 3.98 & 3.02 \\

               
               {Gr-Sp2} & 2013  &  $-0.55^{+0.04}_{-2.98}$ & 0.14$^{+0.88}_{-0.08}$ & 0.83$^{+0.22}_{-0.63}$ & 27.65$^{+1.17}_{-13.40}$ & 49.97$^{+9.82}_{-18.69}$ & 0.12$^{+0.17}_{-0.20}$ & 3.51 & 5.14 \\

               \hline
               
               {Gr-Sp1} & 2014 & $-0.80^{+0.25}_{-4.52}$ & 0.34$^{+0.98}_{-0.23}$ & 1.09$^{+0.16}_{-0.10}$ & 12.93$^{+15.32}_{-2.59}$ & 55.29$^{+11.72}_{-17.45}$ & 2.09$^{+2.59}_{-0.78}$ & 2.23 & 1.74 \\

               

               
               {Gr-Sp2} & 2014 & $-0.61^{+0.08}_{-0.20}$ & 0.13$^{+0.24}_{-0.06}$ & 1.13$^{+0.20}_{-0.27}$ & 27.68$^{+1.11}_{-17.63}$ & 58.98$^{+8.25}_{-12.67}$ & 0.82$^{+0.67}_{-0.24}$ & 1.73 & 1.33 \\
               
               \hline
        \end{tabular}
    \end{minipage}
\end{figure*}

The MCMC model described in Section~\ref{Sect:IRModelling} was also applied to each individual observation season for each combination of optical and IR light curves to test whether the best-fit parameters remained consistent over time and to lessen the impact of the gaps between observation seasons. The models corresponding to the highest posterior distribution are displayed in Figure~\ref{fig:individ_seasons}, and the means and their 1$\sigma$ uncertainties for each parameter are listed in Table~\ref{tab:season_params}. 

Overall, the simulated light curves for the individual seasons are shown to fit the observations quite well, and the best-fit parameters found are largely consistent with each other and the results found from modelling multiple seasons. However, many of the parameters are not constrained well; specifically, the radial power-law indices and the vertical scale height power-law indices are often returned with uncertainties that cover nearly the entire prior range, and the inclination angles are often found with 1$\sigma$ uncertainties that cover 20$^\circ$-- 40$^\circ$. This is thought to be due to the length of the observation seasons not being sufficient to properly constrain a best-fit DTF.

\begin{figure*}
    \begin{minipage}{\textwidth}
        \subfloat[figure][Comparison between the mean optical-IR amplitude conversion factors. \label{fig:comp_weffs}]{\includegraphics[width=0.49\textwidth]{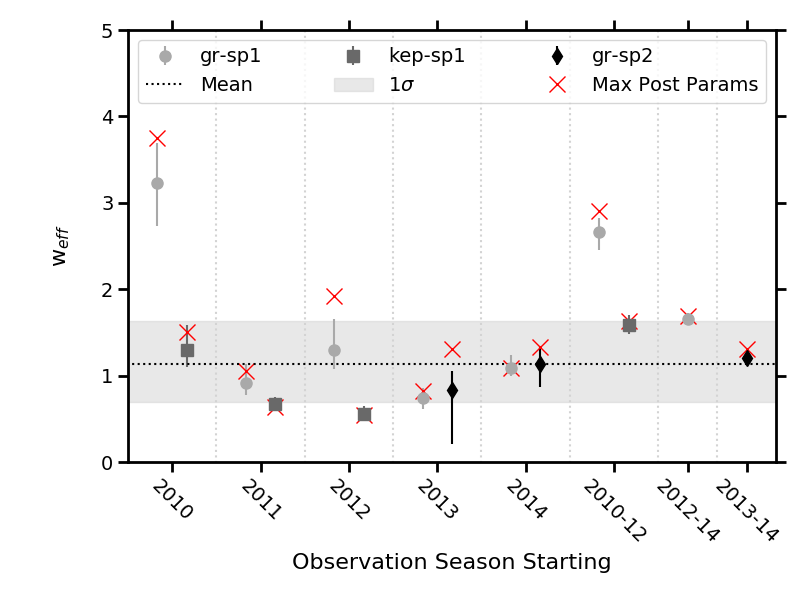}} 
        \hfill
        \subfloat[figure][Comparison between the mean additional offsets. \label{fig:comp_offsets}]{\includegraphics[width=0.49\textwidth]{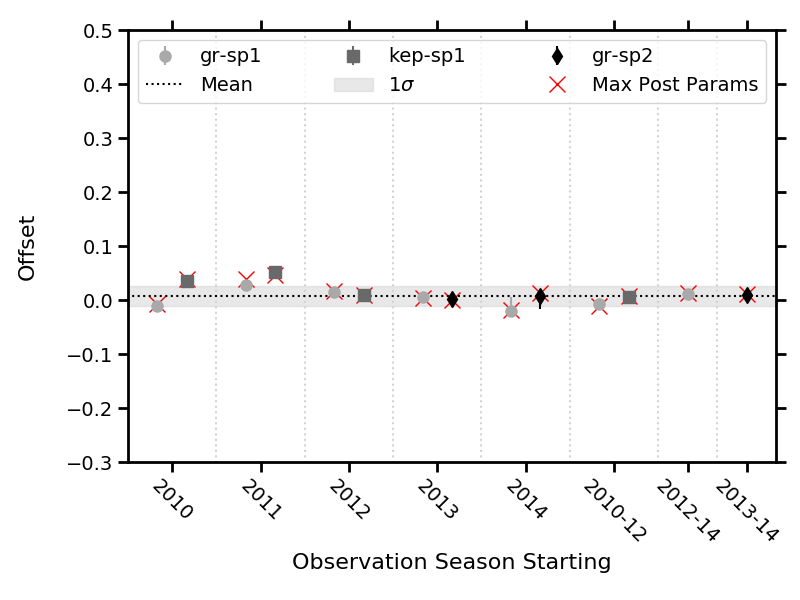}} 
        \caption{Comparison between the mean output parameters of the MCMC modelling of each combination of optical and IR light curves, for the individual observation seasons and multiseason light curves. The values corresponding to the maximum posterior distribution are also plotted in red.} \label{fig:comp_params_extra}
    \end{minipage}
\end{figure*}

\begin{figure*}
    \begin{minipage}{\textwidth}
        \centering
        \subfloat[figure][Distribution of the lag found modelling the ground-{\it Spitzer}~2 light curve in the season starting 2014.]{\includegraphics[width=0.49\textwidth]{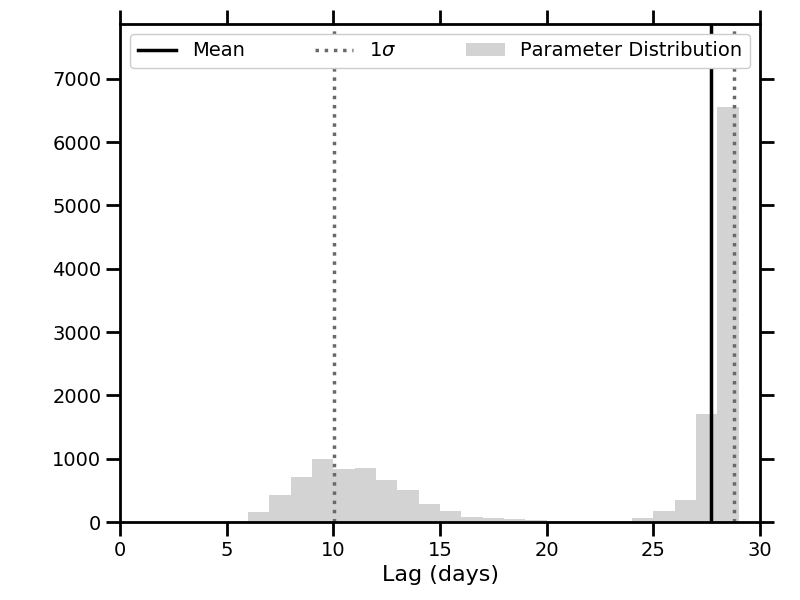}} 
        \\

        \subfloat[figure][Simulated light curves for ground-{\it Spitzer}~2 in the 2014 season, plotted with $\alpha=-0.59$, $\beta=0.17$, $\tau=11.28$\,days, and $i=64.90^\circ$.]{\includegraphics[width=0.49\textwidth]{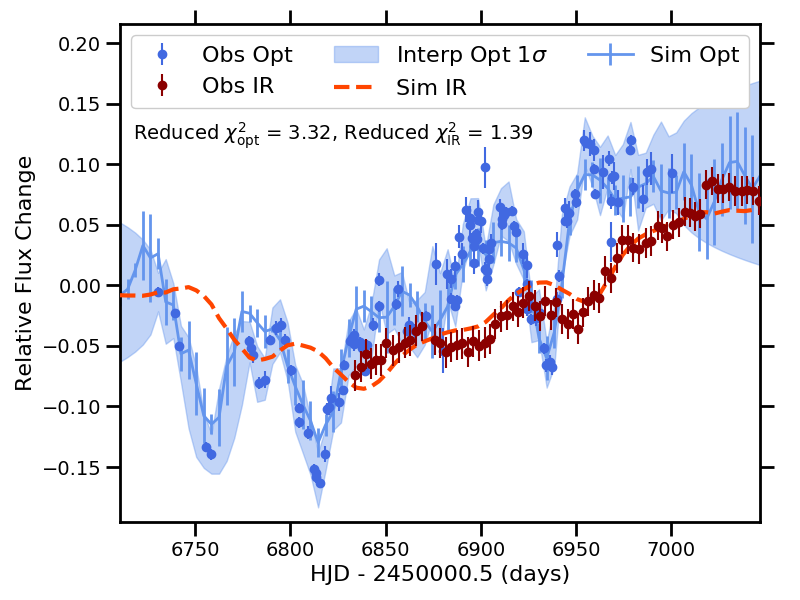}} 
        \hfill
        \subfloat[figure][Simulated light curves for ground-{\it Spitzer}~2 in the 2014 season, plotted with $\alpha=-0.62$, $\beta=0.19$, $\tau=28.73$\,days, and $i=56.22^\circ$.]{\includegraphics[width=0.49\textwidth]{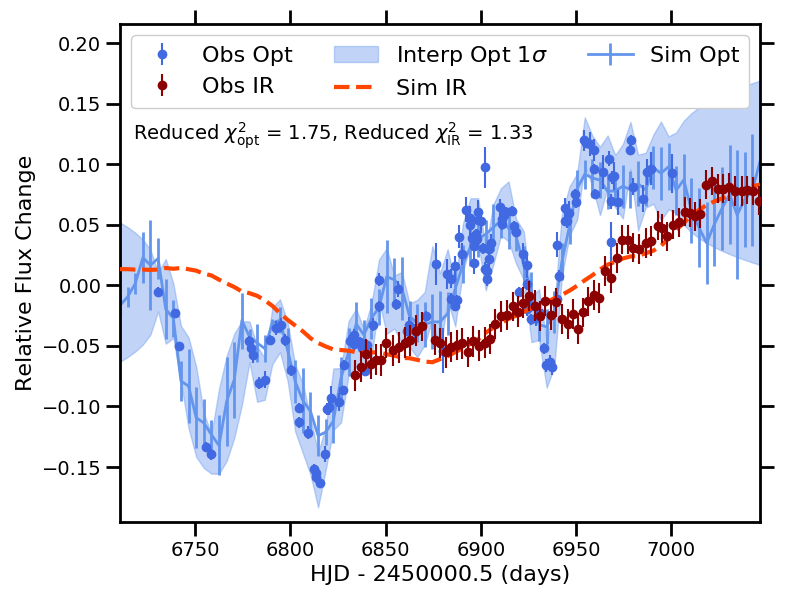}} 
        \\
        \captionof{figure}{The distribution of time lags found modelling the ground-{\it Spitzer}~2 light curve in the 2014 seasons, and the models corresponding to each of the peaks found in the distribution. \label{fig:gr_sp2_S5_lag_distrib}}
    \end{minipage}
    \vspace{0.3cm}
    \begin{minipage}{\textwidth}
        \captionof{table}{Mean output parameters corresponding to the best-fit DTFs found by MCMC modelling the individual observation seasons, separated into the lower and higher values of lag found for the ground-{\it Spitzer}~2 light curves in the individual observation seasons, and the quality of their fits described by reduced $\chi^2$. In this table, values of the offset have been multiplied by $10^{2}$. \label{tab:multi_season_params}}
        \begin{tabular}{C{0.09\textwidth}C{0.07\textwidth}C{0.08\textwidth}C{0.1\textwidth}C{0.08\textwidth}C{0.08\textwidth}C{0.08\textwidth}C{0.08\textwidth}C{0.06\textwidth}C{0.06\textwidth}}
                \hline
                Light-Curve Combination & Season Starting & Radial Power-Law Index & Vertical Scale Height Power-Law Index & Amplitude Conversion Factor & Lag (days) & Inclination (degrees) & Offset & $\chi_\textit{opt}^2$ & $\chi_\textit{IR}^2$  \\
                \hline

               {Gr-Sp2 ($\tau < 20$\,days)} & 2013  &  $-2.57^{+1.089}_{-1.31}$ & 0.80$^{+0.49}_{-0.53}$ & 0.21$^{+0.02}_{-0.01}$ & 14.40$^{+1.81}_{-2.04}$ & 35.74$^{+7.55}_{-13.24}$ & 0.09$^{+0.18}_{-0.18}$ & 4.05 & 8.44 \\

               {Gr-Sp2 ($\tau > 20$\,days)} & 2013  &  $-0.53^{+0.02}_{-0.07}$ & 0.08$^{+0.26}_{-0.02}$ & 0.97$^{+0.12}_{-0.19}$ & 28.47$^{+0.42}_{-1.21}$ & 53.09$^{+8.54}_{-4.76}$ & 0.13$^{+0.17}_{-0.20}$ & 4.00 & 5.14 \\

               \hline
               
               {Gr-Sp2 ($\tau < 20$\,days)} & 2014 & $-0.57^{+0.05}_{-0.52}$ & 0.17$^{+0.35}_{-0.09}$ & 0.89$^{+0.07}_{-0.09}$ & 10.08$^{+2.60}_{-1.74}$ & 60.13$^{+7.38}_{-17.63}$ & 1.51$^{+0.76}_{-0.95}$ & 3.32 & 1.39 \\

               {Gr-Sp2 ($\tau > 20$\,days)} & 2014 & $-0.63^{+0.10}_{-0.17}$ & 0.12$^{+0.18}_{-0.06}$ & 1.24$^{+0.12}_{-0.27}$ & 28.36$^{+0.50}_{-2.86}$ & 58.68$^{+8.41}_{-11.07}$ & 1.13$^{+0.47}_{-0.95}$ & 1.75 & 1.33 \\

               \hline
        \end{tabular}
    \end{minipage}
\end{figure*}

Furthermore, some observation seasons find different results depending on the combination of optical and IR light curves used. Specifically, the ground-{\it Spitzer}~2 light curves differ from the ground-{\it Spitzer}~1 in the seasons starting in 2013 and 2014, as the ground-{\it Spitzer}~1 light curves find a lag of $\sim 10$\,days while the ground-{\it Spitzer}~2 light curves find a larger lag of $\sim 30$\,days and a larger inclination angle. However, the lags in these models are not as well constrained as the earlier seasons, returning 1$\sigma$ uncertainties of $\sim 15$-20\,days. These larger uncertainties occur because the distribution of the lags found are actually double peaked at $\sim 10$ and 30\,days, as shown in Figure~\ref{fig:gr_sp2_S5_lag_distrib} for the ground-{\it Spitzer}~2 2014 season model, for example. When separating the models into those corresponding to lags greater than and less than 20\,days, the model corresponding to the lower lag is shown to better replicate the shape of the variations of the IR observations, but the higher-lag model returns a lower value of reduced $\chi^2$, especially in the optical light curve. 

Additionally, in Figure~\ref{fig:comp_params_extra}, comparison plots of the output parameters of the optical-IR amplitude conversion factor, and for the additional offset are displayed. It can be seen that the amplitude conversion factor typically returns mean values between 0.5 and 1.5, however the ground-\textit{Spitzer} 1 models in the 2010 and 2010-2012 seasons return values greater than 2, which could be a result of the model attempting to fit the IR light curve between HJD 55425 - 55450. In the corresponding \textit{Kepler}-\textit{Spitzer} 1 model, it can be seen that this portion of the simulated IR light curve does not match observations.

\bsp	
\label{lastpage}
\end{document}